%% file: paper.tex
\documentclass[fleqn,usenatbib]{mnras}

\usepackage{mathptmx}
\usepackage[dvipsnames]{xcolor}

\usepackage[T1]{fontenc}
\usepackage{ae,aecompl}

\usepackage{graphicx}	
\usepackage{amsmath}	
\usepackage{amssymb}	
\usepackage{lipsum}
\usepackage{times}

\usepackage{enumitem}

\title[CEEs with low-mass giants: transiting to the slow spiral-in]{Common envelope events with low-mass giants: understanding the transition to the slow spiral-in}

\author[N.~Ivanova \& J.L.A.~Nandez]{
N. Ivanova,$^{1}$\thanks{E-mail: nata.ivanova@ualberta.ca}
J.L.A. Nandez.$^{1,2}$
\\
$^{1}$University of Alberta, Edmonton, AB, T6G 2R3, Canada\\
$^{2}$SHARCNET, The University of Western Ontario, London, Ontario, N6A 5B7 \\
}

\date{Accepted XXX. Received YYY; in original form ZZZ}

\pubyear{2015}

\begin{document}
\label{firstpage}
\pagerange{\pageref{firstpage}--\pageref{lastpage}}
\maketitle

\begin{abstract}

\input{abstr}
\end{abstract}





\input{sect1_intro}


\input{sect2_definitions}


\input{sect3_mapping}


\input{sect4_energies}


\input{sect5_am}


\input{sect6_entr}


\input{sect7_rec}


\input{sect8_ejecta}


\input{sect9_discussion}

\section*{Acknowledgments}

NI thanks NSERC Discovery and Canada Research Chairs Program. 
JLAN acknowledges CONACyT for its support. 
The authors thank their colleagues, C.~Heinke, S.~Morsink, E.~Rosolowsky, and G.~Sivakoff  
for checking the English in the manuscript.
We thank an unknown referee for a number of helpful remarks that helped to improve the manuscript.
Computations were made on the supercomputer Guillimin from McGill University, managed by Calcul Qu\'ebec and Compute Canada. The operation of this supercomputer is funded by the Canada Foundation for Innovation (CFI), Minist\'ere de l'\'Economie, de l'Innovation et des Exportations du Qu\'ebec (MEIE), RMGA and the Fonds de recherche du Qu\'ebec - Nature et technologies (FRQ-NT).

\bibliographystyle{mn2e}
\bibliography{comenv}

\input{appendix_am}

\bsp	
\label{lastpage}
\end{document}

%% file: abstr.tex
We present a three-dimensional (3D) study of common envelope events (CEEs) to provide a foundation for future one-dimensional (1D) methods to model the self-regulated phase of a CEE. The considered CEEs with a low-mass red giant end with one of three different outcomes --  merger, slow spiral-in, or prompt formation of a binary.  To understand which physical processes determine different outcomes, and to evaluate how well 1D simulations model the self-regulated phase of a CEE, we introduce tools that map our 3D models to 1D profiles. We discuss the differences in the angular momentum and energy redistribution in 1D and 3D codes.  We identified four types of ejection processes: the pre-plunge-in ejection, the outflow during the plunge-in, the  outflow driven by recombination, and the ejection triggered by a contraction of the circumbinary envelope. Significant mass is lost in all cases, including the mergers.  Therefore a self-regulated spiral-in can start only with a strongly reduced envelope mass.  We derive the condition to start a recombination outflow, which can proceed either as a runaway or a stationary outflow.  We show that the way the energy of the inspiraling companion is added to the envelope in 1D studies intensifies the envelope's entropy increase, alters the start of the recombination outflow, and leads to different outcomes in 1D and 3D studies.  The steady recombination outflow may dispel most of the envelope in all slow spiral-in cases, making the existence of a long-term self-regulated phase debatable, at least for low-mass giant donors.

%% file: sect1_intro.tex
\section{Introduction}

A common  envelope event (CEE)  is an episode in the life of  a binary
system during  which the outer  layers of one  of the stars  expand to
engulf the companion -- thus  producing an envelope around both stars,
or  a common  envelope (CE).
  The concept was originally
proposed almost 40 years ago by \citet{Pacz76} (this publication
cites private communication with Ostriker as well as \citet{1975PhDT.......165W}, for the origin of this idea).
The concept was later developed into the modern state of the energy formalism in \citet{Webbink84}
  and \citet{livio88}.
This  brief but  crucial episode  leads
either to a complete merger, or to  the expulsion of the CE, leaving a
drastically   shrunk  (short-period)   binary  --   a  likely   future
gravitational wave source,  or X-ray source,  or SN Ia progenitor, etc.
The merged  star can form  a variety of  exotic objects, or  produce a
long gamma-ray burst,  one of the most luminous events  known to occur
in the  Universe \citep[for a recent review on the current understanding of the CE, as well as    overall importance for binary populations and applications,  see][]{Ivanova+2013Review}. It is  widely accepted that a  CEE is the  {\it main
  mechanism} by which an initially  wide binary star is converted into
a very close binary star, or by which two stars merge.

Despite being vital for understanding a vast number of important binary systems, CEE
theory  is presently  very poorly  understood.  Brute-force  numerical
simulations of CEEs are hard \citep{1996ApJ...471..366R,1998ApJ...500..909S,Passy11,Ric2012,2014ApJ...786...39N,2015MNRAS.450L..39N,2016ApJ...816L...9O};
often, key elements of the event proceed
not only on  a dynamical timescale, but also on  a thousand to million
times longer thermal time-scale.  Various physical processes including
radiation transfer,  convective energy  transport and  nuclear burning
can take place during a CEE \citep{Ivanova+2013Review}.  Understanding
the transition  between a fast  CEE that can  only be modelled  using a
three-dimensional  (3D) hydrodynamical  code with  simplified physics,
and a slow  CEE that can only be modelled  with a one-dimensional (1D)
stellar  code,  where various  mechanisms  for  the energy  transport,
nuclear  energy  generation  and   other  physics  are  included,  was
advocated to  be the most important  step for further progress  in CEE
studies  \citep{Ivanova+2013Review}.   Several  research  groups  have
recently restarted  to carry out 1D  studies of slow  CEEs \citep[e.g.,][also
  Podsiadlowski,  priv.  comm.]{2015IAUGA..2250783M,2015PhDT.........5H,2015IAUGA..2255902L,2011HEAD...12.0305L}.   Determining or
constraining the  initial conditions  for 1D  CEEs, starting  from the
moment  when  the dynamical  phase  ended,  is crucially  needed.   In
addition,    1D    simulations,    starting    from    the    pioneers
\citep{Taam+1978,1979A&A....78..167M},  are  naturally  forced  to  use  either
assumptions or  prescriptions, which have not yet been verified  against 3D
simulations. For  example, prescriptions for the angular  momentum transfer or
its distribution, or  how the released orbital  energy is deposited
in the envelope.

Understanding the initial conditions at the start of the slow spiral-in, however, can not
  be done without understanding how the dynamical phases proceeded: 
  how the energy redistribution between the orbit and the envelope
  takes place in 3D, as compared to what 1D treatments we adopt,
  whether any envelope material was lost prior the slow spiral-in,
  and how important quantities (e.g., specific angular momentum and
  entropy of the material) have evolved by the end of the rapid evolution.

In this paper,  we study the transition to the slow spiral-in by analyzing
3D simulations of  CEEs with a low-mass
giant donor and companions of  several different masses.  The outcomes
of those CEEs vary from a rapid  and complete envelope ejection  to the
strong  binary   shrinkage  that  can   be  indicative  of   a  merger
(\S~\ref{sect:classif}).  We are most  interested in understanding the
start of the intermediate regime,  when the binary orbital dissipation
is not dynamical anymore, but the envelope has not yet been ejected --
the regime that  1D codes are best suited to  model.

To provide  a transition  between our results  and 1D  simulations, we
discuss  several ways to  map the  3D CEE  simulations to  1D
(\S~\ref{sect:map}).  We  compare the  results of  this mapping  
 with  the  results  of  the  simplified
assumptions and  prescriptions that 1D  codes have used to study  slow CEEs.
We also use  our mapped simulations to analyze  the physical processes
during a transition to a slow spiral-in, to  find clues about what leads  to the branching of
the outcomes of  CEEs, to consider how strong the asymmetry of
the CE at the  start of a slow CEE is, and to  discuss the challenge with
density inversion that a 1D approach may have.

We  provide a  detailed technical  description of  how energies  are
considered in the 1D and 3D  approaches (\S~\ref{sec:en}). This  leads to
a discussion of how the two kinds  of energies can be  compared between
the approaches and whether the simplified  prescription is valid during the transition from fast  to 
 the slow spiral-in.  We  also provide an equation, and its
limitations, for how  energy conservation
should be used  in the 1D approach after the start of the slow spiral-in.
We analyze the angular momentum redistribution
in 3D simulations and compared the results to  existing   1D   approaches (\S~\ref{sec:am}).

We discuss in \S~\ref{sec:entropy} how entropy generation  is different between the 3D and
1D  approaches.  We argue that the currently-used
1D treatments of energy conservation via ``heating'' lead to a different outcome
compared to 3D simulations. In particular, the usual 1D treatment
  leads to a different entropy profile at the start of the slow spiral-in.
We discuss next how recombination  governs  the  CE  ejecta
(\S~\ref{sec:rec}).   We then  advocate that
artificially  forced entropy  generation, coupled  with recombination,
may produce  unrealistic outcomes in  1D codes.  Finally,  we consider
the  ejecta, identify  several  types  of   ejecta at various  CEE
phases,  and   discuss  the physics  behind   each  type  of  ejecta
(\S~\ref{sec:ej}).   We conclude  with several recommendations  for 1D
codes that can improve the modeling of CEEs in 1D (\S~\ref{sec:disc}).

%% file: sect2_definitions.tex
\section{Classification of CEE outcomes }

\label{sect:classif}

\input{table1}

\subsection{{Definitions of CEE phases}}

In this  paper, we adopt the  definitions of the distinct  phases of a
CEE as  described by \cite{podsi01}, see also the thorough details in
the   review  by   \cite{Ivanova+2013Review}.    For   a  more   {\it
  quantitative} differentiation  of the  phases, we indicate  here the
typical rates of the orbital dissipation:

{\bf I.   Loss of corotation.}  During this  phase, the change  in the
orbital separation $a$ is less than one per cent over the orbital period
$P_{\rm  orb}$, $|(\dot  a P_{\rm  orb})|  /a <  0.01$. The  companion
orbits either outside of the future  CE, or inside the CE's outer expanded
and rarefied layers.

{\bf  II.  Plunge-in.}   This  is  the fastest  phase  of the  orbital
shrinkage, when the rate of change  of the orbital separation is large,
$|(\dot a P_{\rm orb})|/a \ga  0.1$.  The companion plunges inside the
CE, and, at the  end of the phase, most of the  CE mass is outside
the companion's orbit.  During the  plunge-in, the concept of a Keplerian binary
orbit is equivocal. 

{\bf III. Self-regulating spiral-in}. During  this stage the change in
the orbital  energy $  E_{\rm orb}$  is small, $  | (\dot  E_{\rm orb}
P_{\rm orb}) /  E_{\rm orb}| < 0.01$. The companion  orbits inside the
CE.

With a  3D code, a  proper self-regulating  regime is not  possible to
achieve --  a 3D code  typically lacks  consideration of  energy
transport that operates on a thermal timescale
\citep[e.g., convection, see also the recent steps in 
the treatment of convective envelopes in ][]{2016ApJ...816L...9O},
among other effects.
 Therefore in this paper we  will rather consider the {\it initial phase}
of the  self-regulating spiral-in, to which  we will refer here  as 
the {\bf  ``slow spiral-in''}.

We define that the slow spiral-in starts with a similar
  criterion as the self-regulating spiral-in, i.e. when $| (\dot
  E_{\rm orb} P_{\rm orb}) / E_{\rm orb}| < 0.01$.  In our
  simulations, this stage is modelled for the first few
dynamical timescales of the expanded envelope after the
plunge-in has ended.  On the other hand its duration can be
  compared to about several thousands orbital periods of the formed
  close binary.  A well-established slow spiral-in -- thousands binary
  orbits after the start of the slow spiral-in -- could be quite
  different from the start of the slow spiral-in for the region around
  the binary. This is becuase this region's dynamical timescale is
  comparable to the binary period, but has evolved for many thousands of
  the binary periods.  On the other hand, it is still in the initial
  phase of the self-regulated spiral-in as the modelled time is smaller
  than the thermal timescale of the envelope.

For the definition  of phase  III  above (as well as for the slow spiral-in), we  use  the orbital  energy
dissipation  as  an  indicator,  rather than  the  orbital  separation
decrease.  This is  because  in  our code  during  this period  the
orbital energy of  the now close binary is nearly  constant, as expected,
but the  binary has  a non-zero, albeit  small, eccentricity,  and the
decrease of its  average orbital  separation   is  harder to  obtain
numerically. Note that due to the adopted quantitative classification of the
  principal phases as described above, there are also transitional phases:
  between the end of the corotation and the start of the plunge-in,
 and between the end of the plunge-in and the start of the slow spiral-in.

\subsection{{Numerical setup of simulations}}

 {  To study  a  CEE from  before the Roche lobe overflow  and to  a
   well-established  slow  spiral-in,  we  use  {\tt  STARSMASHER},  a
   Smoothed Particle Hydrodynamics (SPH) 3D code \citep[see instrument
     papers][where  the method  and  relevant tests  are described  in
     detail]{2006ApJ...640..441L,2010MNRAS.402..105G,Lombardi11}.  An
   SPH approach is appropriate to  model the interaction between two stars
   without imposing boundary  conditions, and is commonly  used in the
   community  for  this class  of  problems  \citep[e.g., see  various
     implementations                                              used
     in][]{1993A&A...272..430D,1996ApJ...471..366R,2009MNRAS.395.1127C,Passy11}.
   The code {\tt STARSMASHER} was originally
   developed to  treat interactions  between two  stars, and  has been
   frequently   used   in   studies   of  this   class   of   problems
   \citep{2010MNRAS.402..105G,Iva10,2011ApJ...731..128A,Lombardi11,2015ApJ...806..135H,2014ApJ...786...39N,2015MNRAS.450L..39N,2016arXiv160207698P}.
   As  an  example, one  of  {\tt STARSMASHER's} important features,
   developed specifically  for studies  of binaries, is  the specially
   designed   relaxation  setup   of   a  close-to-contact   binary
   \citep{Lombardi11}.  This  relaxation procedure  minimizes spurious
   effects of  artificial viscosity that  may affect the start  of the
   spiral-in.  }

 {\tt STARSMASHER's} internal physics has been recently upgraded to
 take into account recombination processes
 \citep{2015MNRAS.450L..39N}.  { The modification was done by
   replacing the code's default equation of state that includes ideal
   gas and radiation pressure \citep[][]{2006ApJ...640..441L} by the
   tabulated equation of state from MESA that accounts for states of
   ionization \citep[][also see more details on which elements are
     taken into account for recombination in
     \S~\ref{sec:rec}]{2011ApJS..192....3P}.  The version of SPH code
   we use evolves specific internal energy of an SPH particle $u_i$
   and density of an SPH particle $\rho_i$ \citep[see Eq. A18 and A7
     of ][]{2010MNRAS.402..105G}, and pressure then is found from the
   internal energy, density, and the adopted equation of state.}  It
 is this implementation {of the more complete equation of state
   that has enabled modelling of the complete CE ejection , for the
   first time.}
 
{ For this  study  of the  branching  of CEE
  outcomes, we chose for a donor}  a low-mass  red  giant with  a  mass of
$1.8~M_\odot$, a core mass  of $0.318~M_\odot$, and  radius  $16.3~R\odot$ (for
more details on the ambiguity in the definition of the donor radius in
3D,     see       the    discussion     in
\citealt{2014ApJ...786...39N}).
{To create  the initial red giant  donor star,  we use the {\tt  TWIN/EV} stellar
code   \citep[][   recent    updates   are    described   in
  \citealt{2008A&A...488.1007G}]{1971MNRAS.151..351E,1972MNRAS.156..361E}.}

To obtain  different categories of CEE  outcomes, we varied the  mass of
the companion,  $M_{\rm comp}$,  considering $0.36, 0.20,  0.15, 0.10$
and  $0.05 M_\odot$  companions.   At the  start  of the  simulations,
donors  in  all binaries  are  within  their  Roche Lobes  (RLs).   We
consider the  case of non-synchronized donors.   While synchronization
has a small  effect on the outcomes   \cite[for a discussion
  of how  small the effect  of synchronization on the  outcomes is, and
  that  it  mainly affects  the  energy  carried  away by  the  ejecta, 
  see][]{2014ApJ...786...39N}, we chose to start with non-synchronized
binaries   as  it   significantly   speeds  up   the   start  of   the
interaction.

The red giant envelope is modelled with $10^5$ particles.
 The red giant core and the companions are modelled as point masses, where
 a point mass only interacts gravitationally with normal SPH particles.
 Such point masses are also referred to as special particles.
 {In an SPH code, the gravitational potential
   equation contains an extra smoothing term: the smoothing length $h_i$
   \citep[see the Appendix of ][]{1989ApJS...70..419H}.
   In our simulations, for $10^5$ particles, a smoothing length for the red giant core is $0.35 R_\odot$
   \citep[for details on how the smoothing length is determined
     for special particles in {\tt STARSMASHER}, see][]{Lombardi11}.
 }
 
In previous studies, a larger number of particles has been used to represent the donor 
  \citep[e.g., in][ a CEE was modelled using 5 times more SPH particles]{Passy11}.
  However, the modelled {\it phase} of a CEE was substantially shorter.
  For example,  \cite{Passy11},  have   started  their  simulations  with   placing  the 
  companion  on  the surface  of  the  donor,  and have  finished  their
  simulations at  the moment that we  define as the start  of the slow
  spiral-in.  A similar phase in  our simulations takes only between 1
  and 2 per cent  of the computational time.   About  10  per   cent  of  the
  computational time is  spent before the plunge-in  starts, and about
  90 per cent of the computational  time is spent after the plunge-in,
  until  the  slow spiral-in  is  ``well-established.''
  {The simulations take on average $5\times10^6$ time-step integrations.
    This allows us to follow about 15,000-35,000 binary periods after the end of the plunge-in.
   The physical timescale of the simulations is about 1000 days.}
  While it would be great to model a donor at  better resolution, it is not
  computationally feasible yet to get both a long-term evolution of a CEE, and model it 
  with a resolution substantially larger than $10^5$.

 {The resolution test for {\tt STARSMASHER} with the original equation of state 
   was performed by \cite{2015ApJ...806..135H}. As compared to that version of the code,
   only the equation of state has been changed.}
 \cite{2015MNRAS.450L..39N} carried out {\tt STARSMASHER} CEE simulations
with $10^5$ and $2\times10^5$ particles in an attempt to test resolution
effects. The two simulations were done for the version of the code
that includes the recombination physics. The test has shown that the
final orbital separation varies by only a few per cent and that both
simulations produce a similar envelope ejection (the ejected mass mass,
the timescale of the ejection, etc).
While doubling resolution is not sufficient to thoroughly test convergence,
this is the most that can be achieved at present\footnote{{A run with
    $8\times10^5$ particles would require 8-10 GPU years with  NVIDIA driver M2070, as was tested.
    The use of K40, which is currently the fastest NVIDIA driver in the world
    would reduce GPU time by 25 per cent.
    Given the low communication speed
    between the GPU nodes (hardware limitation, and thus the
    scaling is only effective up to 4 GPUs and 16 CPUs),
    and the cumbersome queue setup, this run would require more than 2 years of 
  waiting time in the real world, if started at the available Westgrid GPU clusters.}}. 
  This test suggests that most  phenomena
  that are discussed in this manuscript are not likely to be rebutted
  by a larger resolution run.   We warn however that some results presented in this
  paper should  be taken  with caution,  as future  studies made with a
  substantially larger resolution may negate the phenomenon that
  is  produced by  a small  number of  particles --  specifically, the
  shell-triggered ejection discussed later in \S~\ref{sec:ej}.

The  orbital energies  and the total  energies of  the CE
systems for  each companion at the start of the simulations
are  shown in  Table~\ref{tab1}.   
The donor's
envelope  has an  initial  total  binding energy  $-4.4\times10^{47}$
ergs,  an  initial  potential energy  $-8.8\times10^{47}$  ergs,  an
initial thermal energy $4.4\times10^{47}$ ergs (without recombination
energy),  and additionally  $4.7\times10^{46}$ ergs  is stored  as 
recombination energy (see more details on how the recombination energy is found in \S~7).

We distinguish  the unbound envelope material,  $M_{\rm ej}$, and
the currently bound  envelope material, $M_{\rm env}$.
These masses  are found using the  technique described in  \cite{2014ApJ...786...39N}.
Principally,  if an  SPH particle
has  negative total  energy,  it is  bound.  If  an  SPH particle  has
positive total  energy, it is unbound  and belongs to the  ejecta.
In Table~\ref{tab1}, we provide the final  values  at the end of the
simulations, but  note that the mass  of the ejecta can  not be simply
explained with one  number, and more details about the  ejecta will be
given in \S~\ref{sec:ej}.

During  a CEE,  a  CE system  can  be described  in  terms of  various
energies: the binding energy of  the envelope $E_{\rm bind, env}$, the
internal energy of the envelope  $U_{\rm in,env}$, the thermal energy
of  the envelope  $E_{\rm th,env}$,  the recombination  energy of  the
envelope  $E_{\rm  rec,env}$, the  potential  energy  of the  envelope
$E_{\rm  pot,env}$,  the  kinetic  energy  of  the  envelope,  $E_{\rm
 kin,env}$, the orbital  energy of  the binary  $E_{\rm orb}$,  the total
energy  of the  ejecta $E_{\rm  tot,ej}$, and the kinetic  energy of  the
ejecta $E_{\rm  kin, ej}$. We can  also trace the angular  momentum of
the envelope $J_{\rm env}$, the orbital angular momentum of the binary
$J_{\rm orb}$  and the  angular momentum of  the ejecta  $J_{\rm ej}$.
Details of how  those quantities are obtained from  our 3D simulations
can be found in \S~\ref{sec:3den} and Appendix \ref{ap:am}.  Values of
the most  important quantities  at the  start and  the end  of each
simulation can be found in Table~\ref{tab1}.  Note that in Table~\ref{tab1}, energies
have  the index  ``3D'',  as  they   are  obtained  assuming  3D  energy 
definitions (see  \S~\ref{sec:3den}), and can be  different from those
inferred by the definitions of the 1D approach (see \S~\ref{sec:1den}).
 We clarify that in all our
 simulations presented  in this paper  the total angular  momentum and
 the total energy  are conserved (the error on  energy conservation is
 less than 0.1\% of the initial total energy, and the error on angular
 momentum  conservation is  less  than 0.001\%  of  the initial  total
 angular momentum).

\subsection{{CEE outcomes}}

We classify the outcomes of our 3D simulations of CEEs as:

$\bullet$~{\bf Binary formation}  -- if the CE is  ejected and $|(\dot
E_{\rm orb} P_{\rm orb}) / E_{\rm orb}| < 0.01$.

$\bullet$~{\bf Slow  spiral-in} -- if the CE has not been fully ejected and  no further rapid mass outflow  of the
envelope material is  taking place on a timescale  longer than a few 
dynamical timescales of  the expanded CE  (although,  see  
\S~\ref{sec:ej} about the shell-triggered ejecta). 
During this stage, the orbital energy release is decreased to $|(\dot E_{\rm
  orb}  P_{\rm  orb})  /  E_{\rm  orb}|  <  0.01$.

$\bullet$~{\bf  Merger}  --  if   the  orbital  separation  is  $<0.15
R_\odot$ (see discussion below on the ambiguity of a merger case).

\begin{figure*}
\includegraphics[width=85mm]{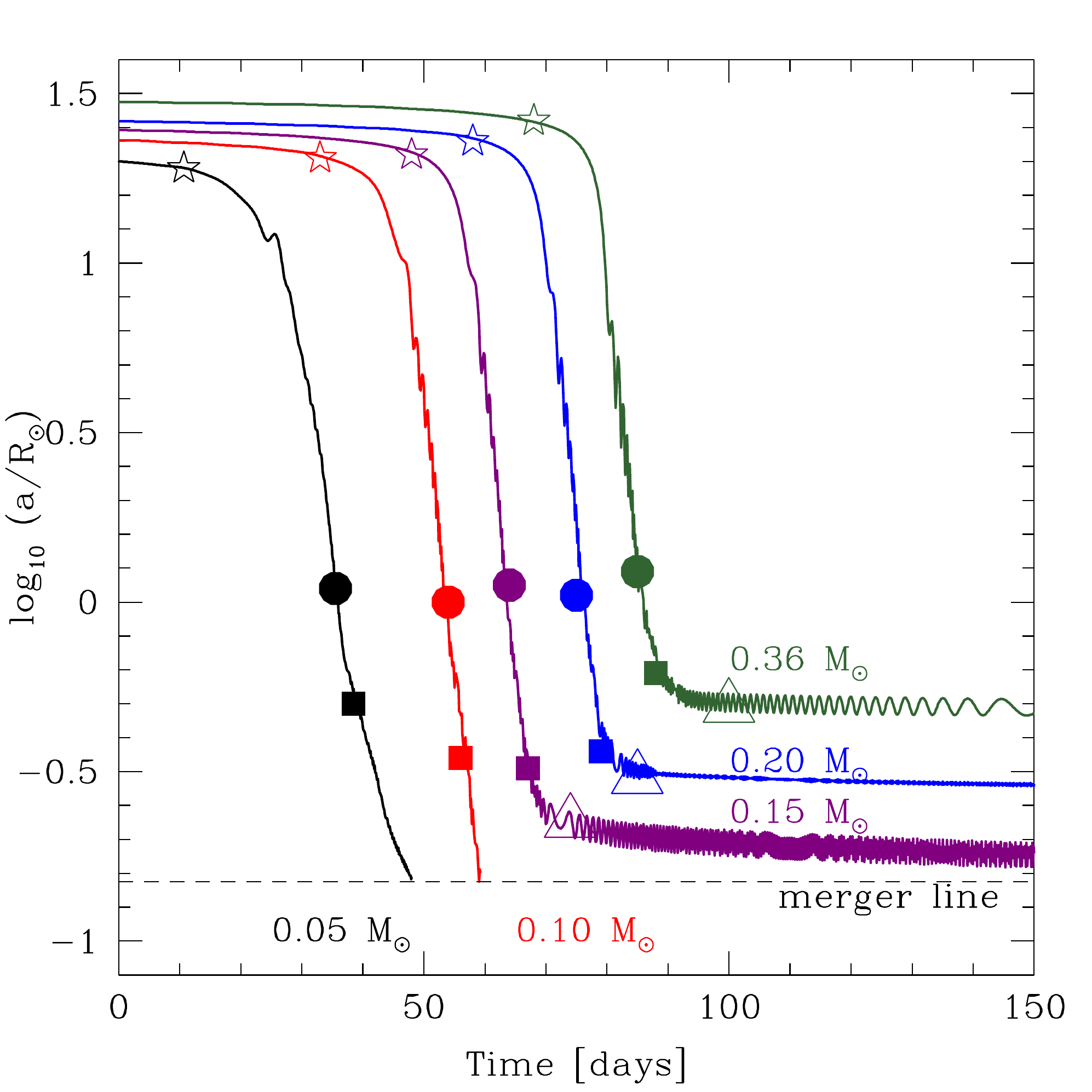}
\includegraphics[width=85mm]{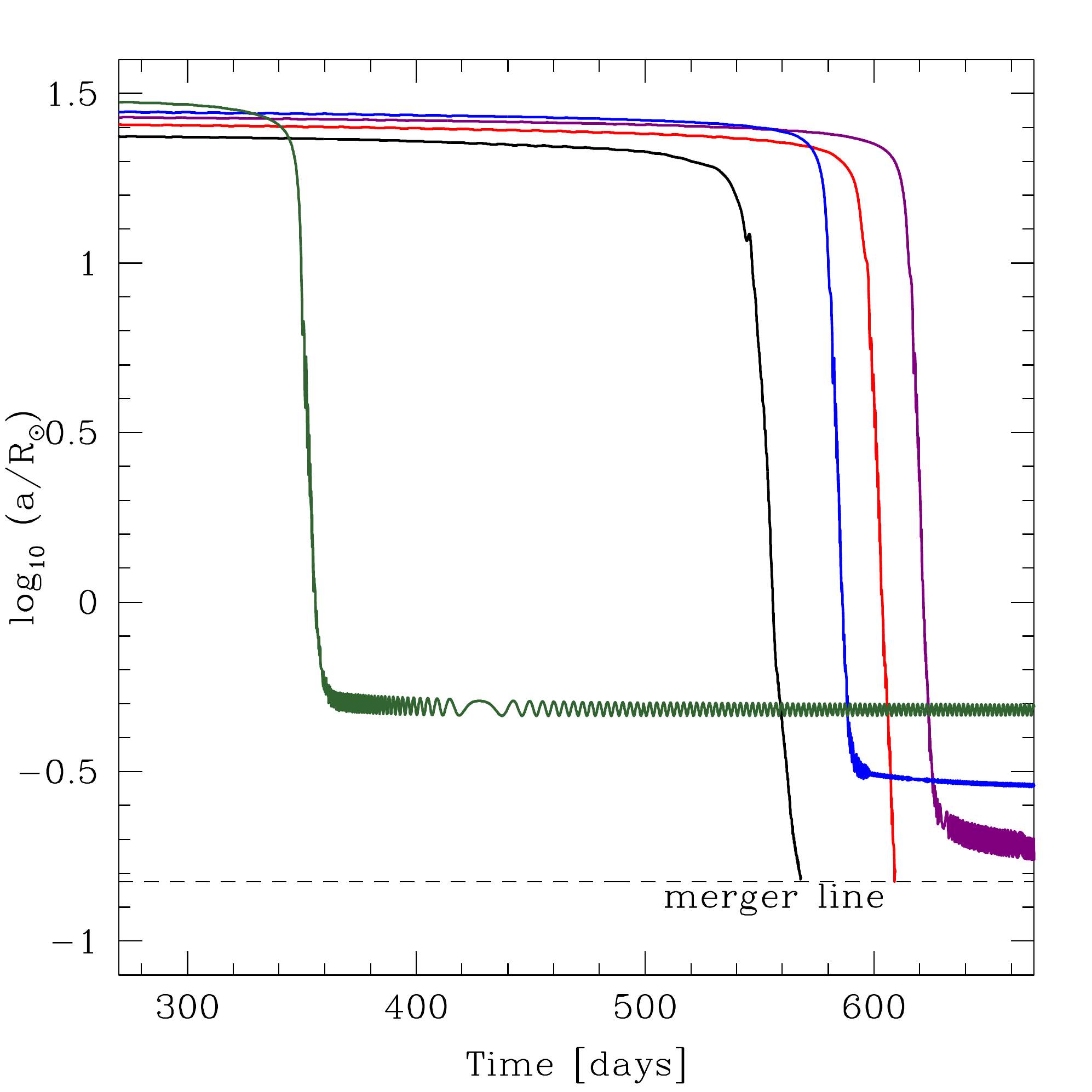}
\caption{Evolution of orbital  separations  in the  binaries with  $0.05,
  0.10,  0.15,  0.20$ and  $0.36 M_\odot$  companions. {On the left panel,} the time axis  is
  shifted to show the relative  orbital evolution in more detail for each system. The
  time shifts are $520, 550,  558, 510$ and $270$ days for $0.05,
  0.10,   0.15,   0.20$  and   $0.36   M_\odot$   companions,
  respectively. {On the right panel, the time axis is not shifted. Please note that the time that has passed
    from the start of the simulation to the start of the plunge-in depends strongly on the degree of the initial Roche lobe underflow, and that quantity is slightly different in the simulations.}
  The dashed line indicates  where the companion and the
  red giant  core   will  definitely  merge.   We  note  that   the  apparent
  non-periodic pattern of the orbital evolution, especially noticeable
  in the case  of the $0.36 M_\odot$ companion,  is because we can  only store
  every 10th model when we run a  long simulation. We have checked that if
  we  store every  model  for  a period  of time,  the
  apparent  pattern disappears,  and  during the  slow spiral-in  each
  binary has constant, albeit slowly
  decreasing,  orbital period,  as expected.  The orbital energy  does not
  oscillate, but due to the nonzero eccentricity, the orbital separation, measured at particular times, shows oscillations.
  {The start of the displayed orbital evolution on the left panel corresponds to the moment of time when the orbital decay is increasing to $|(\dot  a P_{\rm  orb})|  /a =  0.03$ for the  binary with  $0.05 M_\odot$  companion, $|(\dot  a P_{\rm  orb})|  /a =  0.01$ for the  binaries with  $0.10,  0.15$ and  $0.20 M_\odot$ companions, and $|(\dot  a P_{\rm  orb})|  /a =  0.005$ for the  binary with  $0.36 M_\odot$  companion. 
    For each case, the start of the plunge-in, when $|(\dot  a P_{\rm  orb})|  /a$ becomes greater than 0.1, is indicated  with a star symbol.
    The end of the plunge-in, when $|(\dot  a P_{\rm  orb})|  /a$ becomes less than 0.1, is indicated with a circle symbol.
    The start of the slow spiral-in, when $| (\dot  E_{\rm orb} P_{\rm orb}) /  E_{\rm orb}|$ becomes less than 0.01,  is indicated with a square symbol.
    A triangle symbols indicates when $| (\dot  E_{\rm orb} P_{\rm orb}) /  E_{\rm orb}|$ becomes less than $10^{-4}$.
  }
}
\label{fig:orbits}
\end{figure*}

In  Figure  \ref{fig:orbits} we  show  the  evolution of  the  orbital
separation for  all  cases.   We have obtained all  three possible
outcomes:  mergers (with  companions of mass $0.05$ and
$0.1M_\odot$), slow spiral-ins (with  companions of mass
 $0.15$  and $0.2M_\odot$)  and  binary  formation ($0.36M_\odot$
companion).  The quantities that describe the simulations are shown in
Table\ref{tab1}.  We provide these values when the separation
becomes $0.15 R_\odot$, or at 50 days after the end of the plunge-in.
For the binary formation case,  we also list in Table\ref{tab1} values
for  when the  entire  envelope  is lost  (about  700  days after  the
plunge-in has ended).

We also indicate in  Figure  \ref{fig:orbits} the start of the plunge-in,
the end of the plunge-in, and the start of the slow-spiral-in for all the cases.
For example, in the simulation with a $0.15~M_\odot$ companion,  at the end of the plunge-in -- when the orbital separation
stops changing quickly -- the binary separation is about three times larger
than during the ``well-established'' slow spiral-in (50 days after the end of the plunge-in),
and is about two times larger than at the start
of the slow spiral-in. The transition between
the end of the plunge-in and the start of the slow spiral-in lasts for only about 3 days, which is nonetheless about 100 binary orbits.
It takes another 400 binary orbits before the orbital decay decreases to $| (\dot  E_{\rm orb} P_{\rm orb}) /  E_{\rm orb}| \sim 10^{-4}$.
At about 50 days after the end of the plunge-in, or after about 3000 binary orbits, $| (\dot  E_{\rm orb} P_{\rm orb}) /  E_{\rm orb}| \sim 10^{-5}$.

For  the  case when  the  {\it  ``binary  formation''} occurs  in  the
simulation, half of the envelope  is ejected soon after  the end of
the plunge-in (on day 374  after the start of the simulation), and
it  takes about  700  more days  to  steadily eject  the  rest of  the
envelope.  We  note that  before the envelope  is fully  ejected, this
``binary  formation'' case  is  not substantially  different from  the
other two  cases that  we designate here  as {\it  ``slow spiral-in''}
cases,  except for the fact that  the  CEE  evolution  is   faster  with  the
$0.36~M_\odot$ companion. 
 
We call a {\it ``merger''} all simulated systems for which the
 geometrical separation  between the  core and  the companion  is $\la
 0.15  R_\odot$.  For a non-degenerate companion, this limit on mergers is
   naturally consistent with the radii of very low-mass stars.
For a degenerate companion, technically, the
simulations can be forced to run  until the  separation becomes $\la 0.01~R_\odot$.
However, there are other reasons to stop the simulation earlier.
  In an  SPH code, if  the distance between the two point  masses is less than
 two times their  smoothing lengths, then there is  an extra smoothing
 in     the    gravitational     potential    equation     \citep[see,
   e.g.][]{1989ApJS...70..419H}.  Then  the orbital energy is  not the
 same as would be found from their geometrical separation.  In addition,
 the orbital  energy depends  also on  the presence  of the  usual SPH
 particles  within  the  RLs  of  the  special  particles  (see  below
 \S~\ref{sec:3den} for  more details).  Note that  these effects take
 place while the total energy is conserved; the portions of the
 energy assigned  to the  binary, and  to the  envelope, become
 dependent  on  the  smoothing  length.  We  find  that  a  noticeable
 (about a few per cent) mismatch between the geometric orbital separation and the ``energy''-
 derived orbital separation starts to appear when the distance between
 the special particles is less than $\sim 0.15 R_\odot$.
{\citep[see also detailed discussion on this in][]{NandezIvanova15}}.
 Accordingly, the potential energy (by its absolute value) is not as large as 
 would be calculated purely by  the distance between the particles, if
 a Keplerian  orbit were assumed.  As we  have checked,  decreasing the
 smoothing length  by a factor of 2 and increasing the number of particles
by a factor of 8  improves the consistency, but  leads to
 an increase of  the computational time by a factor of 64.   This makes the
 problem  currently  computationally unfeasible.  The  CEEs with
 the companions  of $0.05  M_\odot$ and  $0.1 M_\odot$  have therefore
 been assigned to  be ``mergers'', while in Nature,  if the companions
 are compact, CEEs in these binaries  could result in a slow spiral-in
 with an orbit that is smaller than our cut-off distance  $0.15 R_\odot$.

In  the following  Sections, we  will  discuss the  differences in  the
processes that may lead to these three types of the outcomes.  We
will refer to the binary  formation case with $0.36 M_\odot$ companion
as BF36, to the slow  spiral-in cases with $0.20 M_\odot$ and $0.15
M_\odot$ companions as SS20 and  SS15, and to the merger cases with
$0.10 M_\odot$ and $0.05 M_\odot$ companions as  M10 and M05.

%% file: table1.tex
\begin{table*}
 \begin{center}
 \begin{tabular}{llrrrrrrrrrrrrr}
  Model & $M_{\rm comp}$ &$E_{\rm tot,CE}^{\rm ini}$&$E_{\rm orb,3D}^{\rm ini}$&$E_{\rm orb,3D}^{\rm fin}$
&$E_{\rm tot,ej}^{\rm fin}$&$J_{\rm tot}$
&$J_{\rm ej}^{\rm fin}$
&$a_{\rm orb}^{\rm fin}$&$M_{\rm env}^{\rm fin}$&$M_{\rm ej}^{\rm fin}$
\\
 M05 &  0.05 & -40.19 & -1.40 & -15.75 &  1.07 & 1.98 & 0.34 & 0.15 & 1.45 & 0.02 \\
 M10 &  0.10 & -40.79 & -1.30 & -36.72 &  4.22 & 4.05 & 1.30 & 0.15 & 1.18 & 0.30 \\
 SS15&  0.15 & -41.35 & -1.85 & -43.88 &  6.77 & 6.14 & 3.48 & 0.19 & 1.05 & 0.43 \\
 SS20&  0.20 & -41.87 & -2.38 & -46.26 &  6.90 & 8.24 & 5.46 & 0.29 & 0.93 & 0.56 \\
 BF36&  0.36 & -43.40 & -3.91 & -51.22 &  9.51 & 14.93& 10.77& 0.50 & 0.67  & 0.81\\
 BF36ej&  0.36 & -43.40 & -3.91 & -52.87 &  9.64 & 14.93& 13.82& 0.48 & 0.0  & 1.48\\
 \end{tabular}
 \end{center}
\caption{ $M_{\rm comp}$ is the mass of the companion. $E_{\rm tot,CE}^{\rm ini}$, $E_{\rm orb,3D}^{\rm ini}$, $E_{\rm orb,3D}^{\rm fin}$
 are the initial total energy of the CE system, the initial orbital energy and the final orbital energy, 
respectively (for details on definitions, see \S~\ref{sec:3den}). 
$E_{\rm tot,ej}$ is the total energy carried away by the ejecta.  
$J_{\rm tot}$ and $J_{\rm ej}$ are the total angular momentum and the angular momentum of the ejecta.
$a_{\rm orb,fin}$ is the final orbital separation. 
$M_{\rm env}$, and $M_{\rm unb}$ are the envelope mass and the ejecta, respectively.
All the masses are in $M_\odot$, the orbital separation is in $R_\odot$, all the energies are in $10^{46}$ erg, and all the 
angular momentums are in $10^{51}\, \rm g\, cm^2\, s^{-1}$.
The instants of time for which the final values, including all the quantities for the ejecta, 
are calculated are either when the separation had decreased to  $0.15 R_\odot$ (for M05 and M10 models), 
or 50 days after the plunge-in (SS15, SS20 and BF36 models). 
In addition, for BF36 we also list the values after the entire envelope was ejected (BF36ej).
{All the models are for the same red giant donor with a mass of $1.8 M_\odot$ and with a core mass of $0.318 M_\odot$.}
}
\label{tab1}
\end{table*}

%% file: sect3_mapping.tex
\section{Mapping and symmetry}

\label{sect:map}

\begin{figure*}
\includegraphics[width=85mm]{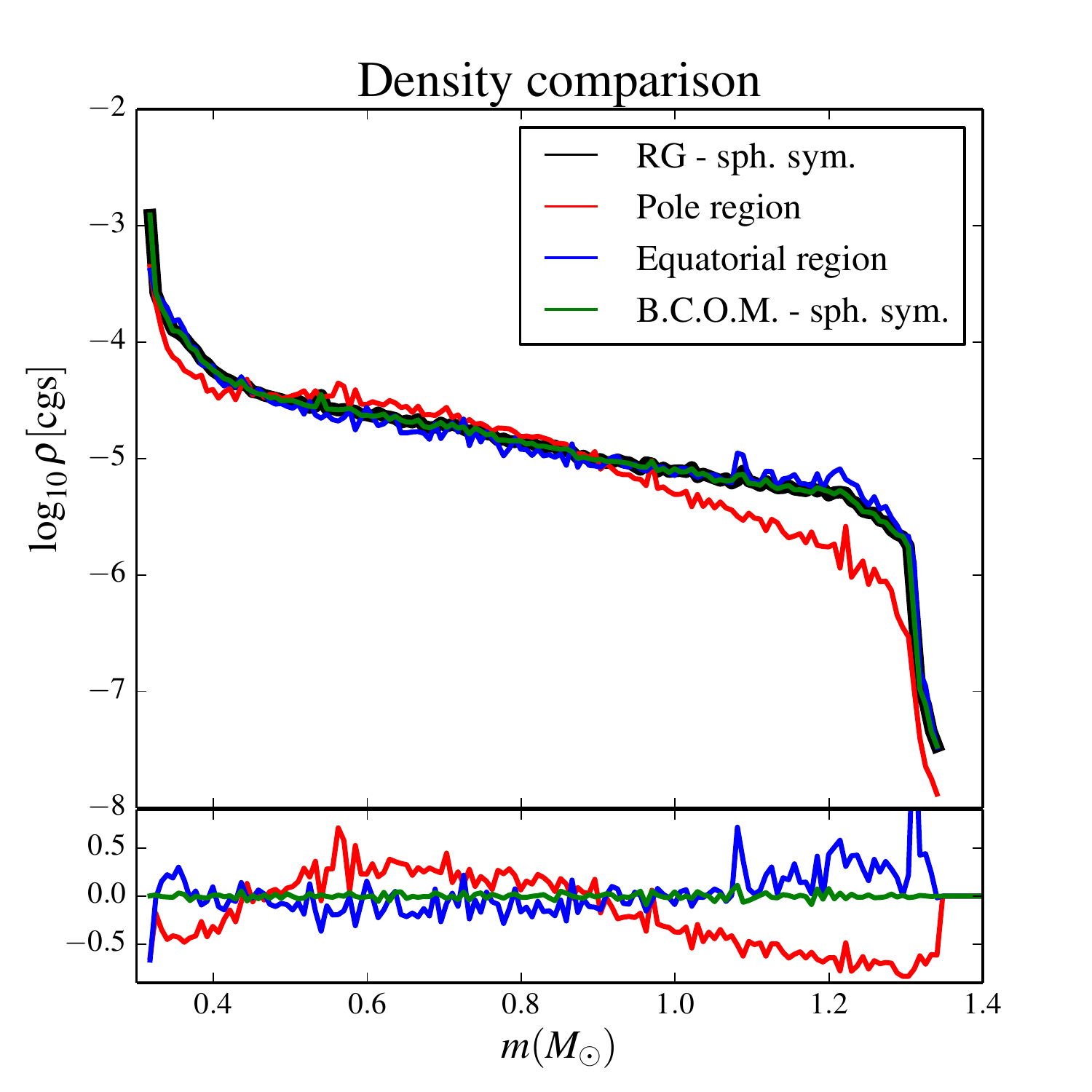}
\includegraphics[width=85mm]{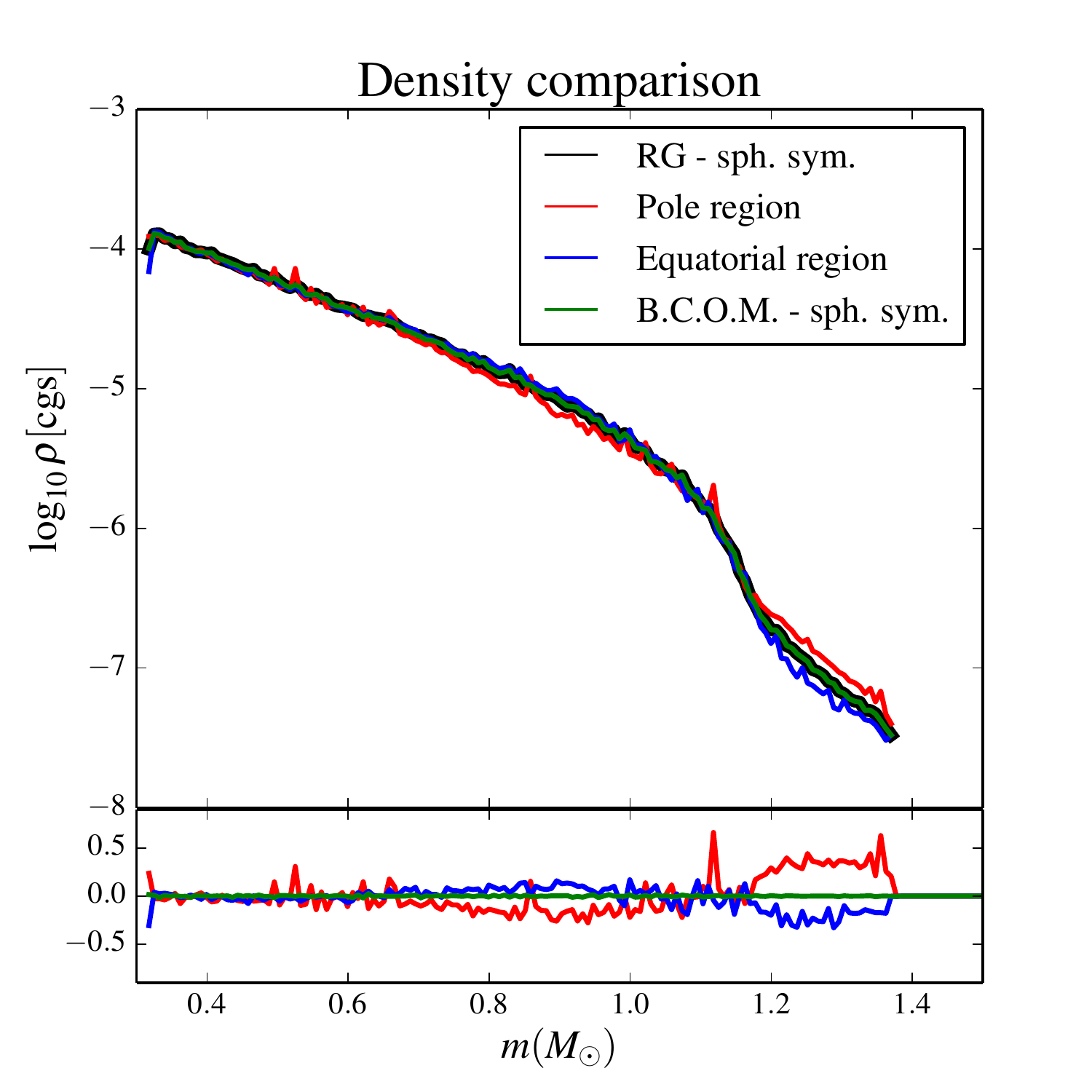}
\caption{Comparison of  the ways to  average quantities, shown  on the
  example of density.  The left panel  shows the case BF36 just before
  the  end  of  the  plunge-in, when  outflows  are  significant  (day
  364).The right  panel shows the case  SS15 at the start  of the slow
  spiral-in, where there is almost no outflows (day 645).  Shown are 4
  ways to average the quantity: with the red giant core at the centre and the
  entire CE object  $\rho_{\rm RG}$ (thick black),  same centre of symmetry and only the polar region
  $\rho_{\rm  pole}$  (red),  same centre of symmetry and only the  equatorial  region  $\rho_{\rm
    equat}$ (blue), and  finally with the centre at the centre
  of  mass  of  the binary,  averaged over  the  entire CE  object
  $\rho_{\rm BCOM}$ (green).  The bottom  panels show the ratios of the
  three    latter    quantities     to    the    first,    $(\rho_{\rm
    pole}-\rho_{\rm  RG})/\rho_{\rm  RG}$ (red),  $(\rho_{\rm
    equat}-\rho_{\rm RG})/\rho_{\rm RG}$  (blue) and $(\rho_{\rm
    BCOM}-\rho_{\rm RG})/\rho_{\rm RG}$ (green).}
\label{fig:diff_den}
\end{figure*}

\begin{figure*}
\includegraphics[width=85mm]{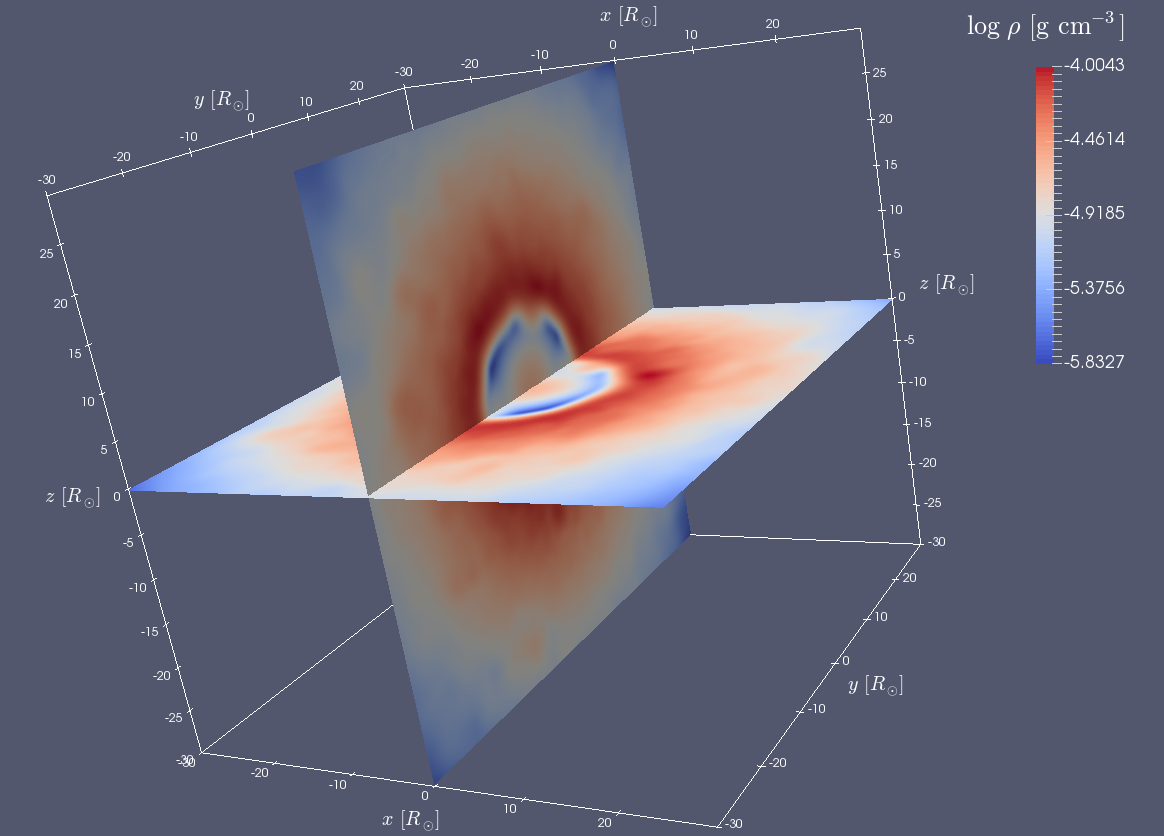}
\includegraphics[width=85mm]{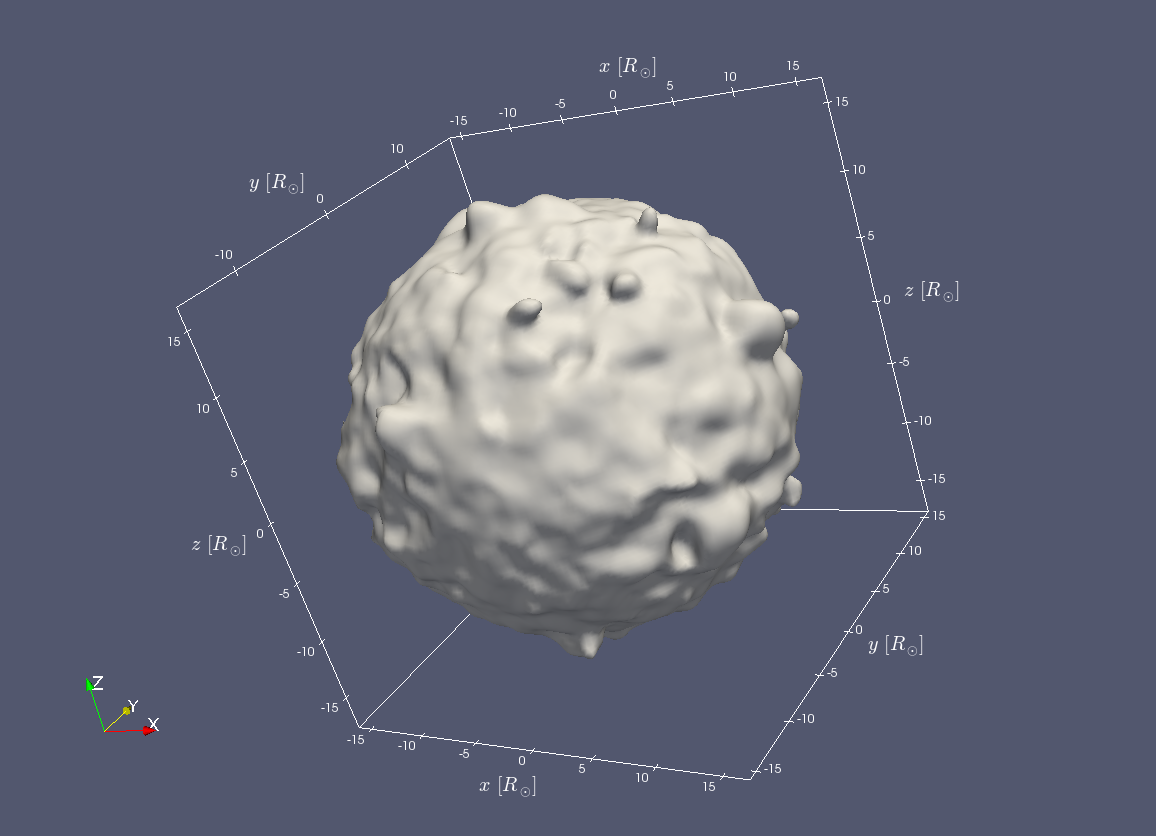}
\caption{Density distribution during  a slow
  spiral-in, the case SS15 (day  747).  
The left panel shows  density slices in the orbital plane  and in the
perpendicular plane. The right panel shows  a contour figure for 
$\log_{10}\rho = -4.3$.
The orbital separation at this
  moment is $a=0.18 R_\odot$, the binary  is not resolved.  This is  a zoom-in,  not the  entire CE  object is
  shown.     This figure shows the system 122 days after the start of the slow spiral-in.  
    By this time, the formed close binary has orbited about 8000 times, and about $0.5 M_\odot$ of the envelope mass is ejected.
    Please note that the densiest part of the hollow shell, while symmetric overall, is still ``clumpy.''
The figures were created using ParaView \citep{ahrens200536,Ayachit:2015:PGP:2789330}.
}
\label{fig:den_spiralin}
\end{figure*}

In  this Section  we  introduce  the mapping  tools  that convert  the
results  of 3D  simulations  to 1D  space, both  for  the analysis  of
physical processes that  will be done in this paper,  and for the possible
future comparisons with 1D simulations of CEEs.

The outcomes of hydrodynamical  simulations are not fully symmetrical,
while a  typical evolutionary  stellar code  deals with  a spherically
symmetric 1D star.  In the 1D case, the centre of symmetry is naturally
located at the centre of the modelled star, in its core.  The centre of
a 3D  object during and after  a CEE is not  that indisputable.  
During the Phase I only the  outer layers are perturbed, and the donor
star mainly keeps  its original symmetry around the  core.  During the
dynamical Phase II, no symmetry can be expected.  Finally, during
the  Phase III,  symmetry --  from  the dynamic  point of  view --  is
expected to  form around  the centre  of mass  of the  binary that
rotates inside the  CE.  However, this phase may  continue long enough
that  the energy  transport starts  to affect the envelope's  temperature and
density profile  (self-regulated spiral-in).  On the one  hand, the energy
transport will be symmetric around  the nuclear energy source, if that
nuclear energy source  (e.g., shell burning) is still  active.  On the
other  hand, the binary's  orbital  energy release  may stem from a different point 
and shift the average centre of symmetry, as the  energy generation rate can outpace the
nuclear  energy  source.  It  is  inevitable  therefore that  no  true
spherical symmetry can  exist, but we can evaluate how the choice
of  the centre of symmetry  can  affect the anticipated  1D  spherically  symmetric
stellar structure.

The mapping of our 3D object to 1D object is done 
by using a mass weighted average of a given quantity over a spherical
shell of radius, $r$ and thickness, $\delta r$:

\begin{equation}
\bar X (r) = \frac{ \sum_{k} x_k m_k }{\sum_k m_k}
\end{equation}

\noindent Here $\bar  X (r) $ is a discrete  (mesh-type, similar to 1D
stellar  profiles) profile  of the  quantity $X$,  the mesh-points   
spaced by  $\delta r$. The variable $x_k$  is the value  of the
quantity $X$ that  an SPH particle $k$  has, $m_k$ is the  mass of the
same SPH particle  $k$, and $k$ is  the index of the  SPH particle. The
summation goes over all particles $k$ such that the distance from each
particle to the chosen centre, $r_k$, is between $r$ and $\delta r$.

We compare  the two choices  for the centre  of symmetry,  the red
giant core and the centre of mass of the binary during a well-established spiral-in, i.e.
the formed binary has made a few thousand orbits.
We find that for the  layers outside  of the  orbit, the  choice of  the centre  of 
symmetry is less significant, and the differences between the averaged
values are within a few per cent (see Fig~\ref{fig:diff_den}).
The large number of the periods is expected to result in a  spin-up of most of the mass that is
 located within a few binary orbits around the centre of mass of the binary. Indeed,
we note that independently on the centre's choice, there is almost no stellar material 
within  a few orbits around the centre of mass of the binary.
For  example, consider the case of simulation SS15
at the moment when 122 days have passed after the the start of the slow spiral-in, or about 8000 binary periods.
The  orbital separation at that time is $\sim 0.2 R_\odot$.
Except  for the only two SPH particles  that are within
the RL of  the companion, and two more particles that are located at about
$6 R_\odot$,  all the other  SPH  particles are located  at least
beyond  $8 R_\odot$  from the  binary, creating  a pronounced  density
inversion  around the  binary for  about 40  orbital separations  (see
Fig. \ref{fig:den_spiralin}).

We note that   the
density inversion  was observed in  previous 3D CEE simulations,
for example see the study by \cite{Passy11}.
In their work, the CEE simulations were done with the two types of
the numerical methods and compared with  an SPH type code and with a
grid-based code.   We note  that there, the density  inversion was
more pronounced in simulations with an SPH code, than with a grid based
code, and we attribute it to the reason we described above.
We should however stress that in our case, we discuss the state of the
  CE system when the formed close binary has revolved for about 8000 orbital periods, affecting its neighborhood,
  while in \cite{Passy11} the density profiles are made for a moment when the formed
  binary has made less than 50 orbital periods.  Using our definitions of the CEE phases as in \S2,
  their binary has just started the slow spiral-in and has had less ``relative''
  time to spin-up its neighboring SPH particles. At the similar stage of the CEE --
  the start of the spiral-in --  our envelope had not yet formed a hollow shell.

 We note as well that  in 1D CEE simulations such
a strong  density inversion and  core stripping have never been reproduced 
\citep[see, e.g.,  Fig 2.8  in][where also a  low-mass giant  case was
  considered]{IvanovaThesis}. This is likely relevant to the fact that
  1D stellar codes, when modelling a CEE, add the energy as a ``heat'' instead of the mechanical energy
  (we will discuss how the form of the added energy affects the outcomes in
  \S~\ref{sec:entropy} and \S~\ref{sec:rec}). 
  Another plausible reason is that  the tidal spin-up of the envelope from the binary is not taken into account
  as it should be, and the spin-up is deposited at an incorrect location.

We anticipate that the decoupling  between the binary and the envelope
might be partially a  numerical problem, due to the small  number of particles
that  are left  near  the binary,  and the  smoothing  length in  that
region.  Unfortunately,   decreasing the mass of SPH particles by a factor of two,
{ increasing thereby a simulations's resolution by a factor of two}, we
run in  the same  problem when decreasing  the smoothing  length to
obtain  a  better  value  for  the  gravitational  potential,  as  was
discussed in  \S~\ref{sect:classif}. Nonethless, we believe that the
  formation of a hollow shell is a physical phenomenon by virtue of a tidal spin-up that operated
  during many binary timescales.

We can also compare profiles of thermodynamic quantities for ``polar''
SPH  particles  (those  particles  that  have  a  polar  angle  within
$\vartheta<25^0$ around the zenith direction  that goes through the red giant
core, and for negative $Z$, $\vartheta>155^0$), and for ``equatorial''
SPH particles  (those that are  located close to the  equatorial plane
and have their polar angle  within $75^0<\vartheta<105^0$). To be more
specific, the mass-radius relation for  the coordinates is the same as
for the  averaging of the whole  3D object, but the  average values of
any quantity  are found only for  the particles that are  within these
polar angles.

We find that,  as expected, at the end of  the plunge-in the asymmetry
is strong -- up to an order of magnitude difference in values that can
be    found    in    polar     and    equatorial    directions    (see
Fig~\ref{fig:diff_den},  the left  panel). The most  pronounced deviations
are  for  the   polar  direction,  where  density   and  pressure  are
substantially lower than  their averages over all  angles in the proximity
of  the  binary   orbit  and  at  large  radii,  while   denser  in  
between. This is likely because  of the active ongoing ``outflows'' in
the  equatorial direction,  which are  compensated by  somewhat slower
``inflows'' in the polar direction.

During a slow  spiral-in, when the binary orbit almost  does not decay
during several dynamical timescales of the expanded envelope (e.g.,
40  days in  case SS15),  non-negligible asymmetry  is still  present.
Qualitatively, the density profiles in  opposite directions are consistent
with deviations  expected from  rotation. Relative differences  for the
density in polar and equatorial  directions in the extended envelope,
still can  reach a factor of a few,  with the density in  the equatorial
direction being  smaller (when taken at  the same distance from  the red giant
core).

%% file: sect4_energies.tex
\section{Energies}

\label{sec:en}

\subsection{Energies definitions in a 1D approach}

\label{sec:1den}

In the 1D approache, the gravitational potential, or the 
specific gravitational potential energy in the envelope during a CEE is  

\begin{equation} 
\label{eq:pot1d}
\phi_{\rm env 1D}(r)= -  \left ( \frac{Gm(r)}{r} + f_{\rm ins} \frac{GM_{\rm comp}}{r}\right ) \ . 
\end{equation}

\noindent Here  $m(r)$ is  the local mass  coordinate within  the star
(excluding the companion), $r$ is the radial coordinate, $f_{\rm ins}$
indicates  the  effect  of the companion,  and  $f_{\rm  ins}=1$  if  the
companion  orbits within  $r$,  or $f_{\rm  ins}=0$  if the  companion
orbits outside $r$. The origin of the second term in this equation 
will be explained in more detail in \S\ref{ss:en_con}.

The potential energy of the envelope $E_{\rm pot,env1D}$ \citep[the
  envelope is everything except what is defined as the core, which can be
  ambiguous, see ][]{Ivanova11}, the internal energy of the envelope
$E_{\rm int,env1D}$ (this energy consists of the thermal energy of the
envelope $E_{\rm th,env1D}$ and the recombination energy of the
envelope $E_{\rm rec,env1D}$), the binding energy of the envelope, and
the kinetic energy of the envelope $E_{\rm kin,env1D}$, are defined as
\citep[see also ][]{Ivanova+2013Review}

\begin{eqnarray} 
\label{eq:enpot1d}E_{\rm   pot,   env  1D}  &=&   -   \int_{M_{\rm   core}}^{M}  \left   (
\frac{Gm}{r} +  f_{\rm ins} \frac{GM_{\rm comp}}{r}\right  ) \, dm\ ;  \\
E_{\rm th, env  1D} &=&\int_{M_{\rm core}}^{M} e_{\rm th}\, dm\ ;  \\ 
E_{\rm rec,  env  1D}  &=&\int_{M_{\rm  core}}^{M}  \varepsilon_{\rm  rec}\,  dm  ; \\ 
U_{\rm int, env  1D} &=&\int_{M_{\rm core}}^{M} u\, dm = E_{\rm th, env  1D} + E_{\rm rec,  env  1D}\ ;  \\ 
E_{\rm   bind,   env  1D}  &=& U_{\rm int, env  1D}+E_{\rm   pot,   env  1D}\ ; \\
E_{\rm kin, env 1D} &=&\int_{M_{\rm core}}^{M} 0.5 V^2 \, dm \ .
\end{eqnarray}

\noindent Here  $M$ is the  total mass of the  star, $m$ is  the local
mass coordinate,  $u$ is  the specific  internal energy  and $u=e_{\rm
  th}+ \varepsilon_{\rm rec}$, $\varepsilon_{\rm rec}$ is the specific
recombination energy, and $e_{\rm th}$  is the specific thermal energy
for  which no  recombination energy  is  taken into  account.  $V$  is
velocity; note that in 1D,  velocities for $E_{\rm kin, env 1D}(M_{\rm
  core})$ do  not include the donor's movement on  its binary  orbit, but
only   the  relative   velocities  in   the  corotation   frame.   The
recombination energy $E_{\rm rec, env  1D}$ is static potential energy,
which is not available immediately and  is only released as a response
of the envelope on its expansion.   This energy release may or may not 
be triggered during a CEE \citep{2015MNRAS.447.2181I}.

\subsection{Energies definitions in a 3D SPH code}

\label{sec:3den}
 
In a 3D approach, the gravitational potential that a particle $i$ has is defined as 
\begin{equation}
\label{eq:pot3d}
\phi_i=\sum_{l\ne i}^N m_l \varphi_{i,l}\ ,
\end{equation}
where $\varphi_{i,l}$ is the gravitational potential between two SPH particles 
of unit mass that have a distance between them $|\mathbf{r}_i-\mathbf{r}_l|$ \citep[technical details on how it is calculated can be found in][]{1989ApJS...70..419H}.

The potential, internal, recombination, and kinetic energies of an SPH particle are: 
\begin{eqnarray}
 \label{eq:enpot3d}
 E_{{\rm pot},k}&=&m_k \phi_k,\\
 \label{eq:appint}
 U_{{\rm int},k}&=&m_k u_k, \\
 \label{eq:appred}
 E_{{\rm rec},k}&=&m_k \varepsilon_{\rm rec}, \\
\label{eq:appkin}
 E_{{\rm kin},k}&=&0.5 m_k (v_{x,k}^2+v_{y,k}^2+v_{z,k}^2) \ .
\end{eqnarray}
\noindent Here $v_{x,k}$, $v_{y,k}$, and $v_{z,k}$ are integrated from the equations of motion
(i.e.,  they are calculated with respect to the coordinate system). 
The variable $u_k$ in our SPH code is integrated over time using the equation of thermal energy change
  \citep[as in ][]{2002MNRAS.335..843M,2010MNRAS.402..105G}, an implementation that
    guarantees conservation of total energy and entropy in the absence of shocks.
$\varepsilon_{\rm rec}$ is a component of $u_k$.
The total energy of a CE system that consists of the envelope 
(the gas that is not yet ejected to infinity), the donor's core and the companion, is: 

\begin{eqnarray}
\label{eq:3dtotce}
 E_{\rm tot,CE}&=&\sum_k^N (0.5E_{{\rm pot},k}+U_{{\rm int},k}+E_{{\rm kin},k})  \\ \nonumber
& = &\sum_k^N 0.5m_k \sum_{l\ne k}^N m_l \varphi_{k,l}+ U_{{\rm int,CE}}+E_{{\rm kin,CE}} \ .
\end{eqnarray}

\noindent Here  the summations  goes over all  SPH particles  that are
still  gravitationally bound  to the  CE system.   The total  internal
energy of  the CE system  is the same as  the energy of  the envelope,
$U_{{\rm   int,CE}}=U_{{\rm    int,env3D}}$.   

The first term  in the  Equation \eqref{eq:3dtotce}  is the  total potential
energy of the CE system:

\begin{equation}
\label{eq:3dpotce}
E_{\rm pot,CE} = \sum_k^N 0.5m_k \sum_{l\ne k}^N m_l \varphi_{k,l} \ .
\end{equation}

\noindent  
The potential  energy of the  envelope consists of such  components as
the potential  energy between the  envelope and the core $E_{\rm pot,e-core}$, 
the  self-gravitating energy  of  the envelope $E_{\rm pot,e-sg}$,  
and the potential  energy between the envelope and the companion $E_{\rm pot,e-comp}$:

\begin{eqnarray}
 \label{eq:envbind}
 E_{\rm pot,env 3D} &= &
\sum_k^{\rm Env}  m_k \bigg (M_{\rm core} \varphi_{k,\rm core} + 0.5 \sum_{l\ne k}^{\rm Env} m_l \varphi_{k,l} \nonumber \\
& +&  M_{\rm comp} \varphi_{k,\rm comp} \bigg ) \ ,
\end{eqnarray}

\noindent where the summation is only for the normal (non-special) particles belonging to the envelope.
The potential energy of the CE system is then:
\begin{eqnarray}
 \label{eq:potCE}
 E_{\rm pot,CE} 
=  E_{\rm pot,env 3D} \nonumber +   M_{\rm comp}  M_{\rm core}  \varphi_{\rm comp,\rm core} , 
\end{eqnarray}

\noindent  and the total energy of the CE system is

\begin{eqnarray}
 \label{eq:etotCE1}
 E_{\rm tot,CE} &=& U_{\rm int,env3D} + E_{\rm pot,env 3D} \nonumber 
\\ &+&  E_{\rm kin, CE} + M_{\rm comp}  M_{\rm core}  \varphi_{\rm comp,\rm core} ,
\end{eqnarray}

The ``orbital energy''  of the binary system is  \cite[see, e.g.,][]{2015MNRAS.450L..39N}:

\begin{eqnarray}
\label{eq:orbenergy}
 E_{\rm orb, 3D}&=& 0.5\mu |V_{12}|^2  + 0.5 \sum_i^{\rm RL1, RL2} m_i\phi_i \\ 
   &-& 0.5 \sum_j^{\rm RL1} m_j\phi_j^{\rm RL1}- 0.5 \sum_k^{\rm RL_2}  m_k\phi_k^{\rm RL_2} \ , \nonumber
\end{eqnarray}

\noindent where $\mu=M_1M_2/(M_1+M_2)$ is the reduced mass, and
$\vec{V}_{12}=\vec{V}_1-\vec{V}_2$ is the relative velocity of the two
stars. The first term gives the kinetic energy.  The second term is
the gravitational energy of the binary, with the sum being over all
particles $i$ that are inside the two immediate RLs, where each RL depends
on the mass of all particles that are bound to the whole mass within
that RL, and are not simply functions of the masses of the RG core and the companion.
The third and the fourth terms correspond to the removal
of the self-gravitational energy of the donor (the sum being over
particles $j$ within the RL of the star 1) and of the companion (the
sum being over particles $k$ within the RL of the star 2),
respectively. The caveat here of course is that during a plunge-in, there
is no orbit, and the orbital energy found using
Equation~\eqref{eq:orbenergy} is not related to the separation between
the particles.

It  can be  seen that  the orbital  energy $E_{\rm  orb, 3D}$  and the
potential  energy  of the  envelope  $E_{\rm  pot,env 3D}$  have  some
similar terms -- the potential energy  between the particles in the RL
of  the companion  and  the  core, the  potential  energy between  the
particles in the  RL of the core and the  companion, and the potential
energy  between the  particles  in  the two  different  RLs.  That  is
because a particle, when it is inside  the RL, is both  part of the
envelope  and  of  the  binary.   Therefore, it  is  not  possible  to
decompose  the  total potential  energy  of  the  CE system  into  the
intrinsic potential energy  of the binary and the  potential energy of
the envelope.

Instead, we can introduce the {\it reduced} orbital energy, where only
the core and the companion are considered:

\begin{eqnarray}
\label{eq:orbCC}
 E_{\rm orb,CC} &=& 0.5 M_{\rm core} v_{\rm core}^2 + 0.5 M_{\rm comp}v_{\rm comp}^2 \nonumber \\
  &+& M_{\rm comp}  M_{\rm core}  \varphi_{\rm comp,\rm core} \ . 
\end{eqnarray}

\noindent Note  that this  energy cannot  in principle determine the
current orbital separation  of the binary.  We stress  that to find
the orbital separation $a$ after the  plunge using energy, one has to
use  $E_{\rm orb,  3D}$;  during the
plunge, there is no orbital separation - energy relation.

We also introduce the kinetic energy of the envelope

\begin{eqnarray}
E_{\rm kin,env3D}& =& 0.5 \sum_k^{\rm Env} m_k v_k^2  
\\ &= &E_{\rm kin, CE} 
- 0.5 M_{\rm core} v_{\rm core}^2 - 0.5 M_{\rm comp}v_{\rm comp}^2 \nonumber \ .
\end{eqnarray}

\noindent Unlike the 1D case, this energy is non-zero even at the beginning, 
since the star's envelope rotates with the binary.

Finally, we can rewrite the energy equation of the CE system as:

\begin{equation}
 \label{eq:etotCE2}
 E_{\rm tot,CE}= E_{\rm orb,CC} + U_{\rm int,env3D} +  E_{\rm pot,env3D} +  E_{\rm kin,env3D}\ .
\end{equation}

\noindent Note  that in a 3D  code, the total energy  of all particles
(including those that are unbound) is conserved but this is not so for $E_{\rm tot,CE}$.
Hence this  equation can serve as the energy  conservation only if
no particles have become ejecta.

\subsection{The energy conservation in 1D}

\label{ss:en_con}

Let us transfer the energy conservation Equation~\eqref{eq:etotCE2} to 1D.
As previously, this is for the case when there is no ejecta.

1. \underline{The internal energy:}

\begin{eqnarray}
U_{\rm int,env3D} &=& U_{\rm int,env1D}\ .
\end{eqnarray}

1. \underline{The orbital energy} can be described using binary orbital 
energy:

\begin{eqnarray}
E_{\rm orb,CC} &=&  - \frac{GM_{\rm core}M_{\rm comp}}{2a} \ .
\end{eqnarray}

\noindent Note that this is {\it not} valid during a plunge, 
where instead Equation~\eqref{eq:orbCC} should be used.

2. \underline{The  kinetic energy.}   There is  no simple  direct link
between $E_{\rm kin,env3D}$  and $E_{\rm kin,env1D}$.  In the first
case, one  measures velocities  in a  stationary reference  frame that
moves with the centre of mass,  and in the second case, velocities are
measured  in  the  corotation   frame.   Before  the  plunge,  $E_{\rm
  kin,env3D}\simeq {GM_{\rm env} M_{\rm  comp}^2}/{(2a (M+M_{\rm comp}))}$, and after the
plunge, if  the angular velocities in  the spun-up envelope  are taken
into account for 1D, $E_{\rm kin,env3D}\simeq E_{\rm kin,env1D}$.

4.  \underline{The potential energy}.  In 1D, an added mass inside the
envelope (an orbiting companion) technically produces the same
potential as a ``thin spherical shell.''  A ``thin spherical shell''
does not create a potential inside of it, nor does it create a
gravitational force that would act on an object inside of it (Newton's
shell theorem).  This is why the potential energy in 1D is written as
in Equation~\eqref{eq:enpot1d} (in \S\ref{sec:comp_pot} we also will
discuss what error is produced by the thin shell approximation in
Equation~\eqref{eq:enpot1d}, i.e., for the CE that is outside of the
companion's orbit).  The difference between 3D and 1D potential energy
should include $E_{\rm pot,e-comp}^{\rm out}$, which is the fraction
of $E_{\rm pot,e-comp}$ that sums only over the SPH particles that are
inside the companion's orbit.  Note that if thin shell approximation is
taken to be valid at any moment of the CEE evolution, it will also
imply that $E_{\rm pot,e-comp}^{\rm out}=0$ when the companion is
still outside of the envelope.  However, at the same moment, the true
value is well approximated by the point mass expression, $E_{\rm
  pot,e-comp}^{\rm out}\simeq - GM_{\rm env} M_{\rm comp}/a$.  As a
result, in 1D, a partially inconsistent approach is usually taken:
for the envelope mass outside of the orbit, the companion is
treated as a thin shell (this is the second term in
Equation~\eqref{eq:enpot1d}); and for the envelope mass inside of the
orbit, $m_{\rm env}(r<a)$, the companion is treated as a point mass,
$E_{\rm pot,e-comp}^{\rm out} = - Gm_{\rm env} (r<a) M_{\rm comp}/a$.

We can now rewrite the energy conservation equation ~\eqref{eq:etotCE2} 
using ``1D values'':
 
\begin{eqnarray}
E_{\rm tot,CE}  &=& 0.5  \left (M_{\rm core}  v_{\rm core}^2  + M_{\rm
  comp}  v_{\rm comp}^2  +  \int_{M_{\rm core}}^{M}  v^2  dm \right  )
\nonumber   \\  
&   +  &   \int_{M_{\rm   core}}^{M}  \left   (  u   -   \frac{Gm}{r}   \right  )\,   dm  
-  \frac{GM_{\rm core} M_{\rm comp}}{a}
\\   
&-& \int_{m(r>a)}^{M}  \frac{GM_{\rm comp}}{r}  dm -  \frac{Gm_{\rm env  }
  (r<a) M_{\rm comp}}{a} \nonumber \ ,
\end{eqnarray}
\noindent where  all velocities are in the  inertial frame,
and  velocities  of  the core  and  the
companion are not  related in a simple way to  the orbital separation,
especially during the plunge.

After the plunge-in, a simplification can be made:

\begin{eqnarray}
E_{\rm tot,CE}  &=& -  \frac{GM_{\rm core} M_{\rm comp}}{2a} +0.5 \int_{M_{\rm core}}^{M}  v^2  dm 
\nonumber   \\  
&   +  &   \int_{M_{\rm   core}}^{M}  \left   (  u   -   \frac{Gm}{r}   \right  )\,   dm  
\\   
&-& \int_{M_{\rm core}}^{M}  \frac{GM_{\rm comp}}{r}  dm  \nonumber \ .
\end{eqnarray}

In conclusion, given the complications with the kinetic energy, the
potential energy, and the non-Keplerian orbit of the  companion during the
plunge, a 1D code cannot self-consistently conserve
energy during the plunge-in.

\subsection{The validity of the thin shell approximation}

\label{sec:comp_pot}

\begin{figure*}
\includegraphics[width=90mm]{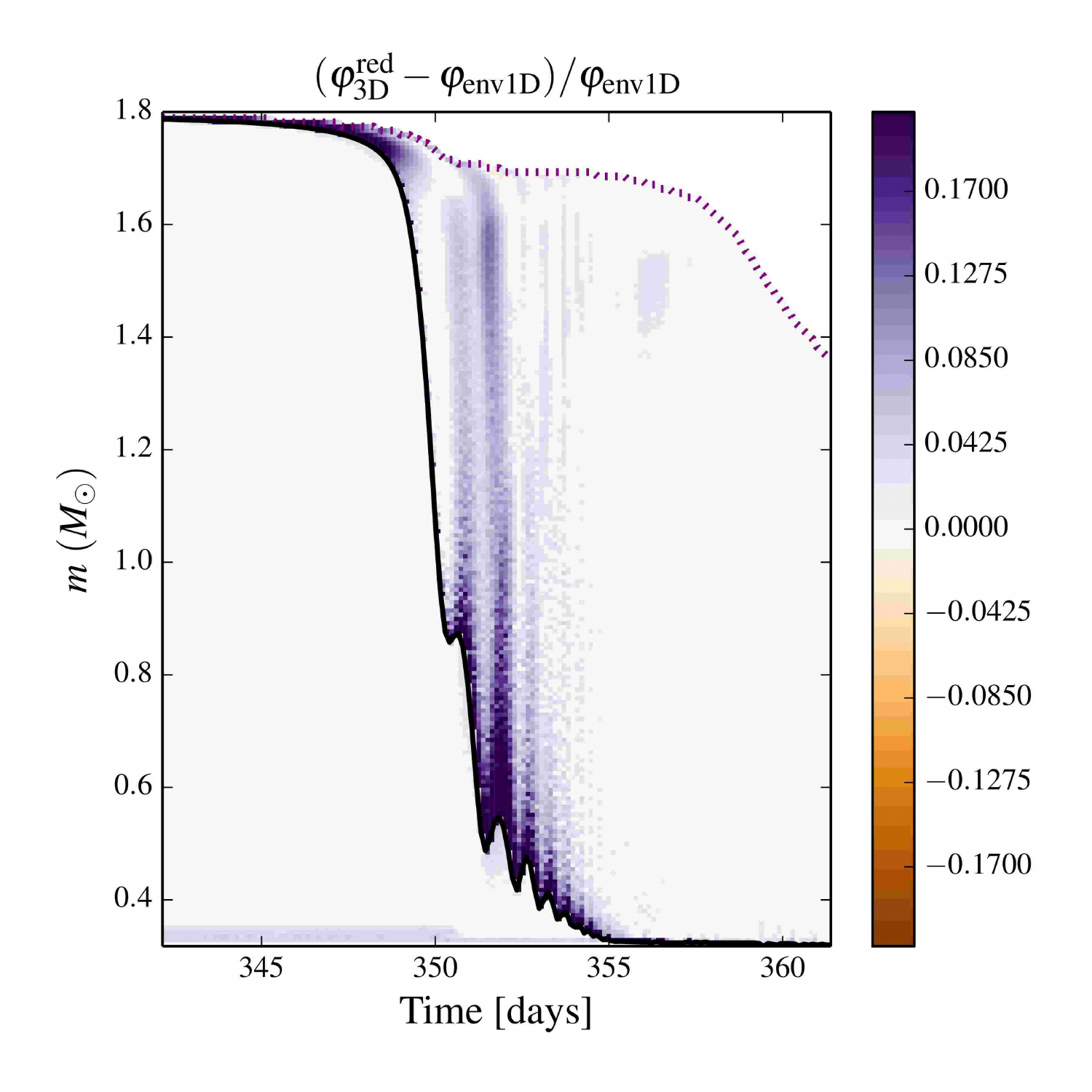}
\hskip-0.5cm\includegraphics[width=90mm]{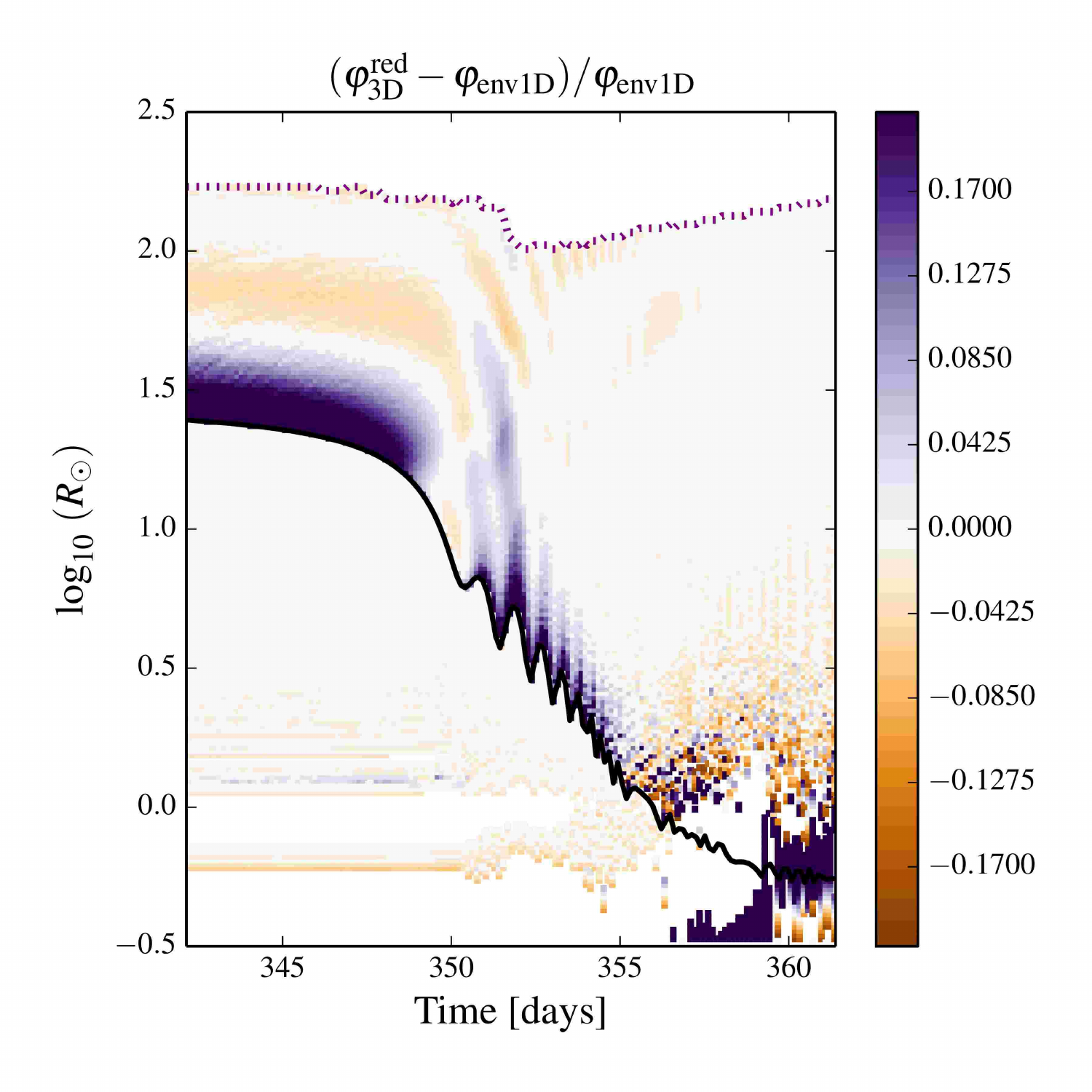}
\caption{Comparison  of  the gravitational  potentials  in  1D and  3D
  approach. Shown  for the example of  BF036. The left panel  shows the
  difference as  a function of mass,  and the right panel  shows the
  same  difference but as a function of  radius. The  solid black  lines
  indicate  the location  of the  companion. To help to distinguish color shades, we indicate with the dashed purple  lines
  the surface of the envelope. As we only show the bound mass, the surface is
    defined as either the mass coordinate  of a bound particle that is farthest
    away from the centre of the symmetry (for the left plot), or 
    the farthest distance from the centre of the symmetry  for bound particles (for the right plot).
    The surface in the plots should not be confused with the stellar photosphere, which is not obtained in our simulations.
  }
\label{fig:diff_epot}
\end{figure*}

To assess the thin shell approximation for the fraction of the envelope  that is outside of the orbit, 
we can  find the ``reduced'' 3D  gravitational potential  for  each SPH  particle  in the  envelope,
$\phi_{i}^{\rm red}$, in a way that will measure the same quantity
as does the 1D definition for the gravitational potential:
 
\begin{eqnarray}
 \phi_{i}^{\rm red} (r) = 
 \sum_{l\ne i}^{\le r} m_l \varphi_{i,l} + M_{\rm core} \varphi_{i,\rm core}  
+ f_{\rm ins} M_{\rm comp} \varphi_{i,\rm comp}  \ ,
\end{eqnarray}
\noindent  where  the  summation  is only for  envelope
particles that have a distance to the chosen  ``centre'' smaller than $r$.
As previously,  for the  numerical mapping from  3D to  1D,
$\phi^{\rm  red}_{\rm 3D}(r)$ is  found by  averaging  the particles  that have  $r$
within some small neighborhood $[r;r+\delta r]$. Summed over all
the envelope particles, this quantity produces the equivalent of $E_{\rm pot,env1D}$.

As  can be  expected,  at the  end of  the  plunge-in, the  difference
between the 3D averaged  value for $\phi_{\rm 3D}^{\rm red}$  and the 1D value
$\phi_{\rm       env1D}^{\rm      red}$       is      small       (see
Figure~\ref{fig:diff_epot}).   However,   during  the   plunge-in,  the 1D
approximation often provides a significantly less  negative potential
well in the envelope near the companion than the 3D model provides for
$\phi_{\rm 3D}^{\rm  red}$.  The deviation can  be as large as  50\% of
the value over a region of  several solar radii.  After the envelope
had decoupled, the gravitational potential within the inner envelope's
``shell'' is also not in good agreement between the two approaches.

By its design,  this  comparison  only  demonstrates  the  effect  of  the
companion  on the  difference  between the  two  potentials above  the
orbit.  However, the full potential  inside the orbit will also depend
on the companion.  Typically,  the 3D approach
provides a  deeper potential  well in the  neighborhood of  the orbit,
while far above the  orbit, the potential  well can be  more shallow
than in the 1D thin  shell approximation (see Figure~\ref{fig:diff_epot}).
The  value  of the  specific  potential  energy  does not  affect  the
structure calculations in a 1D model,  which are affected only by the
value  of the  local gravitational  acceleration.  Note that this is a
post-processed value.  However, it plays a role in what is considered
in 1D simulations  as the immediate energy budget as well as  the rate of the
energy  transfer from  the orbit  to the  envelope, and  therefore may
affect the  rate of the  orbital dissipation.

\subsection{Where and in what form the orbital energy is deposited}

\begin{figure}
\includegraphics[width=90mm]{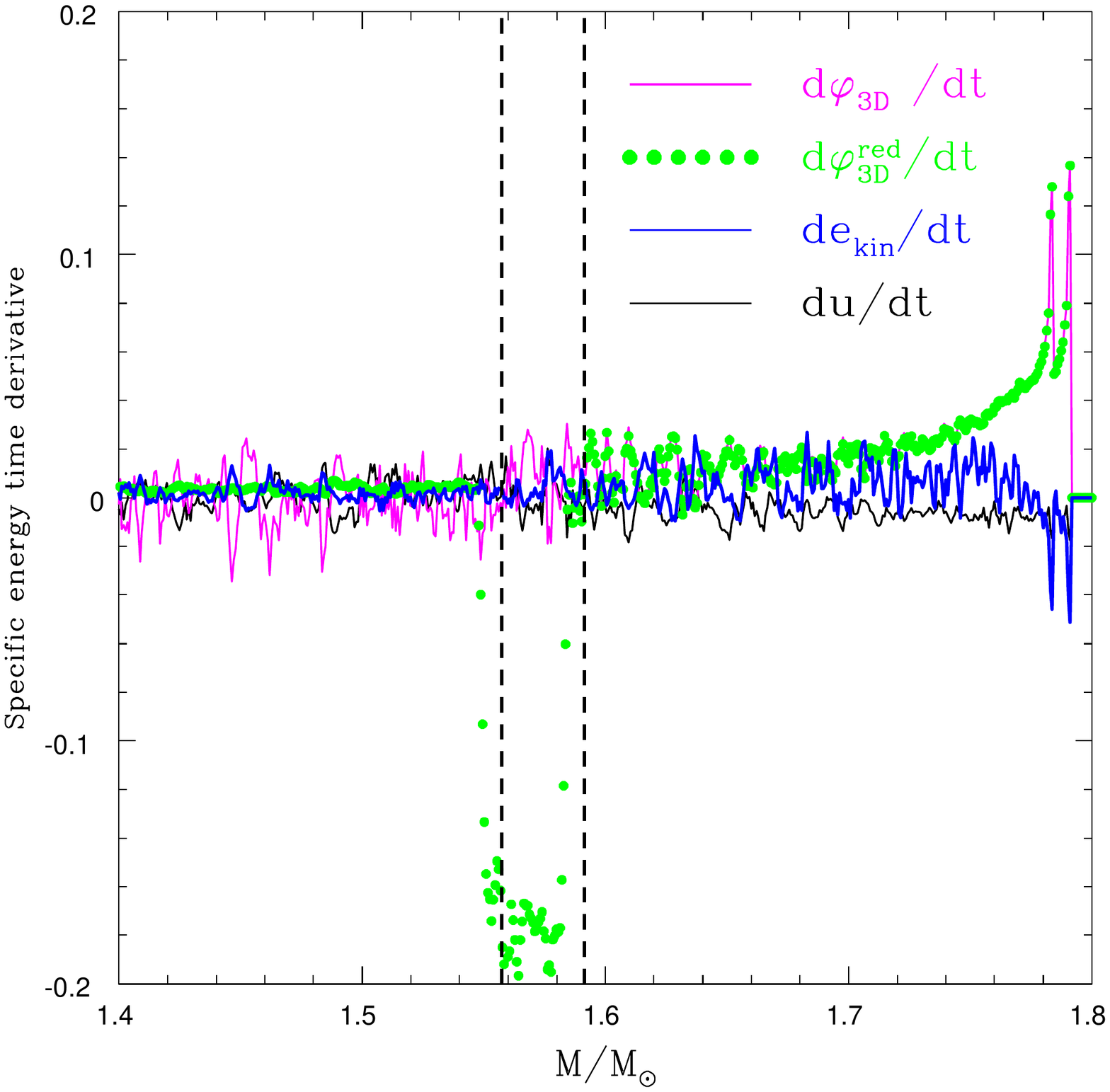}
\caption{Energy changes in case SS15,  day 615. Time derivative of the
  specific energies  are in units  of $10^{15}$ ergs per  sec. $e_{\rm
    th}$    here    includes    both   thermal    and    recombination
  energies. Derivatives are calculated between the two snapshots of 3D
  simulation. The locations of the companion's orbits for the two time
  moments are shown with vertical dashed black lines.}
\label{fig:deriv_en}
\end{figure}

\label{sect:en_comp}

In both 3D and 1D approaches, the recombination energy is deposited
self-consistently where it has been released.  This is not the same
for the orbital energy release. A 1D approach must come up with a
prescription on how to deposit the energy that corresponds to the
independently obtained orbital dissipation.  We can get a hint from
our 3D simulations on where the orbital energy is effectively
deposited, and in what form it is deposited.

In Figure \ref{fig:deriv_en} we show where the local specific energies
are changing while the orbit is shrinking.  It can be seen that the
potential energy and the kinetic energy are clearly increasing
everywhere above the orbit, while the internal energy is changing at a
much smaller pace than that of the potential and kinetic energies.  As
expected, the change in the real potential above the orbit is similar
to that of the reduced potential. In the regions
where the orbit had passed between the two moments of time for which the derivatives are found,
the absolute value of the time derivative of the reduced potential is
a couple of orders of magnitude larger that that of the real potential.  Below the orbit, the 3D
potential derivative  is substantially more noisy than
that of the reduced potential, as it is affected by the entire
envelope above.  The derivatives indicate that we cannot detect the
orbital energy deposition inside the binary orbit, but outside of the
orbit it affects the entire envelope, primarily changing the
mechanical energy of the envelope -- the potential energy and the
kinetic energy (and kinetic energy is converted to the potential
energy with time).  This will become important in \S~\ref{sec:rec},
where we will discuss how the form of the energy deposition may affect
the outcome.

%% file: sect5_am.tex
\begin{figure}
\includegraphics[width=90mm]{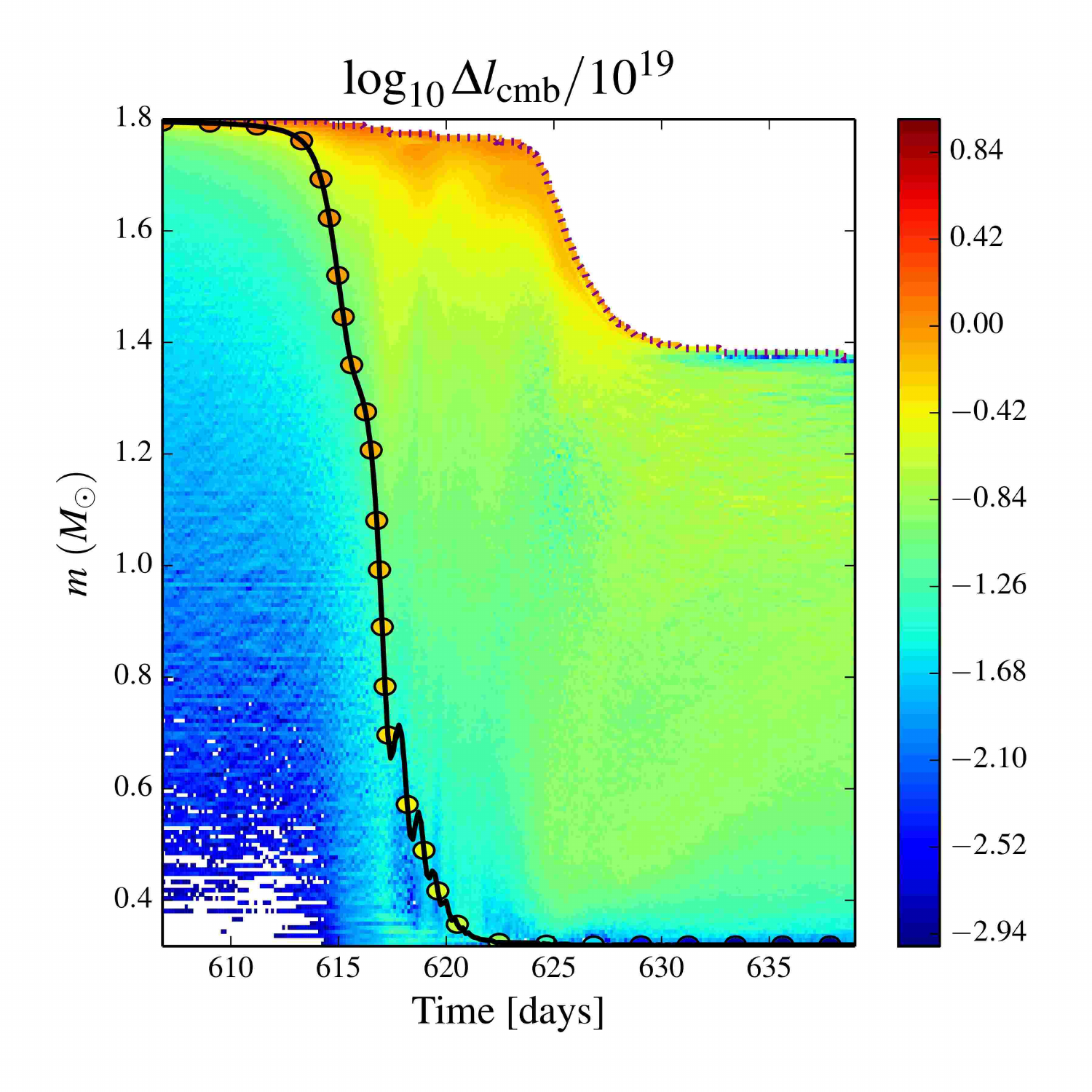}
\caption{Accumulated gain  in the  average specific  angular momentum,
  shown for the case  SS15.  The  black solid line  shows the  instantaneous orbital
  separation between  the core and  the companion.  The colors  of the
  circles indicate  the Keplerian specific angular  momentum that this
  location would have.   The dashed purple line  indicates the surface
  of the envelope. White color implies that that 
the accumulated gain at this location is below the cut-off minimum value, $\log_{10} \Delta l_{\rm cmb}<-3$.}
\label{fig:evol_l}
\end{figure}

\begin{figure*}
\includegraphics[width=58mm]{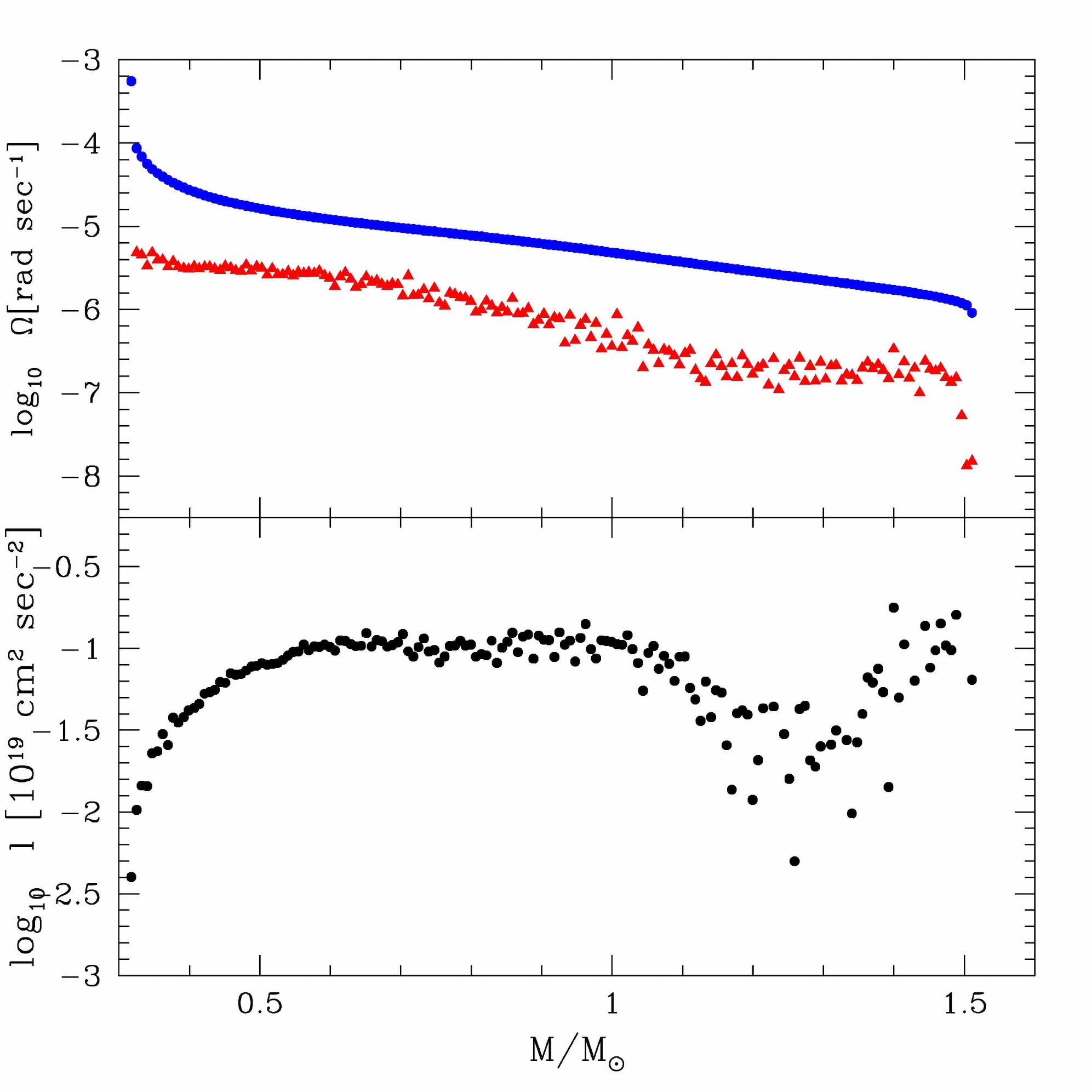}
\includegraphics[width=58mm]{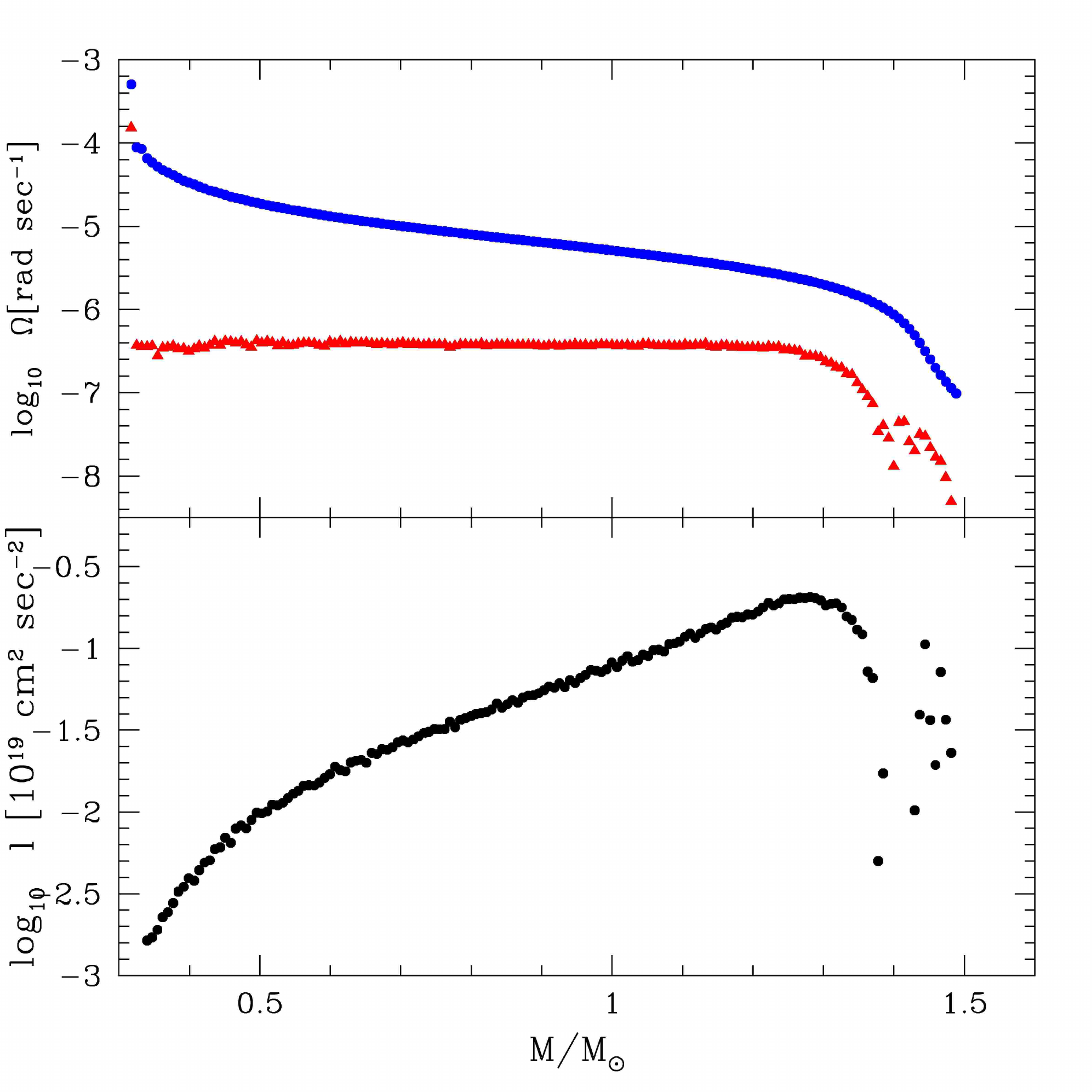}
\includegraphics[width=58mm]{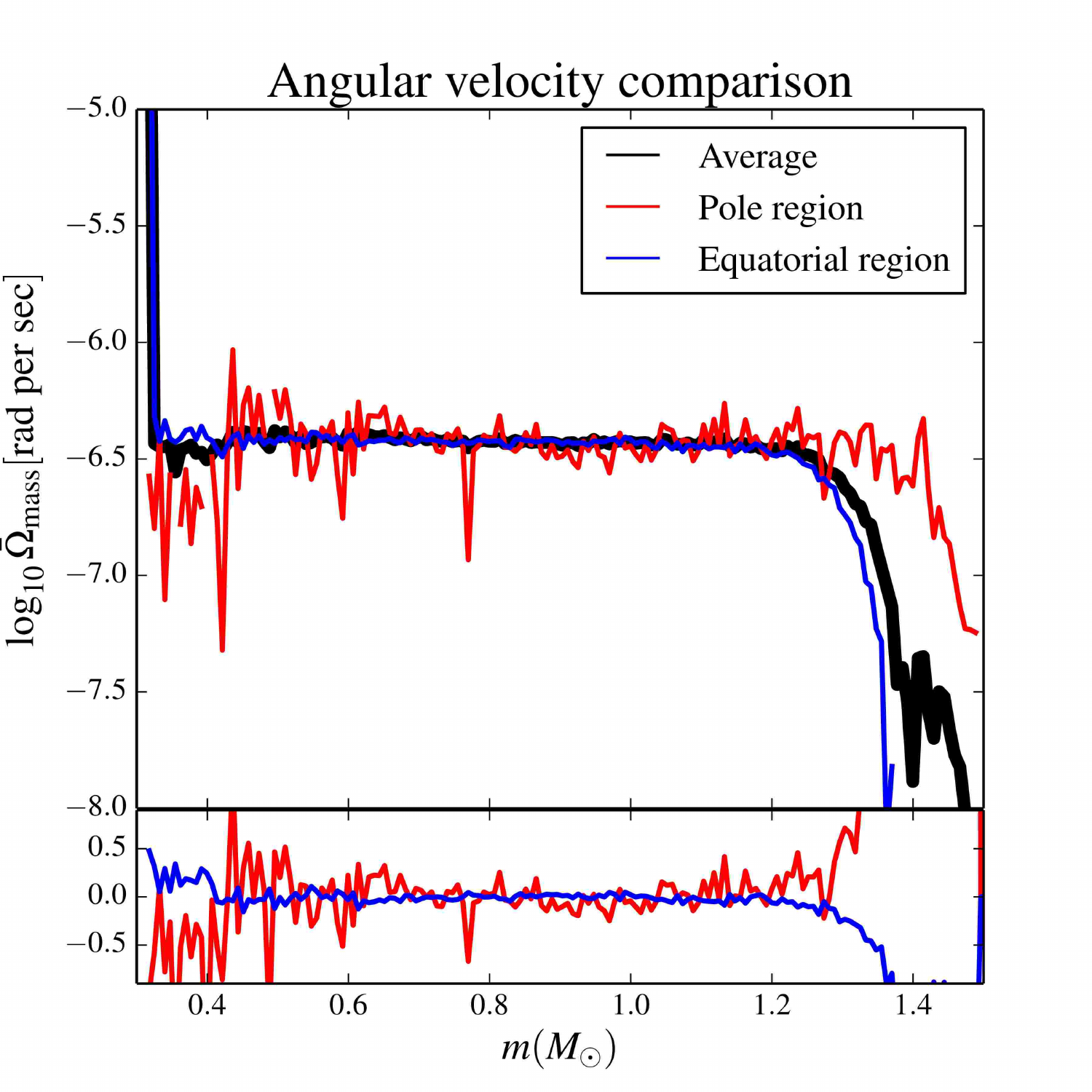}
\caption{The mass-averaged  angular velocity $\bar  \Omega_{\rm mass}$
  (red dots, top panels) and  the specific angular momentum $l$ (black
  dots, bottom panel), for the case M10.
  The  blue dots  on the  top left and middle panels demonstrate  the value  of the
  Keplerian velocity  that this envelope  could have.  The  left panel
  shows that case at the end of the  plunge-in (t=613 days),
  the middle panel shows the profiles  50 days later (t=664 days, this
  is after the  merger took place), and the right  top panel shows the
  comparison of the  polar and equatorial regions for  the same moment
  (t=664 days).  The  bottom right panel shows the ratio  of the polar
  or equatorial angular velocities to the average.}
\label{fig:omega_slowspin}
\end{figure*}

\begin{figure*}
\includegraphics[width=58mm]{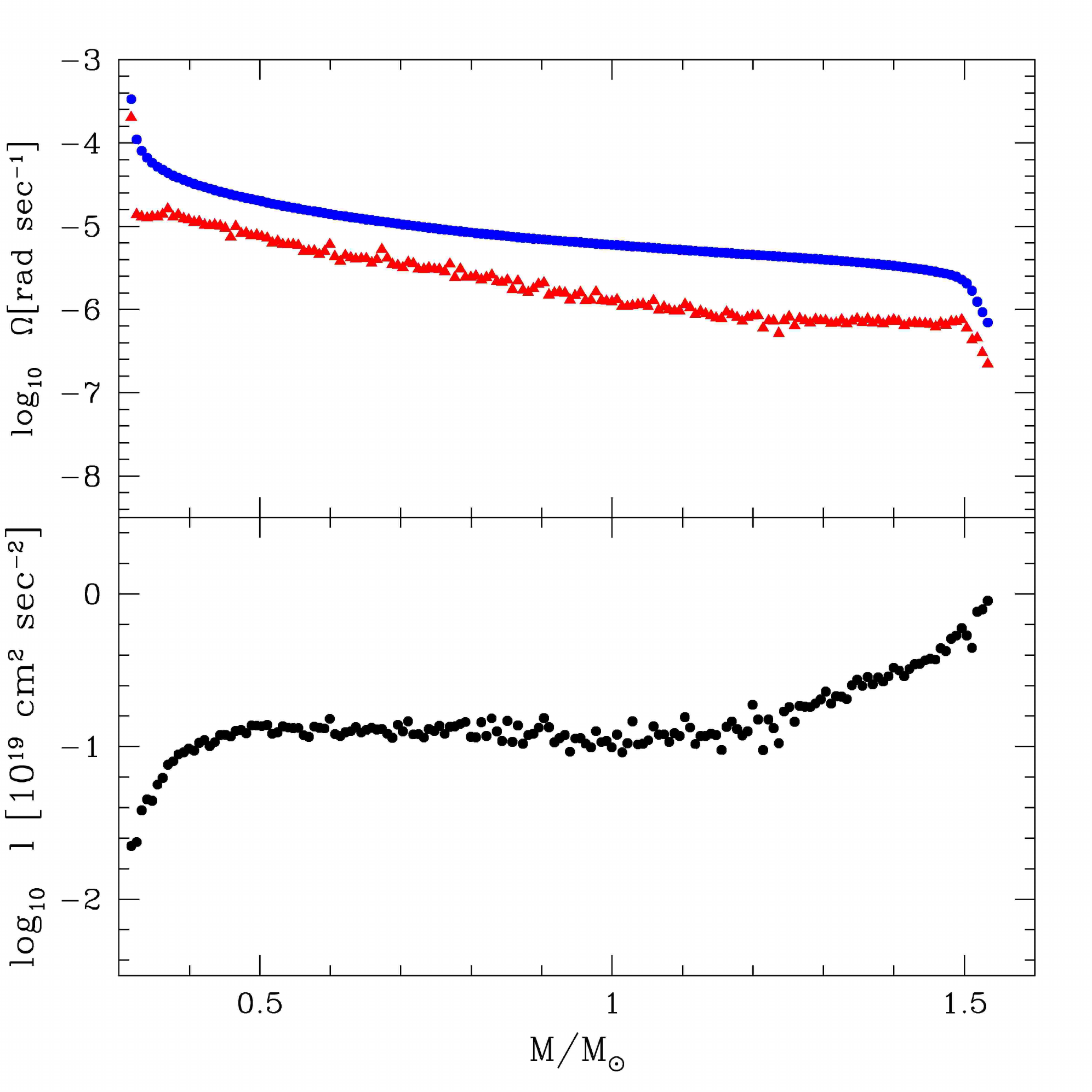}
\includegraphics[width=58mm]{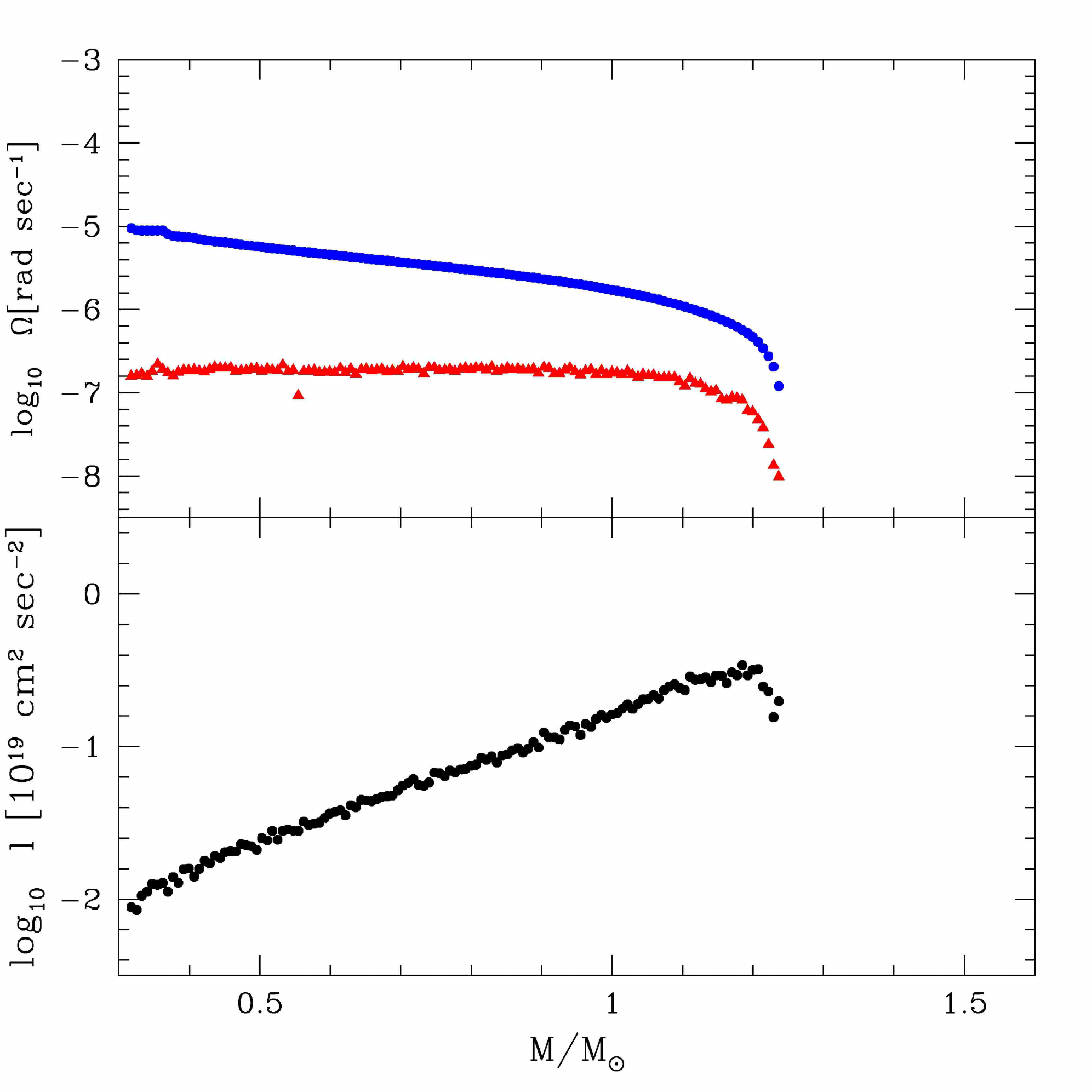}
\includegraphics[width=58mm]{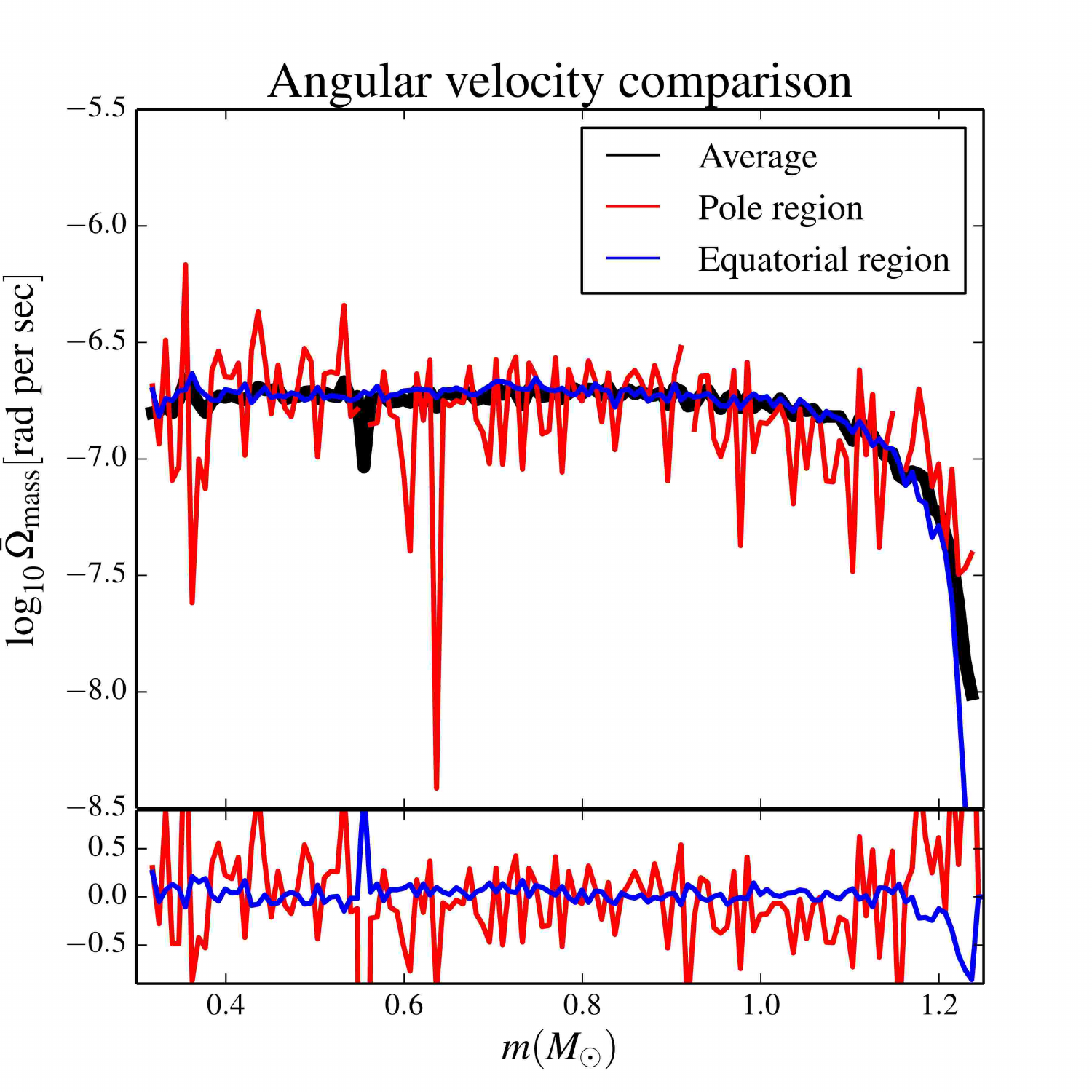}
\caption{  The mass-averaged  angular velocity $\bar  \Omega_{\rm mass}$
  (red dots, top panels) and  the specific angular momentum $l$ (black
  dots, bottom panel), for the case SS15.
   The  left panel
  shows that case  at the end of the  plunge-in (t=626 days),
  the middle panel shows the profiles  175 days later (t=801 days), and the right  top panel shows the
  comparison of the  polar and equatorial regions for  the same moment
  (t=801 days).  Other information as in Figure~\ref{fig:omega_slowspin}.}
\label{fig:omega_slowspins15}
\end{figure*}

\section{Angular momentum}
\label{sec:am}

\subsection{Average angular momentum in 3D case}

\label{sec:3dam}

The $Z$-component of the angular momentum  for each SPH particle is found
with  respect to  a particular  point  that should  be moving  without
acceleration (for technical details,  see Appendix \ref{ap:am}; below,
whenever we  talk about an  angular momentum,  we will imply  only its
Z-component).  However,  the core  of the red  giant --  our preferred
centre  of  symmetry  for  other   CE  system  properties  --  is  not
stationary, nor is it moving at  a nearly constant velocity. On the contrary,
the core is orbiting around a  point which, at least during the slow
spiral-in,   coincides  well   with  the   centre  of   mass  of   the
core-companion binary system.  The  angular momentum therefore cannot
be calculated  with respect to  the core,  and if such  a calculation is
carried out, the value of the angular momentum oscillates.

We choose  therefore to calculate  the averaged angular  momentum with
respect to the  centre of the core-companion binary.   The caveat with
this approach is that the centre of mass of the binary can also be
moving with  an acceleration.   Therefore, during the  plunge-in phase
this averaged  quantity is  also not  entirely meaningful,  and becomes 
most   useful    only   at   the    end   of   the    spiral-in.    In
Figure~\ref{fig:evol_l} we  show the accumulated gain  in the specific
angular momentum  in the envelope, computed  from the start of  the CEE
simulation.

There  are  two  important  features  that  are  present  in  all  the
simulations:
\begin{enumerate}
\item We  do not see that  the region inside the  orbit gains angular
momentum or  spins up.  While we have said  that the value  of the
average angular momentum is not  fully self-consistent before the slow
spiral-in,  the almost complete  absence  of any  spin-up is  nonetheless
important.
\item Most of the initial binary angular momentum is absorbed by
the outer layers of the envelope.  It is those layers that get ejected
during the plunge-in, leaving the remaining CE with a smaller fraction
of  the total  initial  angular momentum  (this can  be  seen both  in
Figure~\ref{fig:evol_l} and from data shown in Table~\ref{tab1}).
\end{enumerate}

\subsection{Angular velocity for 1D star}

To understand how  to relate our 3D  star to a 1D star,  we will briefly 
review how the  treatment of rotation is done for  1D stars in stellar
codes.  Generally,  angular  momentum transport is calculated using
a diffusion equation. A diffusion equation
  is meant to trace and to be written implicitly  for the angular momentum, 
  while in practice it is usually written  using the  angular velocity  variable.  Currently,  there are
several ways  to write the  diffusion equation, however, the
basis for   all    of    them   is   \citep{1978ApJ...220..279E}:

\begin{equation}
\frac{\partial \Omega}{\partial t} = \frac{1}{\rho r^4} \frac{\partial }{\partial r}
\left ( \rho r^4 D \frac{\partial \Omega}{\partial r}
\right ) \ .
\label{eq:diffusion}
\end{equation}

\noindent  Here $D$  is some  diffusion coefficient,  and $\Omega$  is
angular   velocity.    The    differences   between   several   modern
modifications are then based on  what kinds of instabilities are taken
into account to  find the diffusion coefficient $D$.   A more complete
form of the diffusion equation can also take into account the flux due
to  non-zero radial  velocity, and  the temporal  loss of  the angular
momentum   \citep[e.g.,    see   for   technical    details   Appendix
  in][]{IvanovaThesis}.

It is  important that the  transition from  the angular momentum  as a
``diffusing quantity''  to the angular  velocity as the  main variable
can  be done  only by choosing  and fixing  the relation
between angular velocity and specific angular momentum.  Specifically,
the above diffusion equation \eqref{eq:diffusion}  adopts the assumption that
the angular  velocity is only a function of radius.  This  allows a
simplification  in  which the  star  is treated  as if it is 
composed  of thin  spherical  shells, where  each  shell has  constant
angular velocity,  and has  specific  moment  of inertia
$i=2/3 r^2$.

At the same time, it is anticipated that the centrifugal force affects
the local effective  gravity. Hence mass shells need  to correspond to
isobars  rather  than  to  spherical shells  \citep[for  methods,  see
  e.g.][]{1976ApJ...210..184E,1998A&A...334..210H,2000ApJ...528..368H,2013ApJS..208....4P}.
As a result, in stellar  codes where isobars are considered, stellar
structure variables are  taken as being constant on  isobars, while the
angular momentum diffusion is still based on spherical symmetry.

In this  paper, we  chose to reduce the 3D angular  momentum distribution
into a 1D  angular velocity distribution  by adopting the  same relation
between the specific  angular momentum and the angular  velocity as was
done  implicitly  for  the  angular  momentum  diffusion  equation
\eqref{eq:diffusion}.  Therefore,  we provide  values of  $\Omega(r)$ as
averaged  on spherical  shells.

There  are two  ways to  do the  averaging for  the angular  velocity.
First, we can find the total angular momentum in a thin shell that is located
at the  radius $r$ and has a thickness $\delta r$,  
and then  use the  relation for  a specific 
moment of inertia of a thin shell:

\begin{equation}
\bar \Omega_{\rm spher}(r_{\rm shell})) = \frac{3}{2} 
\frac {\bar l (r)}{\bar r^2} =  \frac{3}{2} \frac {1 }{\bar r^2} 
\frac {\sum_{i\in \rm shell} l_{i} m_i}{\sum_{i\in \rm shell} m_i}   \ .
\end{equation}
Here the summation is only done for the particles which belong 
to the thin shell located at $r$ with a thickness $\delta r$.
On the other hand, we can also find the mass-averaged angular velocity at each shell:

\begin{equation}
\bar \Omega_{\rm mass}(r_{\rm shell}) = 
 \frac {\sum_{i\in \rm shell} \Omega_{i} m_i}{\sum_{i\in \rm shell} m_i} 
=  \frac {\sum_{i\in \rm shell}  l_i m_i/ r_{z,i}^2}{\sum_{i\in \rm shell}  m_i} \ .
\end{equation}
\noindent Here $r_{z,i}$ is the distance from the particle to the rotation axis.

We did  not find a
substantial difference between $\bar  \Omega_{\rm mass}(r)$ and $\bar
\Omega_{\rm  spher}(r) $.   There is  some discrepancy  for  small
values of the angular velocity,  where more numerical noise is present
in $\bar \Omega_{\rm spher}(r)$.


Like the case with the specific angular momentum, we do not observe a
spin up of the regions inside the orbit, and the Keplerian (binary)
angular velocity always exceeds greatly that of the surrounding
matter. {There is only a short period of time when the matter
  around the orbit approaches close to the value of the binary angular
  velocity.}  This moment takes place just before the envelope
material expands beyond the binary.
It is close to the same time when the
envelope  and the  inner binary  have effectively  decoupled, and  the
hollow  shell  structure  seen  in  Figure~\ref{fig:den_spiralin} 
started to form.

\subsection{Angular velocity distribution during the slow spiral-in}

At the start  of the slow spiral-in, the specific  angular momentum, unlike
the case  of a typical rotating  star, does not rise monotonically
from the centre to the surface.  We can separate
the  envelope  in two  regions:  the  inner  region, with  an  almost
constant value of specific angular momentum, and the outer region,
where the  angular momentum  was quickly transferred from the companion to the envelope  during the
plunge-in  (see Figures \ref{fig:omega_slowspin} and 
\ref{fig:omega_slowspins15}).  Between  the two regions the  value of
the specific angular  momentum at the start of the  spiral-in
drops (see Figure \ref{fig:omega_slowspin}).

The angular velocity profile and the specific angular momentum profile
evolve very  quickly to ``stable''  profiles after the plunge-in
ends. In a stable situation,  the specific angular momentum increases
with the  distance from  the centre.  This is established after  about a
few dynamical timescales of  the expanded CE, where $\tau_{\rm  d, CE}\approx 50$
days.  The angular velocity profile become flat in most of
the envelope (by mass).  The outer part of the  envelope is small by mass but is
more than  80\% of the  envelope radius.  There, the  angular momentum
transport is not efficient, and the angular velocity drops.

We note that either at the start  of the slow spiral-in, or during the
slow spiral-in, the angular velocity in the envelope is  about an order
of magnitude less than its local Keplerian velocity (see Figures
\ref{fig:omega_slowspin} and \ref{fig:omega_slowspins15}).  In none of our 3D simulations
does the  angular velocity  of the  inner part  of the  envelope approach
the binary's angular  velocity.

%% file: sect6_entr.tex

\begin{figure*}


\hskip-0.2cm\includegraphics[width=90mm]{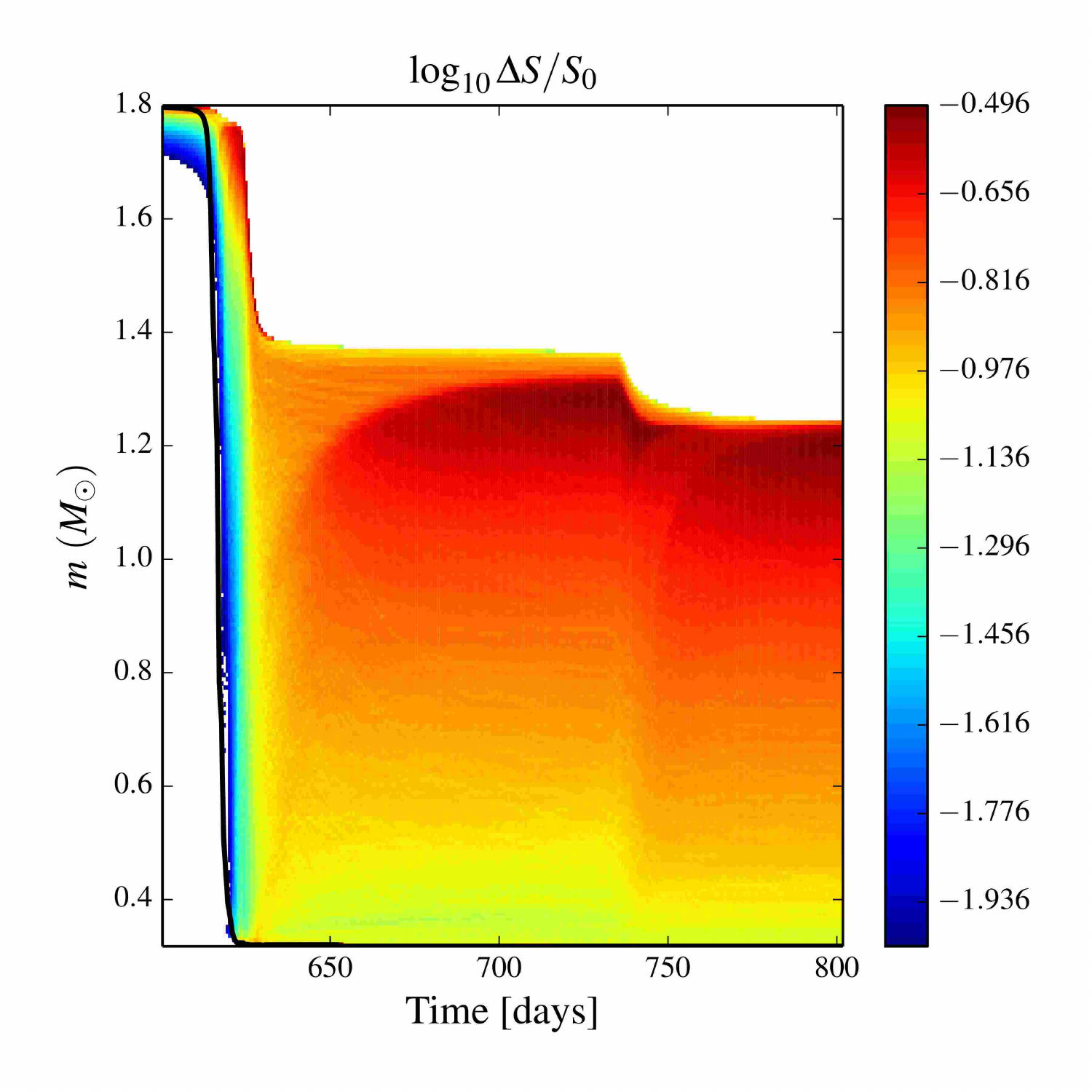}
\hskip-0.2cm\includegraphics[width=90mm]{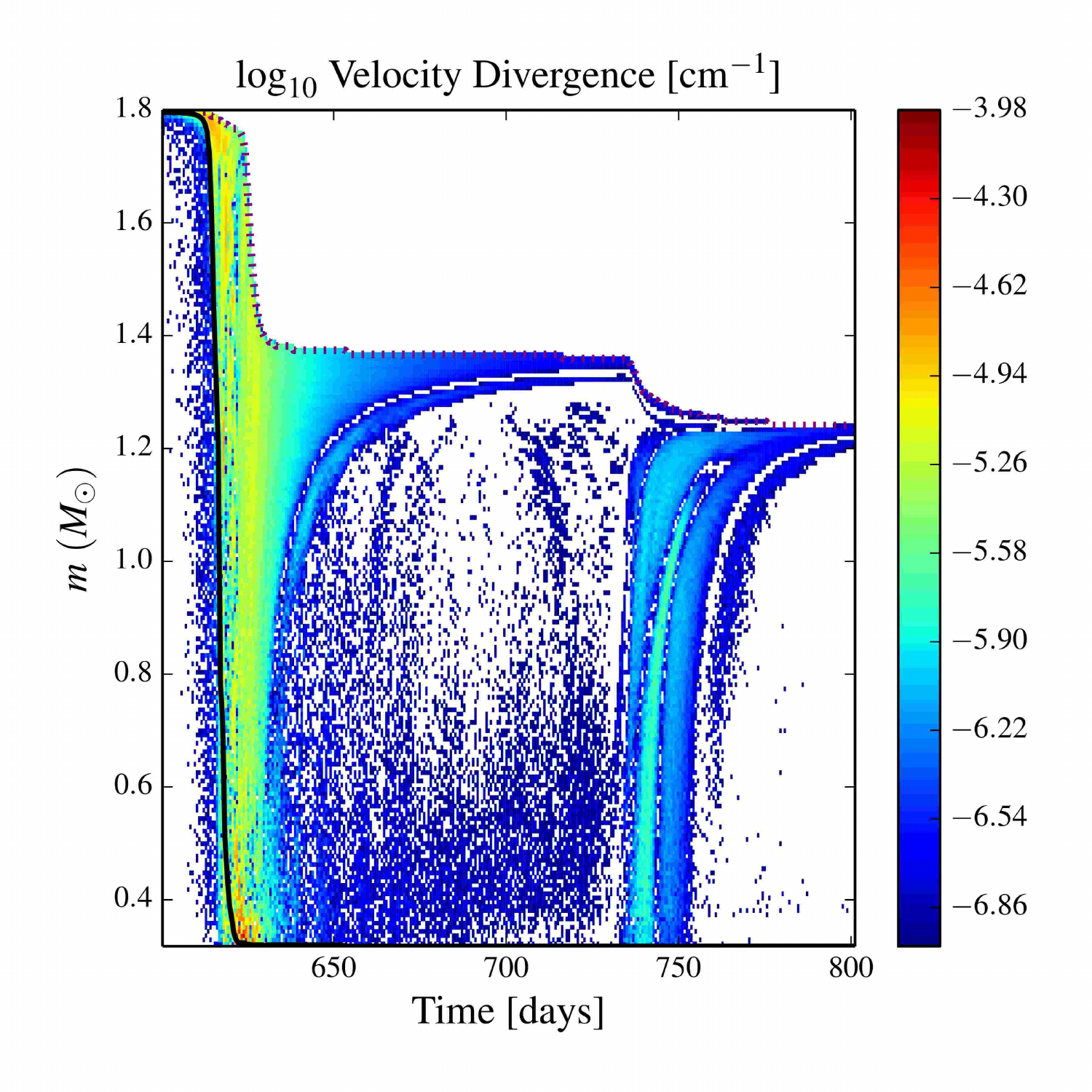}

\hskip-0.2cm\includegraphics[width=90mm]{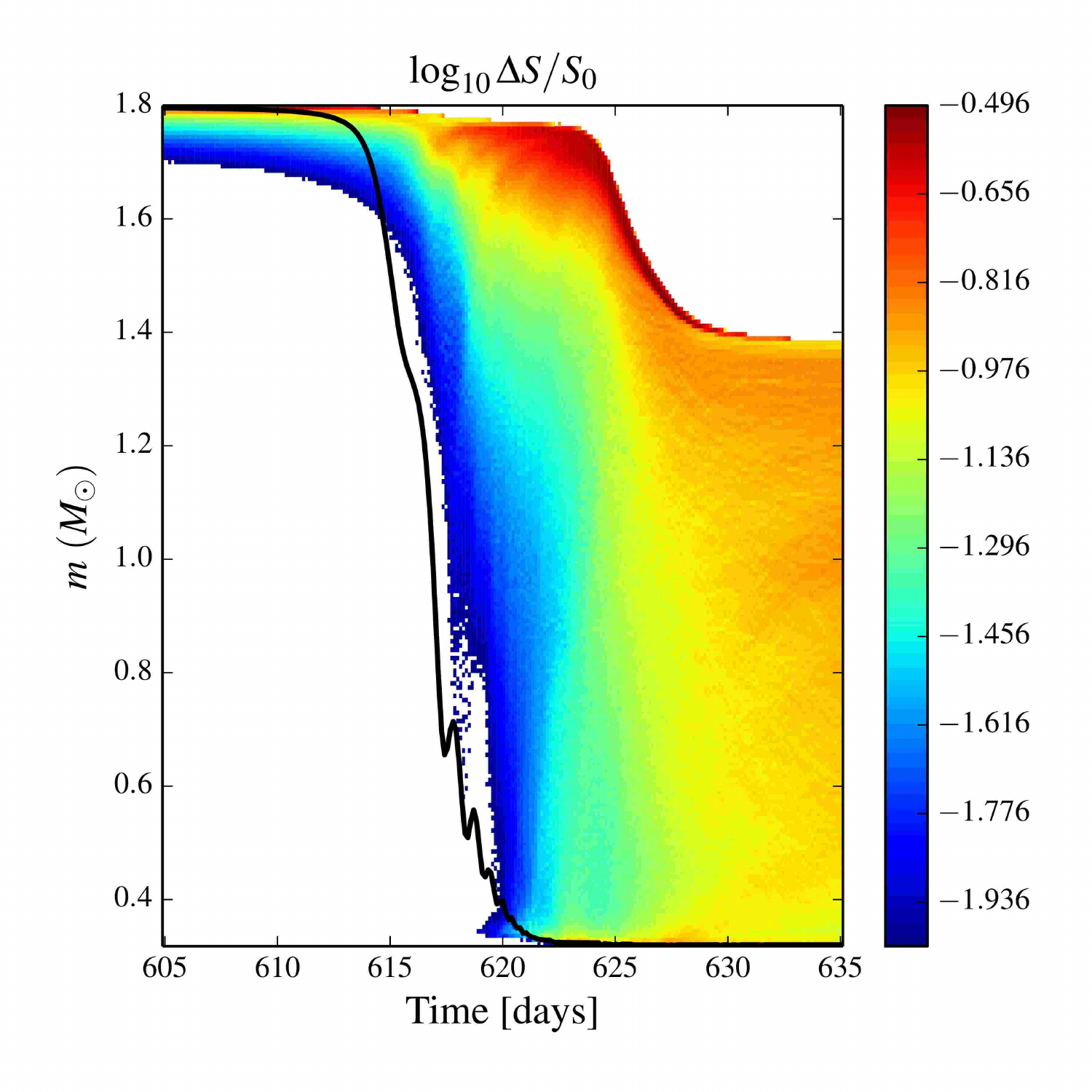}
\hskip-0.2cm\includegraphics[width=90mm]{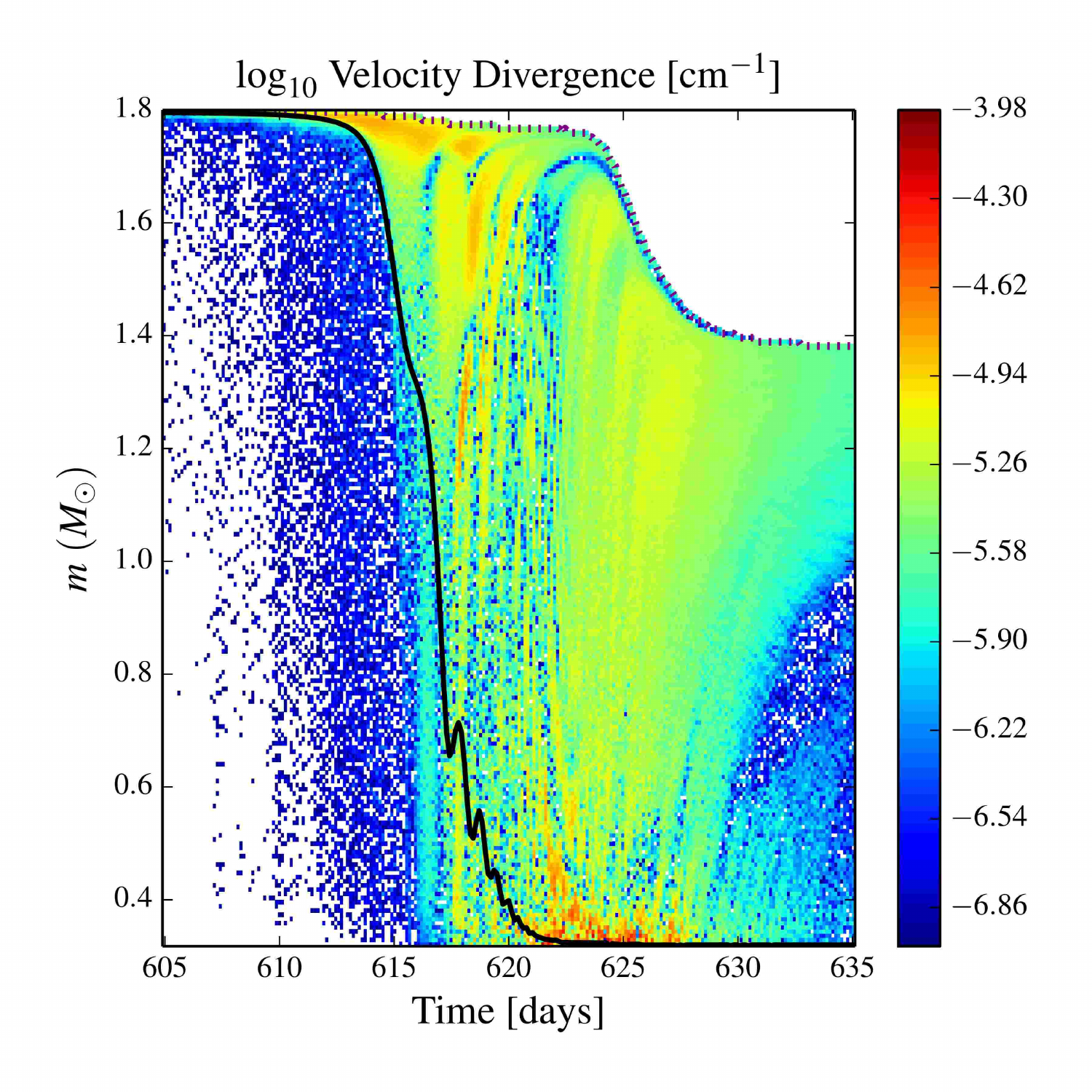}

\caption{Entropy  generation (shown  as  the ratio  of the  cumulative
increase of the entropy from the start of the simulation, over its  initial value, left panels) and 
velocity divergence 
(right panels).   
We show the case  SS15.  The top panels show long-term evolution, 
and the bottom panels show a zoom-in of the plunge-in phase.
Black solid lines show the location of the companion.
Dotted purple lines on the plots for velocity divergence indicate the surface of the envelope. }
\label{fig:entrss15}
\end{figure*}


\begin{figure*}

\hskip-0.2cm\includegraphics[width=60mm]{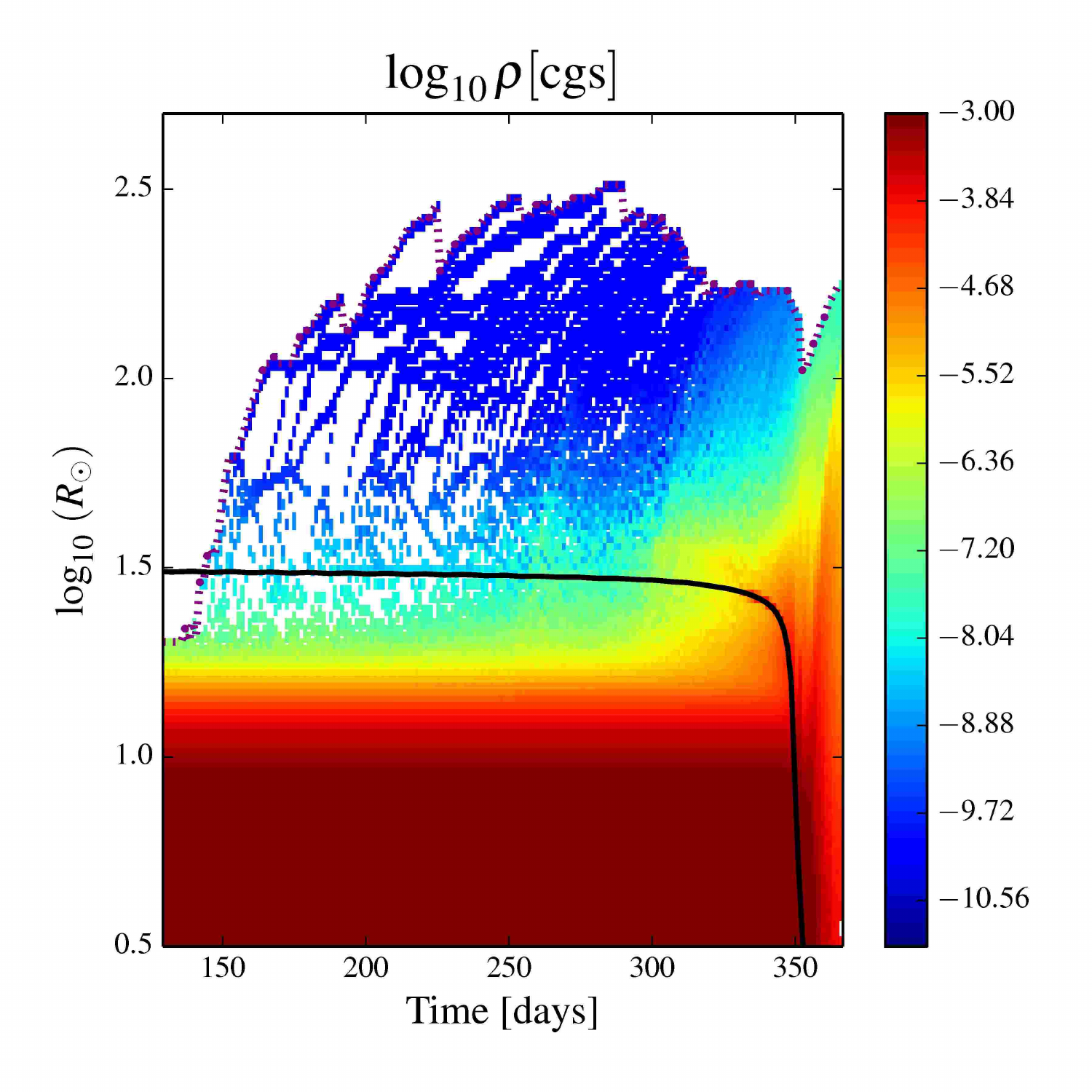}
\hskip-0.2cm\includegraphics[width=60mm]{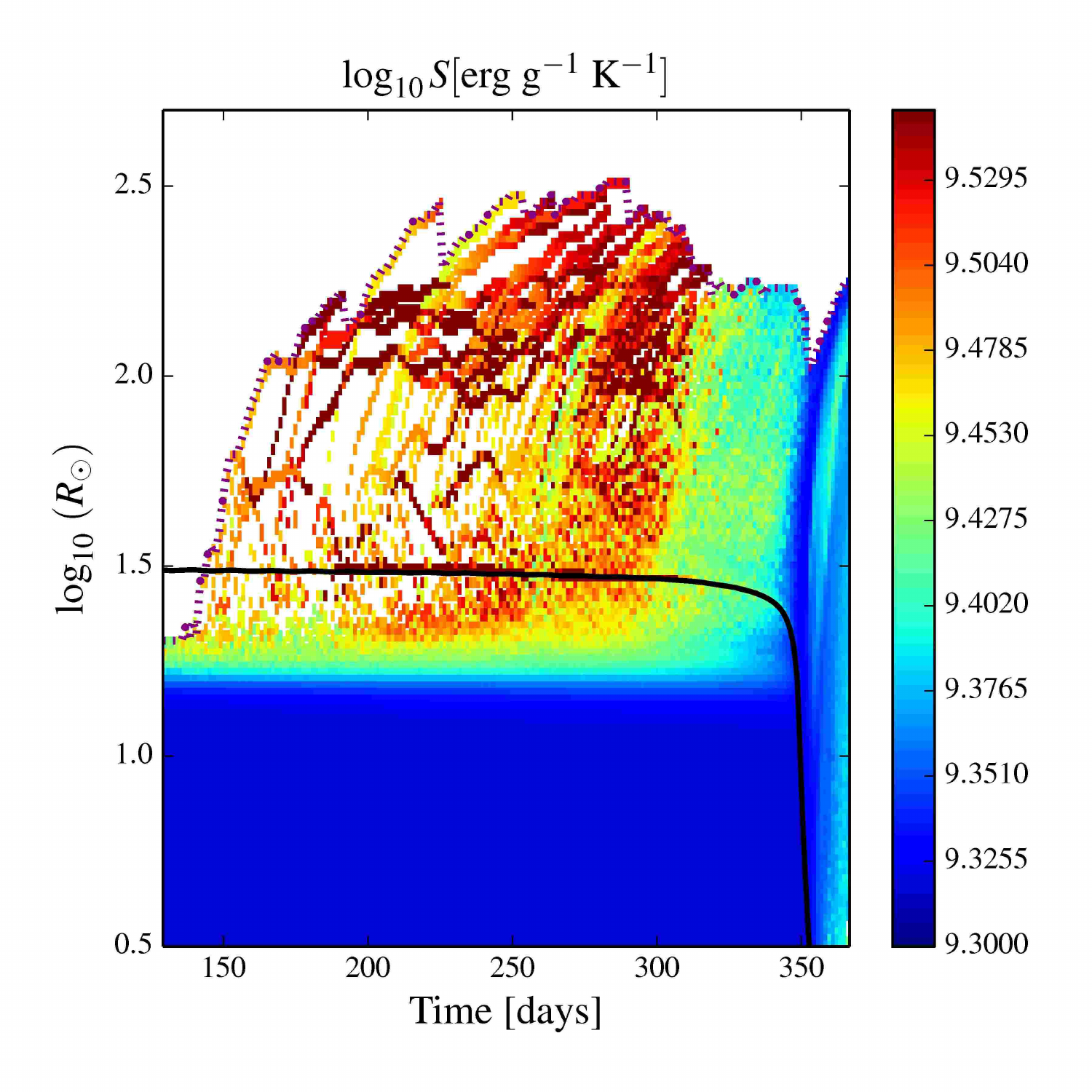}
\hskip-0.2cm\includegraphics[width=60mm]{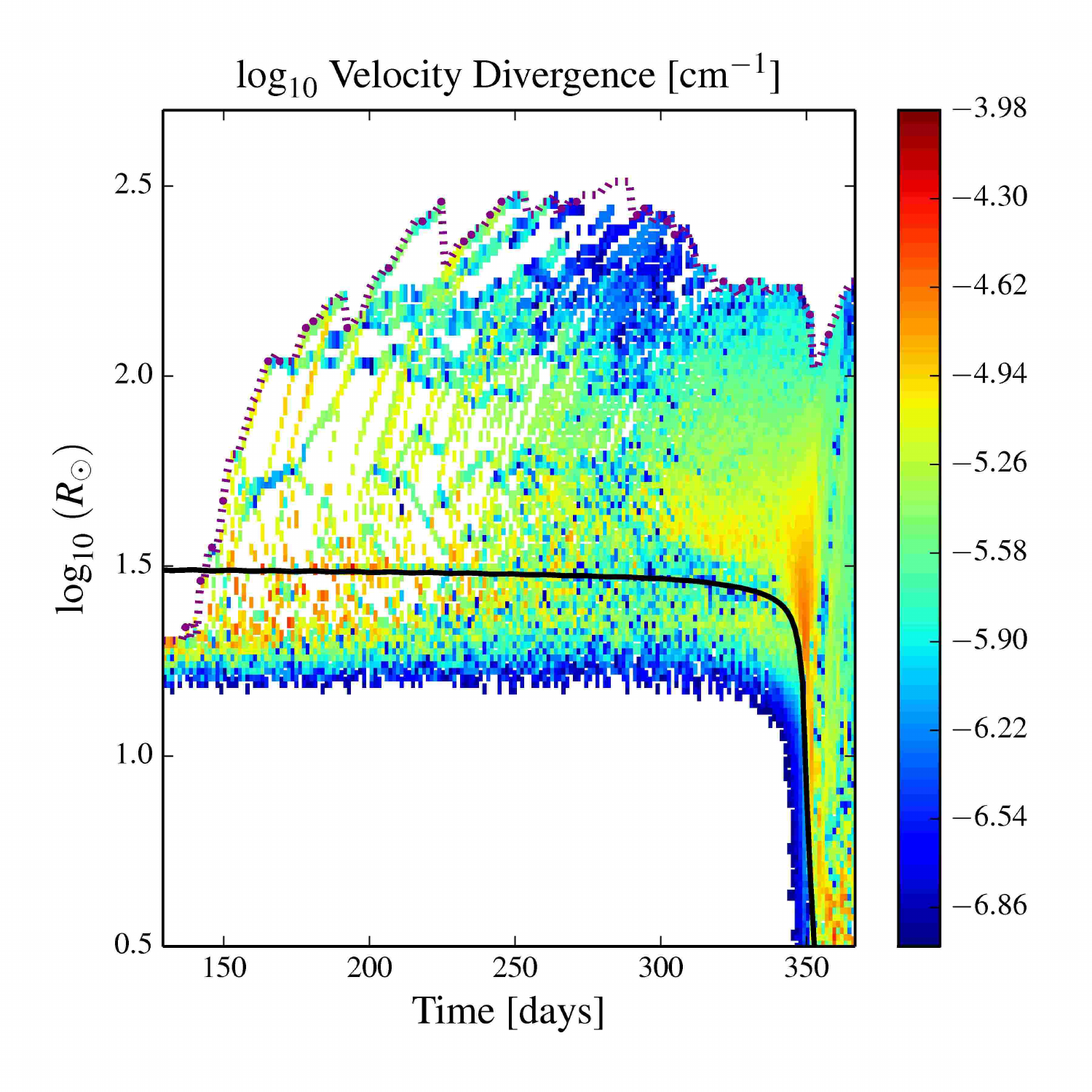}

\hskip-0.2cm\includegraphics[width=60mm]{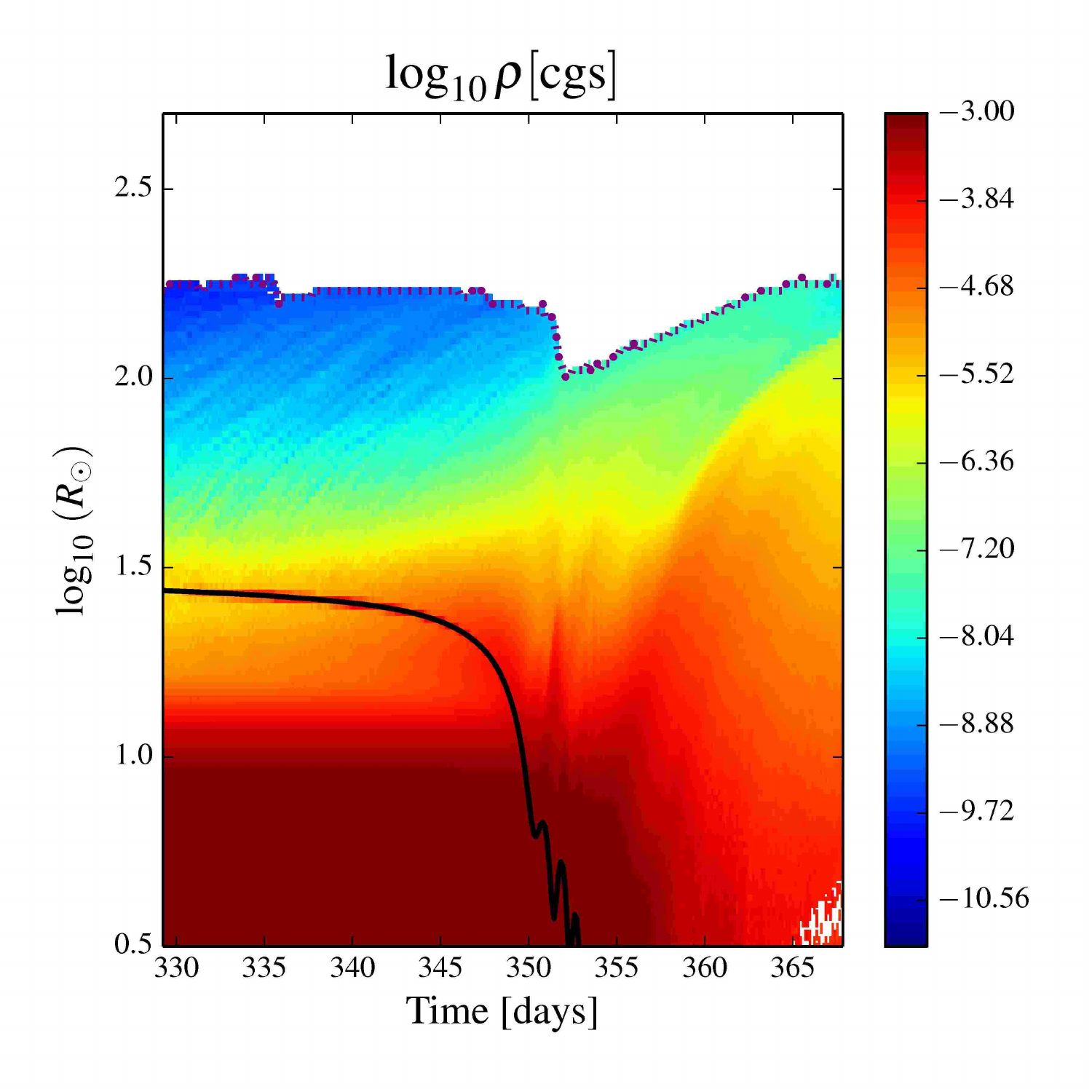}
\hskip-0.2cm\includegraphics[width=60mm]{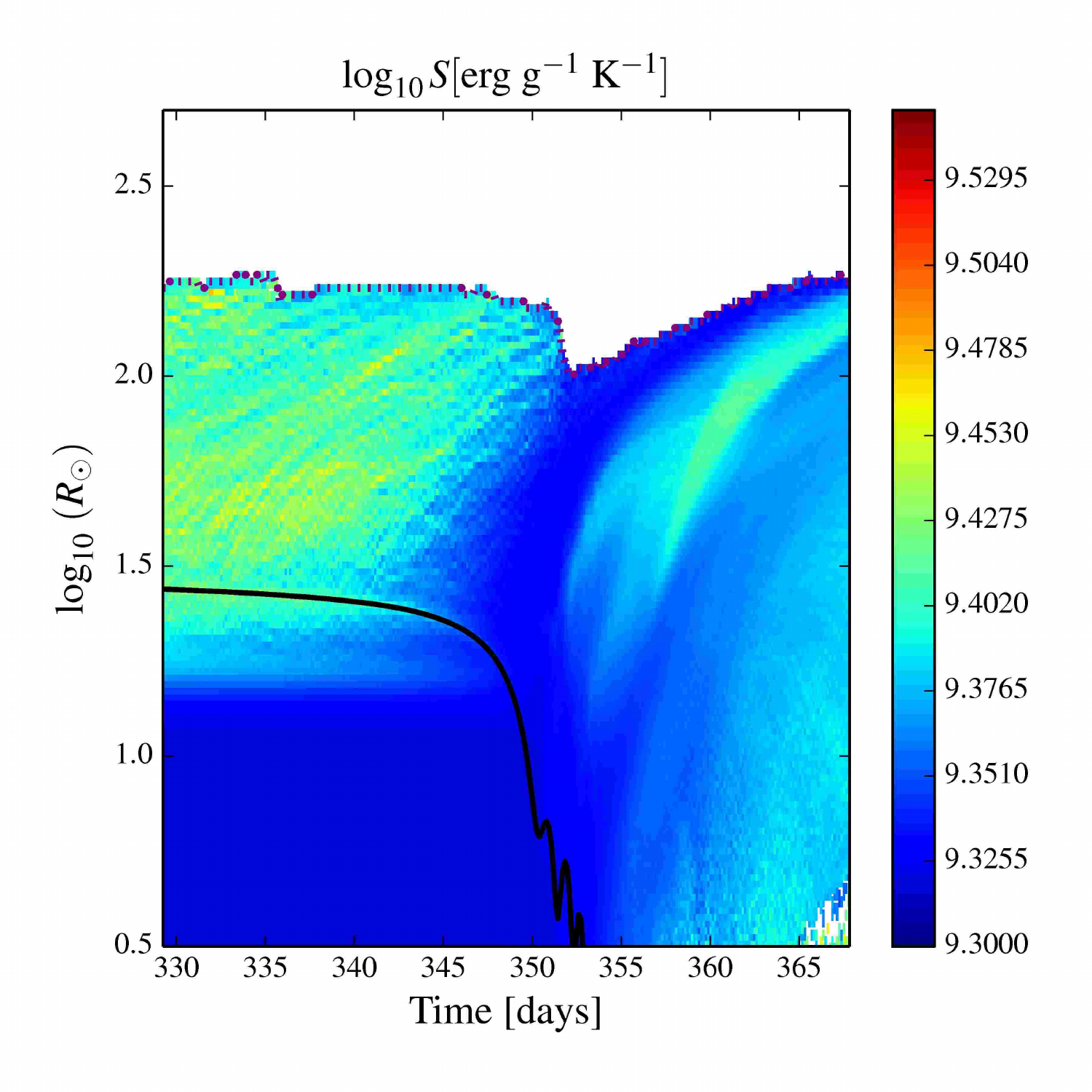}
\hskip-0.2cm\includegraphics[width=60mm]{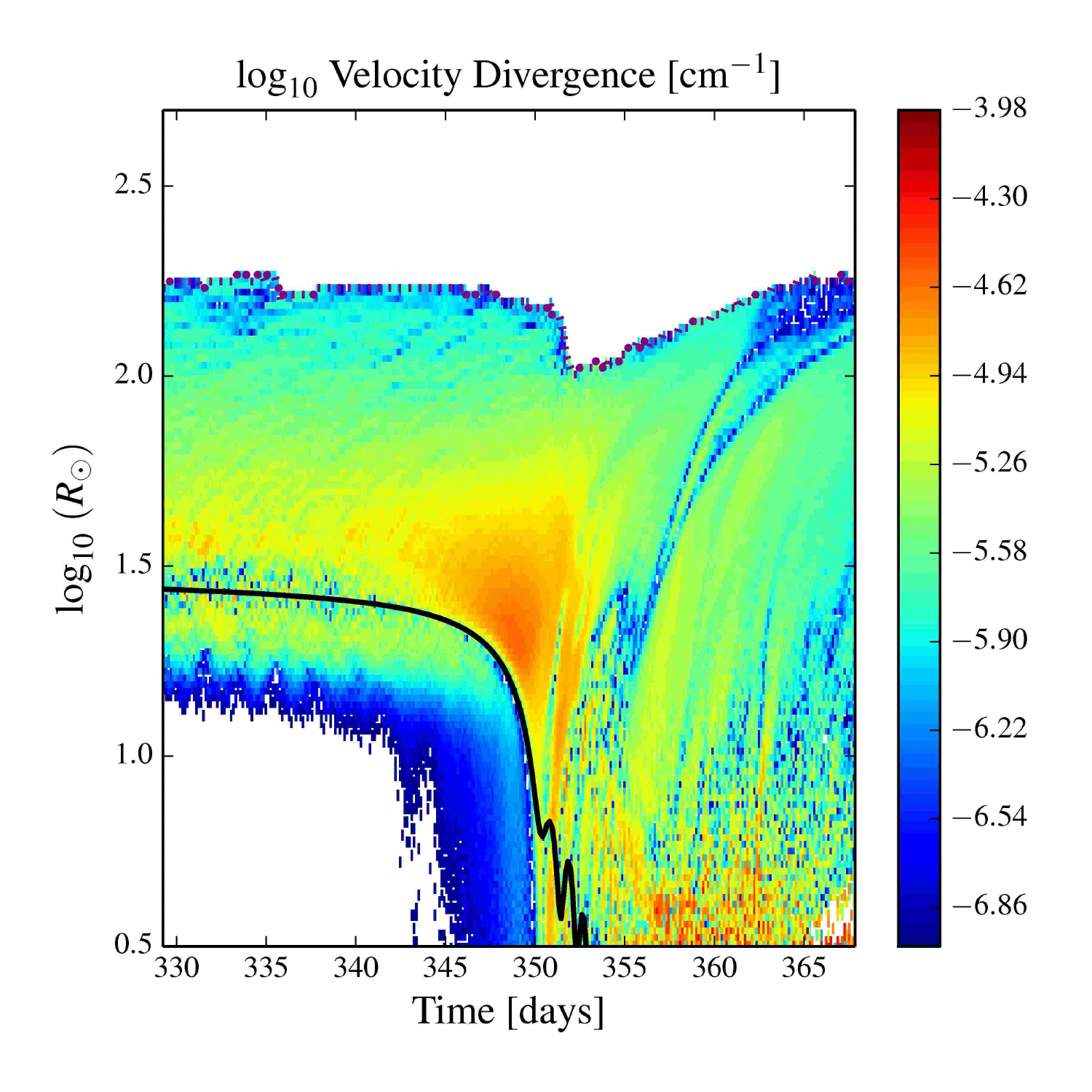}

\caption{{Evolution of the outer layers of the common envelope. We show the case  BF36.  The figure shows density (left panels), entropy (middle panels) and velocity divergence (right panels), as functions of the radial coordinate.
    The top panels show long-term evolution, 
and the bottom panels show a zoom-in to the plunge-in phase.
Black solid lines show the location of the companion. The dotted purple lines indicate the surface of the envelope.
Note that the plots show only the material that remains bound, not the ejecta. }}
\label{fig:entrbf36}
\end{figure*}

\section{Entropy}

\label{sec:entropy}

In 1D studies, a common feature in  the results is the formation of an
``entropy bubble'' in the envelope \citep[as can be seen in figures
  in, e.g.,][]{IvanovaThesis,2015MNRAS.447.2181I}.  An entropy bubble is when
  part of  the material  of the  envelope acquires  a significantly
higher  entropy  than it  had  before  the  CEE  (note also, in
convective envelopes, the pre-CEE entropy  profile of a low-mass giant
envelope is close to flat, with  the exception of the surface layer).
The entropy  can increase by a  factor of a few compared  to its initial
value.   This high  entropy region  also is  often separated  from the
surface by a  region where the entropy did not  change strongly from
its initial value, forming therefore an internal entropy bubble.

The reason for this entropy bubble formation is the way that the energy
conservation  is  implemented in  1D  codes.   To conform  with  energy
conservation, the  energy that is  released from the  shrinking orbit 
has to be added to the envelope. And the commonplace way to do that is
by adding a new ``luminosity''  term in the energy equation \citep[see
  e.g.][]{Taam+1978,1979A&A....78..167M}.   This consideration,  while
using the  term ``luminosity'',  is equivalent to  adding the  net heat
$dQ$ to the internal energy of the envelope material, and as a result 
increases the entropy.

Recently, it has been shown that {\it where} the energy is added, does
matter, and can  define the outcome. The same  overall energy input
to the envelope, but added at different locations, can result
either   in  a   slow   spiral-in,  or   in   the  envelope's   ejection
\citep{2015MNRAS.447.2181I}.  In \S\ref{sec:rec} we  will show that it
matters in {\it what form} the energy is added, and the entropy is the
key to understanding the difference.

In our 3D simulations, entropy is generated due to shocks and shear friction,
where artificial viscosity leads to  dissipation of local velocity differences by
converting them into heat. As a sanity check, we have checked that our code provides
perfectly adiabatic evolution when material recombines, if no shocks or other
matter interactions between the particles are involved, and there is no artificial heating term. 
The velocity divergence is a good indicator of both shocks and a strong shear
\citep[see, e.g., discussion in][]{2010MNRAS.408..669C}.
Therefore one of the best ways to trace the
entropy  generation is  by  looking at  the  velocity divergence  (see
Figures \ref{fig:entrss15} and \ref{fig:entrbf36}). The  velocity divergence is one of the main variables
in the SPH code we use and is calculated using Equation (19) from \citet{2006ApJ...640..441L}.
We  can distinguish how the entropy is generated during the three phases.

During phase I, prior to the plunge-in, when the companion's orbit is
still outside the radius of the donor, we find that the surface layers
are ``shock heated'' and obtain high entropy.  Those surface layers ``overheated'' by
shocks are quickly ejected.

A similar behavior continues even while  the companion is orbiting inside
the envelope, which had expanded to several times its initial size.  While the companion
is already orbiting inside the envelope (but before the plunge-in), very little of the donor's mass 
($\ll 1\%$) is outside the companion's orbit.  Both the entropy
generation  area   and  the  region  with   high  velocity  divergence
propagate inside the  orbit, down to the initial  donor's radius, but
do not  affect much of  the envelope's mass {(this can be seen best
  in Figure \ref{fig:entrbf36})}.

The situation changes  when more than 1\% of the  donor's mass surrounds
the orbit, and the companion plunges  inside the layers that were not
shocked previously.  Rapid  plunge-in takes place, and  the entropy is
mainly generated outside the orbit.   The material with high entropy and
high angular  momentum becomes  quickly unbound, leaves  the envelope,
and  cannot  any  longer be  seen  in Figures \ref{fig:entrss15} and \ref{fig:entrbf36}.

During the  slow spiral-in phase,  we detect entropy  generation due  to two
processes: quick expansion provides shocks once more, and some entropy
possibly comes from  the flattening of the rotational  profile. 
{As a result, the entropy is increased in the internal part of the envelope, while 
 in the outer part of the envelope the entropy is almost unchanged (see Figure~\ref{fig:entrss15},
 where this entropy generation can be  seen during the two episodes of mass  ejection 
 at the top panel showing the long-term  evolution of  SS15).
 This entropy generation can appear to be similar to what was described above as an ``entropy bubble''
 observed in the past 1D studies.
It is noteworthy}, however,
that the entropy  is increased  inside
this bubble only by about 30\% as compared to its initial value. This is 
much less than predicted  by 1D  simulations (a factor of several times).
The importance of this lack of entropy generation in 3D as
compared to 1D will be shown in \S\ref{sec:rec}.
 

%% file: sect7_rec.tex
\section{The role of the recombination in driving the CE ejection}
\label{sec:rec}

\begin{figure*}

\hskip-0.2cm\includegraphics[width=60mm]{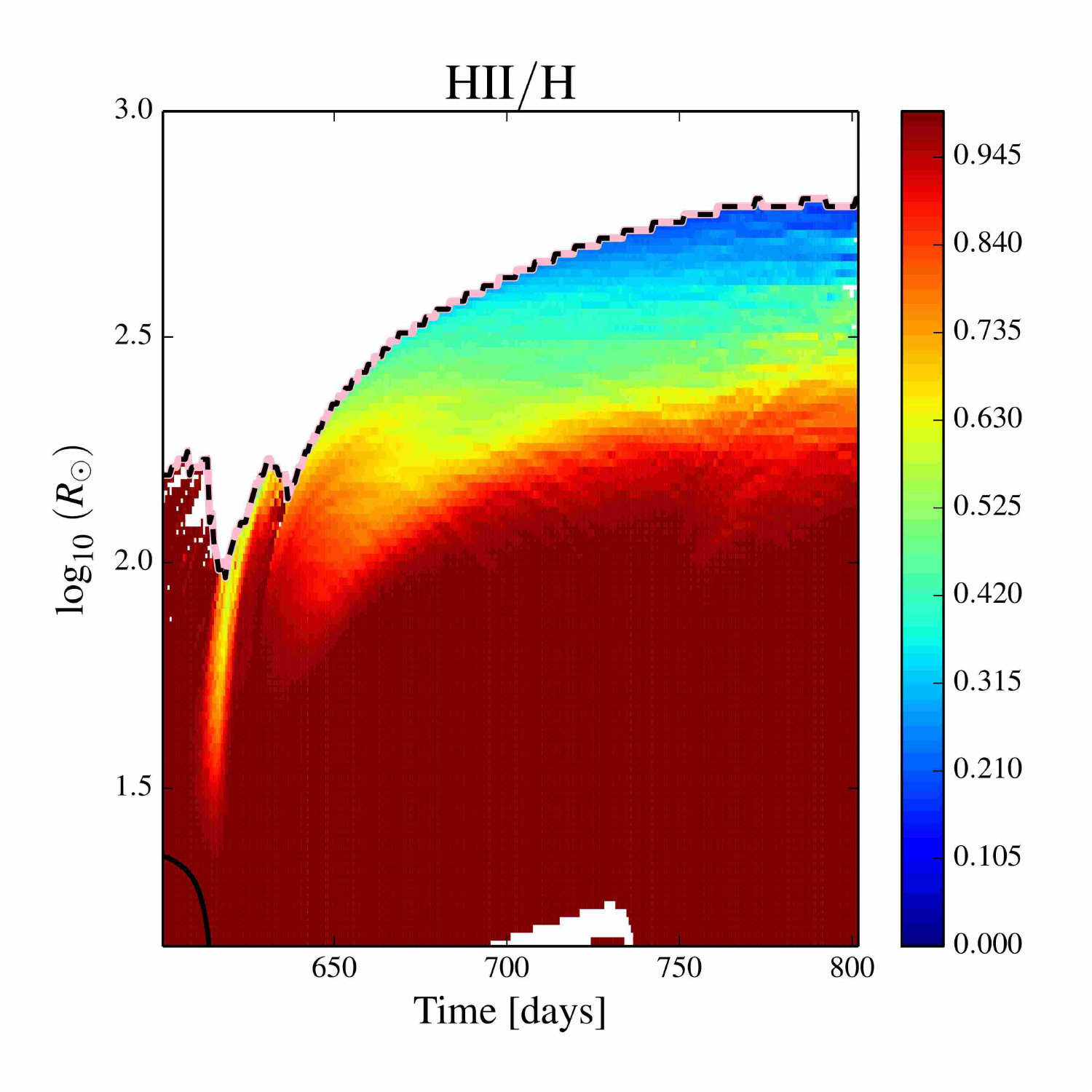}
\hskip-0.2cm\includegraphics[width=60mm]{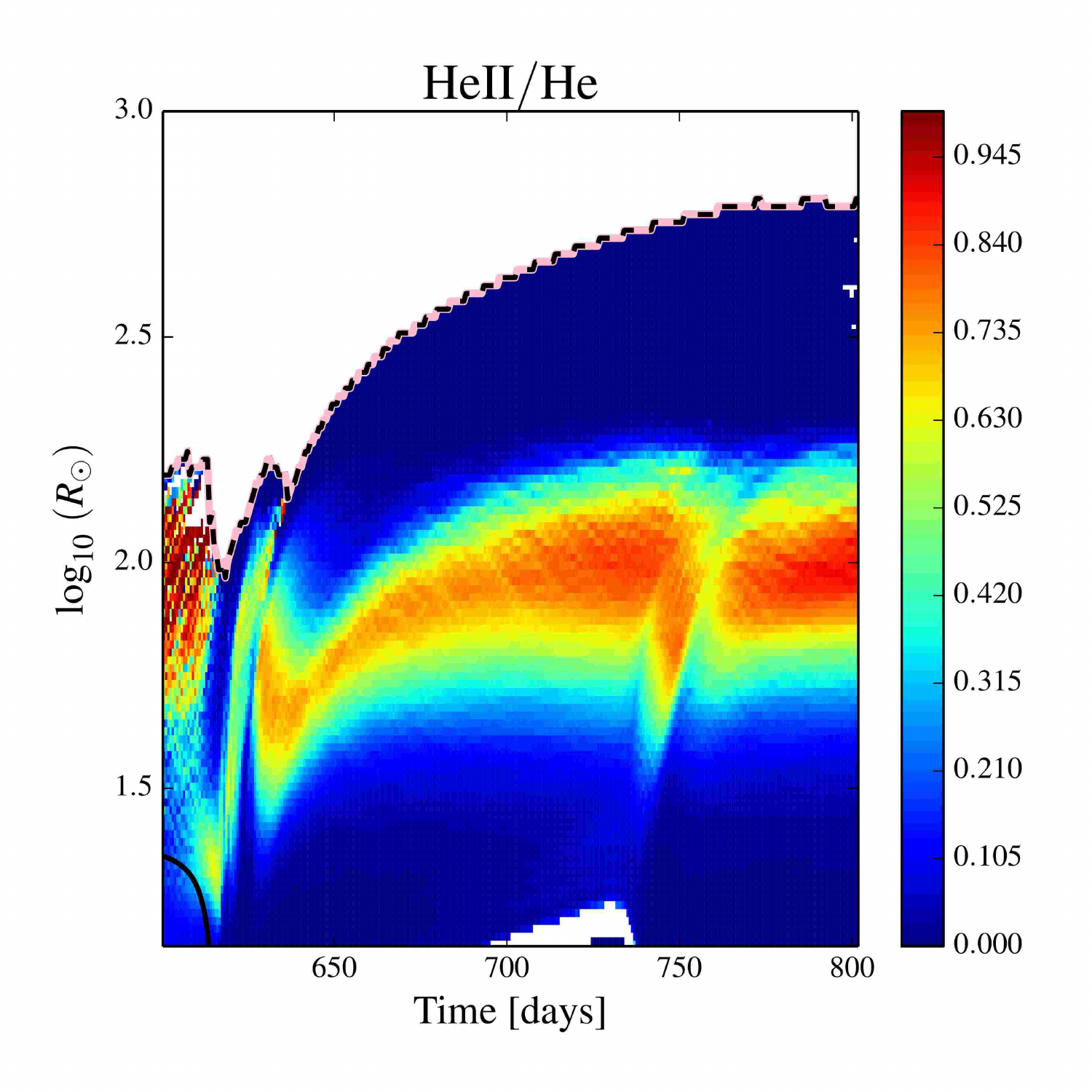}
\hskip-0.2cm\includegraphics[width=60mm]{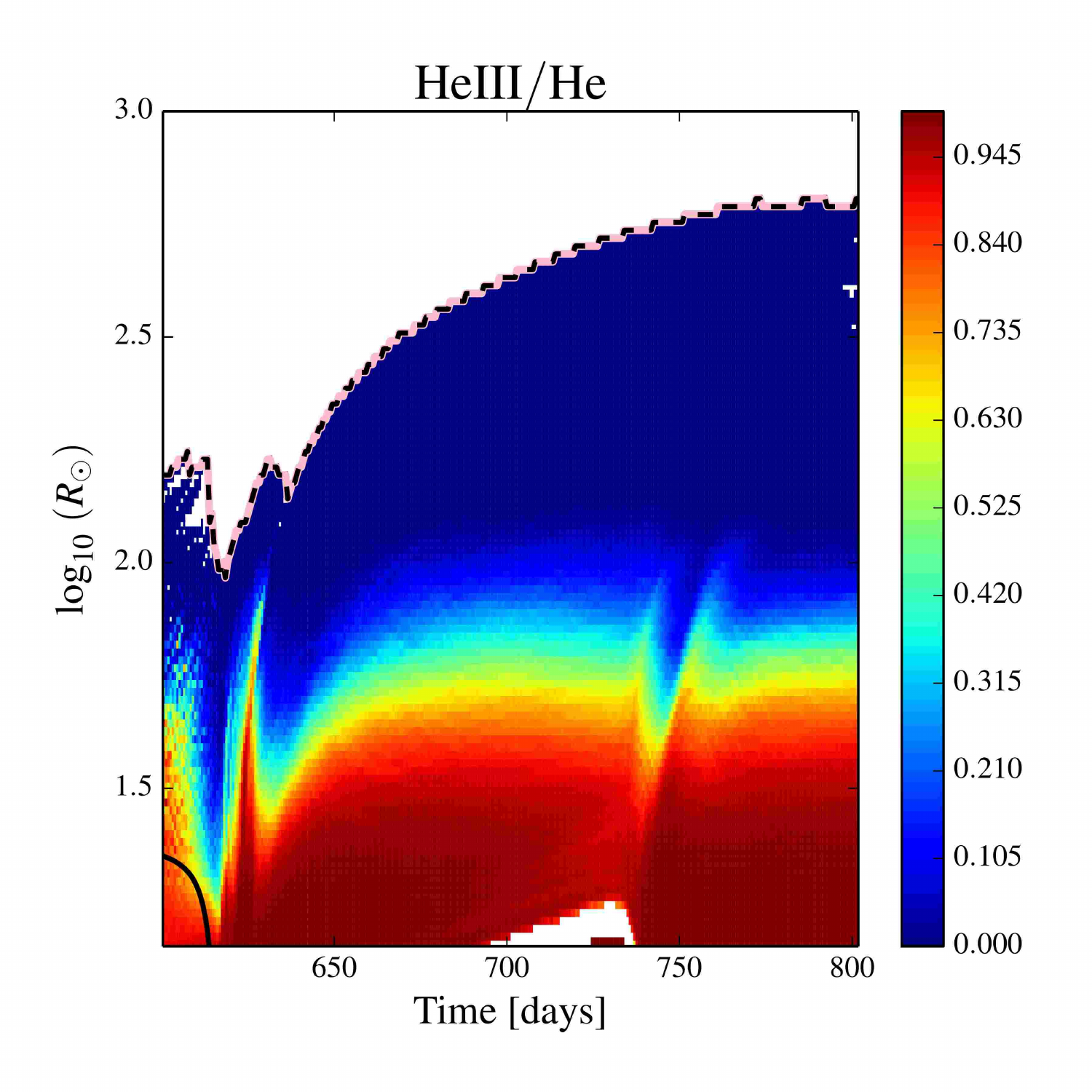}

\hskip-0.2cm\includegraphics[width=60mm]{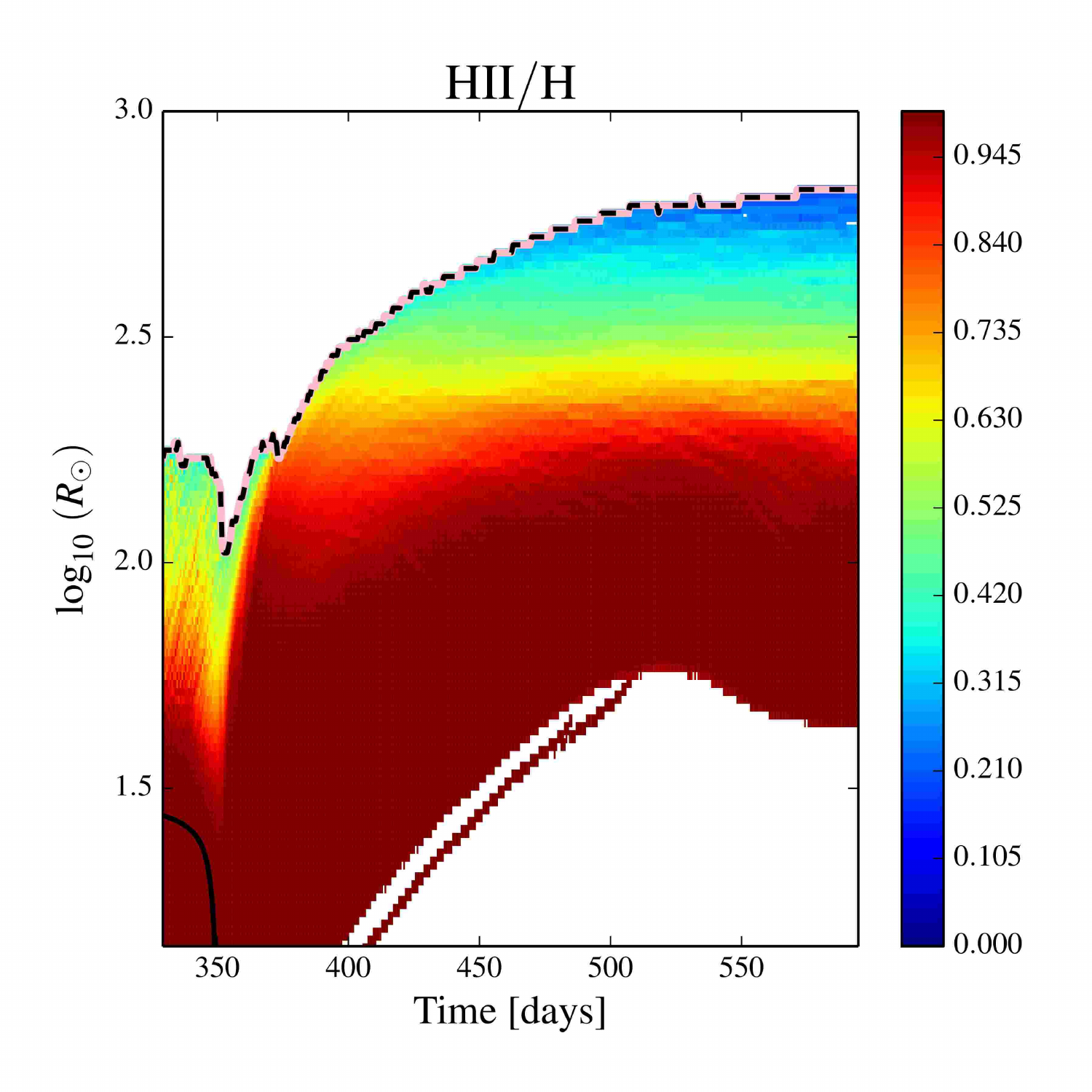}
\hskip-0.2cm\includegraphics[width=60mm]{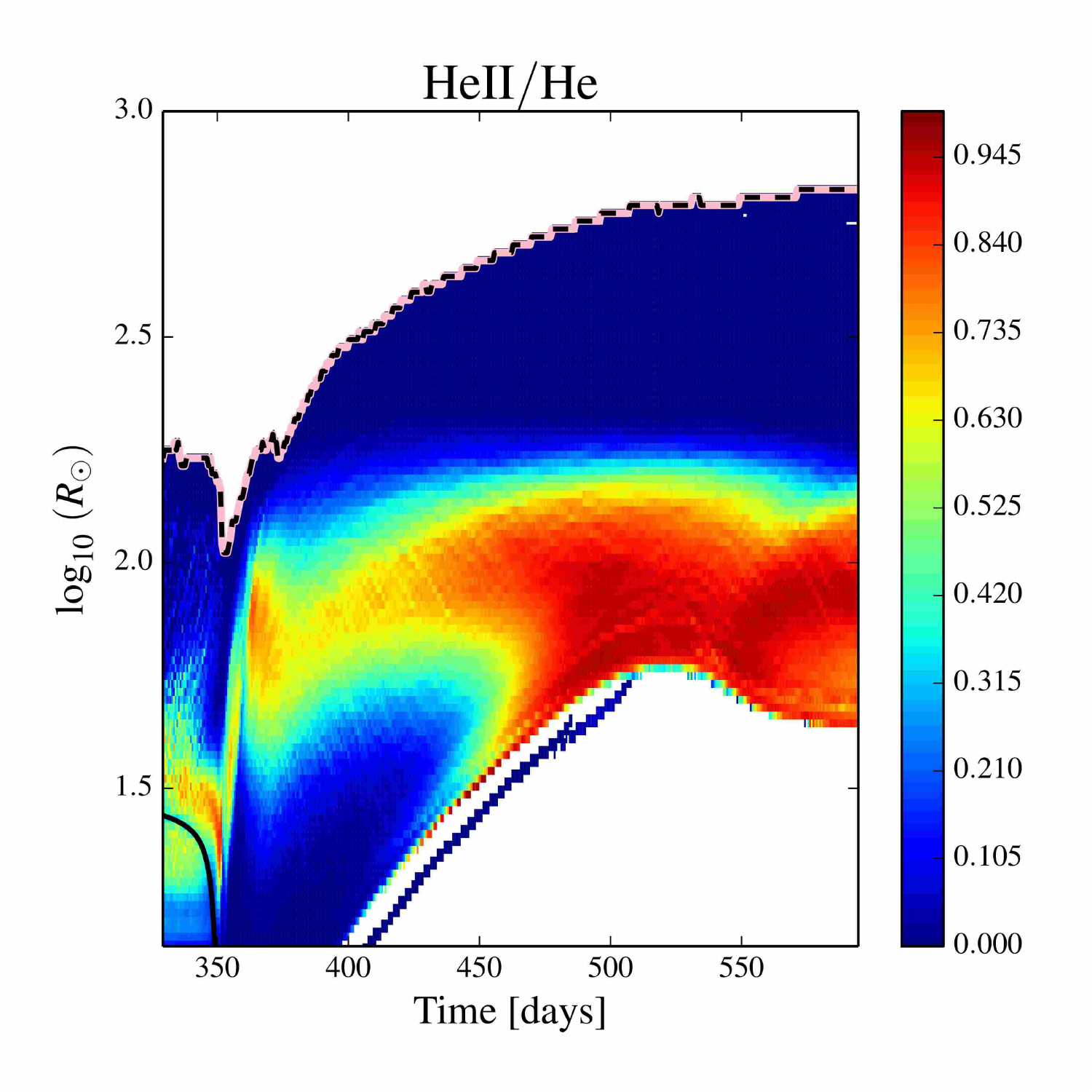}
\hskip-0.2cm\includegraphics[width=60mm]{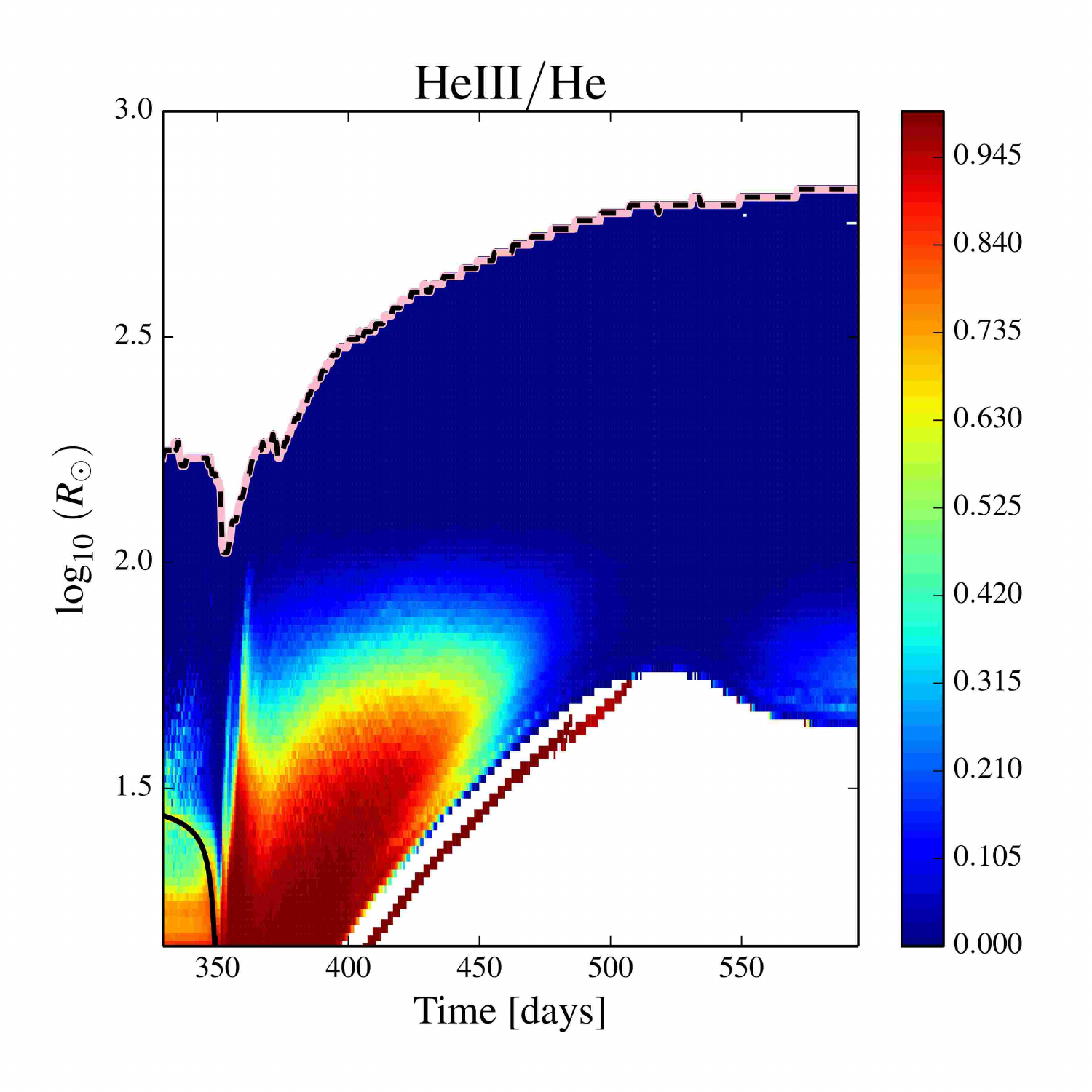}

\caption{  Locations  where  hydrogen  and  helium  are  in  different
  ionization  states  during the  CE  evolution.  In each  panel,  the
  relative fraction of the considered ion  is plotted (from 0 to 1.0).
  We show the  cases SS15 (top panels) and BF36  (bottom panels).  The
  plunge-in  in the  SS15  case ends  at about  622  days. During  the
  illustrated evolution,  the envelope loses $\sim0.4  M_\odot$ (right
  after the plunge-in).  At about  735 day, a shell-triggered ejection
  takes place  (see \S~\ref{sec:ej}).  The  plunge-in in the  BF36 case
  ends at about 360 days, and during and immediately after this plunge
  the envelope  loses $0.7M_\odot$. After  the plunge-in in  BF36, the
  envelope  loses $0.4  M_\odot$ on  a longer  timescale. Black  solid
  lines indicate the location of  the companion. Dashed lines indicate
  the surface of the envelope.}
\label{fig:rec1}
\end{figure*}

The specific   energy  that  is  released  due  to
recombination  of hydrogen  and helium,  in the case that they were  initially
fully ionized, is


\begin{eqnarray}
\varepsilon_{\rm rec} \approx 1.3\times 10^{13} {\rm erg\ g}^{-1} \times \left ( X f_{\rm HI} + Y f_{\rm HeII} +  1.46 Y f_{\rm HeI}  \right )
\end{eqnarray}

\noindent  Here $X$  is the hydrogen  mass  fraction, $Y$  is the helium  mass
fraction,  $f_{\rm  HI}$  is the fraction of  hydrogen  that becomes
neutral,  $f_{\rm HeI}$  is the  fraction  of helium  that becomes
neutral, and $f_{\rm HeII}$ is the  fraction of helium that becomes only
singly ionized.   In the giant star  that we have  modelled, the  envelope during  the giant
stage has $X=0.673$  and $Y=0.306$.   In the case  of 
complete  recombination,  the  released   energy  is  $\sim  1.5\times
10^{13}$ ergs per gram.  The equation of state that we use in our SPH
code  also  takes  into  account ionization  of  other  elements.
In more details, we use the tabulated  equation of state  incorporated  from {\tt
  MESA} \citep[see \S4.2 of][]{2011ApJS..192....3P} and implemented as
described  in \citep{2015MNRAS.450L..39N}.   This  tabulated  equation of state includes  recombination
energy for H, He, C, N, O, Ne, and Mg.
The dominant contribution to the total recombination energy comes from hydrogen and helium, and the other
  elements only provide 3\% of the total recombination energy.
  When averaged over the whole envelope, the available specific recombination energy per gram for our donor is
  $\simeq 1.6\times10^{13}$ erg/g, providing  $4.7\times10^{46}$ ergs in total.

We can compare  the efficiency of this energy release  to the specific
potential  energy of  the same  matter,  assuming that  the matter  is
located at  a distance $r$ from  a gravitating mass to  which it is
still bound, $m_{\rm grav}$:

\begin{equation}
\varepsilon_{\rm pot} = - \frac {G m_{\rm grav}}{r} = - 1.9 \times 10^{15} 
{\rm erg\ g}^{-1}  
\ \frac{m_{\rm grav}/M_\odot}{r/R\odot} 
\end{equation}

Next we use the standard assumption that in a stellar envelope,
the binding energy is about half of the potential energy (note that
here we use  the binding energy in its 1D  definition).  Comparison of
$\varepsilon_{\rm rec}$  and $\varepsilon_{\rm  pot}$ shows  that {\it
  if}  the material  is cooled  down to the point when it can start recombination,  and the
material at the  same moment is located at $r_{\rm  rec}\ga 65 R_\odot
\times  m_{\rm grav}/M_\odot  $,  then {\it  recombination alone}  can
eject the  material with  no other energy  sources needed.   Note that
this radius, depending  on $m_{\rm grav}$, is smaller  than the radius
of evolved stars  (e.g., asymptotic giant branch  stars), however, the
material in  the envelope  of an  evolved star is  usually too  hot to
start  the recombination.   For  the hydrogen  recombination alone  to
trigger  the material  outflow, $r_{\rm  rec,H}\ga 105  R_\odot \times
m_{\rm grav}/M_\odot $.  As more of the envelope matter is lost, the
radius at  which the  recombination can  act as  an energy  source powering the outflow, gets
 smaller.  When the upper layers  of the envelope are lost, the
internal  layers  start their  expansion  towards  a new  equilibrium.
Together, these  two effects  produce a  recombination runaway  of the
envelope that can take place  once the recombination starts \citep[see
  also][]{2015MNRAS.447.2181I}.

Indeed, let us have a look at the case BF36, where the entire envelope
is lost.   After the  plunge-in, during the  slow spiral-in,  when the
envelope is  constantly outflowing, the recombination  takes places at
about the same  radius continuously  (see Figure~\ref{fig:rec1}).  The
inner  hollow shell has expanded  so much,  that there  is no  double
ionized helium left.  Hydrogen recombination starts at about $\sim 155
R_\odot$ (this  is the  radius where  5\% of  H is  recombined), while
$r_{\rm  rec,H}\la 115  R_\odot$  (and it  decreases  as the  envelope
outflows). This case is a clear recombination runaway.

Now let us consider the case SS15. Hydrogen recombination starts a bit
lower.  E.g.,  on  day 720,  5\% of hydrogen is  recombined at $130
R_\odot$.    Since  the   post-plunge-in  mass   below  the   hydrogen
recombination zone  is larger than in  the case BF36, we  find $r_{\rm
  rec,H}\sim  160  R_\odot$.   Hence, hydrogen  recombination  is  not
capable of rapidly ejecting the envelope in this case.

We find the case SS20 to be intermediate; similarly to the case BF36,
 the recombination is  established to  take place  at the  radius above  $r_{\rm rec,H}$.
However, the  mass loss rate  is noticeably  lower than the  mass loss
rate in the case  BF36. It is hard to classify this  now as a runaway,
as  the timescale  to lose  the entire  envelope is  longer than  the
dynamical  timescale.  We  classify  this as  a  steady  recombination
outflow.

Why, in the  case of a more massive companion,  does the recombination
start at  a larger radius?  In  the case of a  more massive companion,
deeper, initially  hotter envelope  layers gain  the energy  to expand
enough to start  recombination.  E.g. in the SS15  case, the immediate
post plunge-in expansion leads to helium recombination starting at the
$0.8 M_\odot$ mass coordinate, while in the BF36 case, it reached down
to  $0.45 M_\odot$.  In the  latter  case, the  disturbed layers  were
initially  hotter, and  hence this  material had  to expand  to larger
distances before it was cold enough to recombine.

We can  consider adiabatic  expansion of gas  that initially  had some
specific  potential energy  and  specific internal  energy, when  some
(specific) energy  is added to the  gas. The sum of  these energies is
the boosted  total specific energy of  the gas.  The layer  will start
expanding,  to find  itself at  a new  ``equilibrium'' radius  $R_{\rm
  eq}$, where its  total energy, assuming that the gas  is not moving,
is the sum of its new potential energy and new internal energy, and is
equal  to the  boosted total  specific energy  of the  gas at  its old
location.  In that  new location, the gas will  have $T(r)/T_0 \approx
(R_0/R_{\rm eq})^{2}$, where  $T_0$ is the gas  temperature before the
expansion, and $R_0$ is the location  of the gas before the expansion.

Now it can be seen why the difference in the entropy generation between 3D and 1D models,
found in \S~\ref{sec:entropy},  plays a role in the outcomes.
In the  case when mechanical  energy is  added, as in  3D simulations,
$T_0$ is the initial temperature of the (almost) unperturbed star.  In
the case when the  energy is added to the internal  energy in the form
of heat (resulting in a substantially larger entropy increase
  in the shared envelope),  as in  1D simulations,  $T_0$ is  increased compared  to the
initial temperature of the same layer in the same unperturbed star.

Let us  compare the effect  of adding  the energy mechanically,  or as
heat, on  a specific example  -- the 1D profile  of the star  that was
used to create our giant in the  3D CEE study presented here.  To each
mass mesh point in  this 1D profile, we add the  same amount of energy
via the  two ways described,  and find  the temperature when  the mass
shell  reaches its  equilibrium radius.   Indeed, we  find numerically
that the second case (where the energy  is added as heat) results in a
smaller  equilibrium  radius than  the  first  case (where  mechanical
energy is added).  Adding heat energy  also results in matter having a
hotter temperature  when it reaches  the equilibrium radius.   When we
compare the temperature of two mass  shells that have reached the same
equilibrium radius  after the energy  was added,  we find that  in the
added heat energy  case, the temperature at the  equilibrium radius is
also hotter.   While no  deterministic conclusion can  be made  on the
trend for  all possible  cases, it  is clear that  when the  energy is
added  directly  to the  internal  energy,  the  radius at  which  the
recombinations will start is different.

%% file: sect8_ejecta.tex
\section{The ejecta}

\label{sec:ej}

 In our simulations we distinguish four types of ejection processes:

 \begin{enumerate}[label=\arabic*.,leftmargin=2.0em,itemindent=0em,listparindent=2em]
\item\underline{\itshape``Initial ejection''}  of the outer  layers of
  the  original  envelope,  $\delta  M_{\rm  ej}^{\rm  start}$.   This
  low-mass ejection  process takes place before  the plunge-in starts,
  and is seen in all simulated CEEs.
  {A hint of this initial low-mass ejection is suggested from 
    the evolution of the expanding  outer layers of the envelope
    (prior to the start of the plunge-in at about 350 days) shown in Figure~\ref{fig:entrbf36}. 
    }
  
  \cite{Ivanova+2013Science}  have  argued,  based  on  the  V1309~Sco
  merger, that the  ejecta that accompanies a merger  can be estimated
  by comparing the orbital energy release $\delta E_{\rm orb}(r)$ with
  the local  binding energy of  the envelope $\delta E_{\rm  bind, env
    1D}(r)$.  Indeed,  we  observe  that  in all  of  our
  simulations  the  orbital energy  does  not  influence the  envelope
  inside the  orbit.  This supports  the idea that the  orbital energy
  release  can be  compared with  the binding  energy of  the material
  during the spiral-in, and not only at the final orbit.

The initial  low-mass ejection,  $\delta M_{\rm ej}^{\rm  start}$, can
then be found by using:

\vspace{-\baselineskip}
\begin{equation}
\delta E_{\rm orb} (r) + \delta E_{\rm bind, env 1D}(r) =0.
\label{eq:enven}
\end{equation}

From  the  estimates  that  use  the binding  energy  profile  of  the
unperturbed star,  we find that  the expected $\delta  M_{\rm ej}^{\rm
  start}$  is 0.010,  0.022, 0.034,  0.047, and  $0.090 \,  M_\odot$, in
cases  for companion  masses of  0.05, 0.1,  0.15, 0.2,  and $0.36  \,
M_\odot$, respectively.   In 3D  simulations using the  same companion
masses, we  find that $\delta  M_{\rm ej}^{\rm start}$ is  0.01, 0.02,
0.03,     0.045,     and      $0.1     \,     M_\odot$.      Therefore
Equation~\eqref{eq:enven}  provides  a   good  method  for  estimating
$\delta M_{\rm  ej}^{\rm start}$. However,  this is the  least massive
ejection process, even in cases of mergers.

\item \underline{\it``Plunge-in  ejection''} of the  envelope, $\delta
  M_{\rm ej}^{\rm plunge}$.  This ejection  process takes place at the
  end of the  fast plunge-in, when the circumbinary  envelope has just
 formed and the fast orbit depletion has ended.  Like the ``Initial
  ejection'', it takes place in all of our simulated CEEs.
  {An example can be seen in Figure  \ref{fig:entrss15}: this is
    when the mass of the envelope is sharply decreasing  to $1.4 M_\odot$,
    on the timescale of only a  few days.
    The process of the ejection can also be observed in Figure~\ref{fig:entrbf36} for the case BF36, where
    it takes place after the plunge-in starts (approximately between 355 days and 365 days).
}

This ejection process is likely  powered by the mechanical energy that
the  envelope has  absorbed  by the  time that  the  orbit has  become
decoupled from the envelope. We did  not find a good method to predict
the associated energy  release given the initial  energy profile.  For
$\delta M_{\rm ej}^{\rm  plunge} + \delta M_{\rm  ej}^{\rm start}$, we
find  from  the  3D  simulations  that  0.03,  0.3,  0.43,  0.53,  and
$0.75\,M_\odot$ were ejected, for companion masses of 0.05, 0.1, 0.15,
0.2 and $0.36 \,M_\odot$, respectively.
In the case of BF36, half of the initial envelope was ejected during this stage.
  It is about the same fraction as was found in the
  simulations that were performed without recombination energy taken into account
  (about 40\% in \cite{2015MNRAS.450L..39N} and about 25\% in \citet{Ric2012}).
  We cannot rule out however that the inclusion of the
  recombination might have {\it enhanced} the plunge-in ejection,
  as in the case BF36 it led to 50\% of the envelope being ejected during this stage.

\item \underline{\it``Recombination runaway''} is the process that led
  to  the complete  envelope  ejection in the case BF36 \citep[as well as  in all  of  the  CEE  simulations
  presented in][]{NandezIvanova15}. It takes
  place when the post plunge-in  envelope has cooled down sufficiently
  to start hydrogen recombination, while at the same time the envelope
  size  has expanded  beyond $r_{\rm  rec,  H}$ (see  more details  in
  \S~\ref{sec:rec}).
  We find  that  in SS20  and  SS15 models,  this
  process does not start promptly after  the plunge-in has  occurred, and
  that the  material is ejected at  a slower rate during  this process
  than in the case of the  BF36 case {(see Figure~\ref{fig:rec1})}.  Therefore, for these cases this
  process can be thought of as a ``steady recombination outflow''. The
  average  mass loss  rates  of these  outflows are  2,  0.25 and  0.15
  $M_\odot$ per  year, for  BF36, SS20  and SS15  models, respectively
  (the  dynamical  timescale  of  the expanded  envelope  is  about  a
  year). In case of BF36, the recombination driven outflow slowed down
  to  about  0.4 $M_\odot$  per  year  at  about  200 days  after  the
  plunge-in.

\item
\underline{\it``Shell-triggered ejection''} is  the sudden ejection of
a substantial  part of the  envelope during the slow  spiral-in.  This
process takes place about one dynamical timescale of the expanded envelope
after the plunge-in process has ended. {This phenomenon
  can be seen for the case SS15  at about 735 days (Figures
  \ref{fig:entrss15} and \ref{fig:rec1}).}

This ejection process is partially  powered by the energy release from
the re-collapsing hollow shell, and partially powered by the triggered
recombination.   Recall, that  this hollow  shell is  bound after  the
plunge-in, and while it initially  expands, it starts to recollapse on
its dynamical timescale.  This recollapse results in redistribution of
the  energy in  the  envelope, where  the  contracting shell  provides
energy for the  outer layers, shedding away part of  these layers.  It
is not  clear at the moment  what total fraction of  the envelope this
process  can eventually  remove.  First,  this process  is accompanied
with  a   steady  recombination   outflow,  making  it   difficult  to
disentangle  the  two  processes.    Second,  while  this  process  is
continuous, it is very slow (even  slower than the outflow in the case
of  the SS20  model),  and  thus the  model  becomes a  self-regulated
spiral-in.   Third, we  cannot  yet compute  the contraction  of this
shell all  the way to the  binary orbit; thus  we can not see  if this
process can  extract more energy  from the binary and  produce several
similar re-expansions.  In the latter  case, no self-regulated
spiral-in situation can  be expected. We find however,  that the first
episode of the  shell-triggered ejection takes away almost  all of the
angular momentum  that was remaining  in the  CE, leaving the  CE with
less than  0.02\% of  its initial angular  momentum. In  addition, the
envelope that remains after the  shell-triggered ejection is only {\it
  marginally}  bound  ---  its   binding  energy  is  $3\times10^{44}$
ergs.  If there  are no  bouncing re-expansions,  steady recombination
outflow can eject the remaining envelope.

\end{enumerate}

This shell-triggered process does not  always happen even for the same
initial  binary.  We  ran  the  initial SS15  binary  twice, with  the
different Courant numbers (for gas SPH particles only). In particular,
we  changed  the second  Courant  number  $C_{N,2}$ \citep[defined  as
  described  in][]{2006ApJ...640..441L}.   Both simulations  conserved
energy and angular momentum extremely  well.  The larger value for the
Courant number  $C_{N,2}$ was 0.6  (used in the simulation  SS15), and
the  smaller value  was 0.3  (used  in the  simulation SS15sC).   Both
values that we used for $C_{N,2}$ are  in the recommended range, between 0.1
and 0.6,  for the version of the code we use  (Lombardi, priv. comm.).

We find  that in  the run  with the smaller  Courant number,  SS15sC, the
prompt ejection does not occur, and  the evolution is similar to SS20,
just   with   a   bit   slower    mass-loss   rate   of   the   steady
recombination-driven  outflows. However,  we  cannot attribute  this
divergence  between the  two runs  purely  to a  numerical error.   We
traced the difference in the outcomes to the differences in the properties  
  of  the SPH  particles that are currently in  the hollow  shell,
  where those properties were acquired during the plunge-in interaction
  with the special particles that represent the binary. 
In  the SS15sC  case, a
fraction of SPH particles in the inner region of the hollow shell have
acquired  a  higher entropy,  likely  due  to  more shocks  they  have
experienced.  These  particles have  a higher  internal energy,  and a
higher total energy, if compared to SPH particles located at a similar
distance from  the binary in the  SS15 case.  The total  energy of SPH
particles in  the shell  is about  the same,  but the distribution  of the
energies between the  particles is slightly different, and  in a wider
range than  in the case  SS15.  We cannot determine one of the 
  energy  distributions between  the  shell particles as to be more proper
  than the other.

The natural expectation for SS15sC high-energy particles is that their
``fallback'' will start  when the shell expands more that  in the case
SS15.   However,  while the  shell  was  still expanding  (beyond  the
distance it  had maximally  expanded in the  case SS15),  the material
starts He recombination  (from the double ionized state  to the single
ionized state).   This recombination gives an  additional energy boost
to the hollow shell particles, and prevents its fallback, at least for
the duration  of the simulation  we obtained  for SS15sC (for  40 days
after the fallback  had started in the case SS15).   A small deviation
in the energy for a fraction of SPH particles in the hollow shell have
resulted in  a qualitative  change of  the envelope  ejection picture.
This indicates  that even during the  established steady recombination
outflow, a  CE is recombinationally-dynamically unstable  with respect
to  its  inner  shell  recollapse.   The  condition  to  prevent  this
dynamical instability seems to start the fallback after Helium started
its recombination.  As a confirmation, we  find that a weak version of
this instability takes place in all  the models at the moment when the
inner shell started its fallback. As  a weak version, we mean that the
envelope changes  its mass loss rate  at about the same  time when the
inner shell contracts or re-expands, but this mass loss rate change is
not  nearly as  noticeable  as in  the  case of  SS15.   In all  those
``weak'' cases, Helium is not doubly ionized in the inner shell at the
time of  fallback.  To  answer on  whether the  strong shell-triggered
ejection process  is a  numerical artifact  due to  a small  number of
particles near the core {(recalling that the total number of SPH
  particles in this  simulation is 100,000)}, or is  the real physical
effect  that  occurs   if  the  fallback  starts   before  the  Helium
recombination, a study with  significantly more particles is required;
{this is unfeasible at this  time, due to the long computational
  times needed to  model a well estbalished  slow spiral-in \citep[see
    \S2,  also  for  a  description of  the  computational  timescales
    see][]{NandezIvanova15}.  }

We stress in this Section that  usually under the ``CE ejection'' term
only  complete  ejection  ---  the  one that  leads  to  naked  binary
formation ---  is considered.  Such  complete ejection is  an implicit
assumption of the energy budget  formalism.  However, it is important,
that even when a CEE fails to eject the entire envelope, a significant
fraction of the mass of the envelope is lost during plunge-in ($\delta
M_{\rm ej}^{\rm plunge}$),  and that during the slow  spiral-in the CE
mass will decrease further.

%% file: sect9_discussion.tex
\section{Discussion}

\label{sec:disc}

We discussed how to average different quantities obtained in 3D CEE
simulations to produce information useful for 1D studies.  As could
have been expected, we find that the asymmetry of the profiles is
extreme during the plunge-in, and is non-negligible during a slow
spiral-in. For example, the angle-dependent deviation of thermodynamic
quantities at the start of the slow spiral-in can reach 50\%. Second,
the typical assumption of spherical symmetry (i.e., the ``thin-shell''
approximation for the companion) underestimates the depth of the
potential well near the orbit.  Finally, the decoupling of the
binary and the envelope during a slow spiral-in, with the formation of
a dense hollow shell around the binary, provides challenges for 1D
codes (where it would appear as a sharp density inversion).

We reviewed the energy balance and discussed how energy conservation
can be treated in 1D codes.  We outlined three problems when one tries
to keep the energy conserved self-consistently in 1D, and provided the
energy conservation equation that can be used in 1D codes during a
slow spiral-in, once the post-plunge-in configuration is known.

We find that most of the angular momentum is quickly lost from the CE
system with the ejecta.  In our simulations, there is no corotation between the binary and
any part of the CE during any phase of a CEE.  In some cases, after
the hollow shell formation, only a few particles could have been left
within a RL of either the core or the companion; only these particles
can be considered as having been in ``corotation''.  During a slow
spiral-in, the angular velocity becomes constant in most (by mass) of
the CE, the spherical symmetry approximation works well, and the value
of the angular velocity is significantly smaller than the local
Keplerian velocity anywhere in the envelope.

The flat profile for the angular velocity we derive from our 3D
simulations differs strongly from the results of typical 1D CE
evolution. In the latter, the region around the binary is always in
corotation, as well as a region inside the binary orbit.  The angular
velocity is then obtained with the diffusion equation, and is found to
decrease steadily with the distance from the binary \cite[e.g.,][note
  however that this was in part by the design of the chosen
  prescriptions in the 1D
  codes]{Taam+1978,1979A&A....78..167M,IvanovaThesis}.  It
can be argued that the difference between the 3D and 1D profiles may
arise from the different ways that viscosity is treated in the two
approaches.  It is hard to evaluate how well artificial viscosity in
SPH matches the convective viscosity \citep[e.g., see discussion in
][]{2014ApJ...786...39N}.  It is clear however that 3D SPH
simulations, when compared to 1D studies, at the same time, are less
efficient in transferring angular momentum to the regions near the
orbit, and are more efficient in the distribution of the angular
momentum throughout the envelope.  Therefore neither too low, nor too
high artificial viscosity in SPH, as compared to what is taken as
viscosity in 1D simulations, can work as an explanation of these
profile differences.  In addition, the significant loss of angular
momentum with the ejecta does not lead to as significant a spin-up
of the envelope as 1D codes predict.  Frictional heating from the
differential rotation is therefore not expected for most of the
envelope, except during the start of the slow spiral-in, when the
angular velocity profile quickly flattens.

In addition, the evolution of the entropy of the CE material differs
between 3D simulations and 1D results.  In 3D, there is no entropy
generation within the orbit during the plunge-in.  While we can see
entropy generation near the surface and a small ``entropy bubble''
formation, the entropy increase is much smaller than what 1D studies
predict --- 3D simulations predict a 30\% increase, while 1D results
predict a factor of a several increase. In 1D, the entropy is generated
because the energy was added as heat.

We have identified four types of ejection processes: the initial
ejection, the plunge-in outflow, the recombination runaway outflow,
and the shell-triggered ejection.  The initial ejection and the
plunge-in ejection take place in all the CEEs we have considered,
including those that end up with a merger.  We provide a simple way to
find the mass of the initial ejecta. The prompt plunge-in ejection
carries away substantially more mass, but there is no easy way to
estimate the magnitude of this ejection.  The shell-triggered ejection
takes place during the slow spiral-in and is caused by the hollow
shell fallback; this provides another prompt ejection process as part
of the CE. The recombination runaway outflow starts during a slow
spiral-in, once the expanded envelope is cooled down to start hydrogen
recombination above $r_{\rm rec,H}$.  In this case, the rest of the
envelope can be removed within several dynamical timescales of the
expanded envelope.  Since the radius at which the recombination energy
release overcomes the potential well depends on the entropy of the
material, the entropy generation observed in 1D codes will likely
predict different outcomes of 1D CE evolution. The combination of a
difference in the entropy profile and the entropy's effect on
recombination runaway was likely the reason why recombination runaway
found in recent simplified 1D studies of CEEs
\citep{2015MNRAS.447.2181I} have happened in models where either a
lot of heat was added to the entire envelope, or less heat was added,
but in a region that was more confined to the bottom of the envelope (thus
keeping the upper envelope's entropy unchanged).

The slow timescales of recombination runaway can be up to several
hundreds day, and are getting longer as the mass of the companion
decreases. For these cases, 3D simulations are no longer
self-consistent, as this is a timescale on which radiative losses can
become important, and hence 1D codes must be used. Note that the
envelope is not stationary at any moment, and instead exhibits
``stationary'' outflows.  Similar recombination driven outflows were
considered for the case of simplified hydrogen envelopes
\citep{1976AZh....53..742B}, where several possible modes of
instabilities were discussed.  A recent study of simplified CEE in 1D
has shown that the CE is prone to dynamical instabilities
\citep{2015MNRAS.447.2181I}, and a runaway recombination was found,
but a connection to steady recombination outflows was not made then.
The energetic consequences for CEE outcomes in the case of a stationary
outflow instead of a prompt dynamical ejection were discussed in
\cite{Ivach11}, but, further studies of the instabilities
of envelopes with steady outflows powered by recombination
are highly needed.

Using the tools developed for this paper, we inspected the models presented in \cite{NandezIvanova15}, where all of the modelled
CEEs resulted in binary formation. There, the models were only analyzed for their final states --
  binaries parameters and energy taken away by the ejected mass. We find that the length of the
envelope removal via recombination runaway is increasing with the
potential well of the donor. The donor that we consider now had the
longest envelope removal time (700 days) out of all the models
considered in \cite{NandezIvanova15} (the next longest timescale of 250 days occurred
in a model that used a $1.6~M_\odot$ red giant with a $0.32~M_\odot$
core).  In the current study, we find that the ejection timescale
increases as the companion mass decreases.  This implies that the
model that provides the timescales and plateau luminosities of the
Luminous Red Novae powered by CEEs \citep[as proposed
  in][]{Ivanova+2013Science}, in its future developments, should take
into account CEE recombination runaway features, such as the initial
envelope binding energy and the mass ratio of the companions.
 
For a  proper treatment  of a self-regulated  spiral-in, one  needs to
know the orbit  at which the companion slowed down its fall, the mass
that remained in  the envelope after the companion's  plunge, how much
angular  momentum remained  in the  envelope, and  how much  energy was
carried away by the ejecta. This  can be done only in conjunction with
the preliminary 3D  simulations that are performed until  at least the
end  of the  plunge-in.  For  the cases  when it  is not  possible, we
outline several important points:

\begin{itemize}
\item  The  plunge-in  takes  place  on a  timescale comparable  to  a  freefall
  timescale.  No energy conservation during the plunge can be properly
  treated in 1D. Instead a CE structure should be constructed assuming
  almost adiabatic envelope expansion as a result of the plunge.
\item No heat should be added to the envelope --- we stress that it is
  important to produce an adiabatic envelope expansion.  Despite
  changing the energy of the envelope by the same amount when the heat
  is added, or when the mechanical energy is added, the change of the
  CE material after these two ways to add energy alters which envelope
  layers would start recombination. We strongly recommend the use of kinetic
    energy injection instead of heat injection.
\item An effort should be given to  model a hollow shell outside
  of the inner binary in 1D.
\item The envelope unavoidably loses a substantial fraction of its
  mass.  While the minimum (initial) ejecta can already be estimated,
  future studies are needed to determine how to estimate mass involved
  in the plunge-in ejection.
\item The expanded envelope should be checked for the condition of 
  recombination runaway,  as in this case  no long-term self-regulated
  spiral-in can take place. All 1D codes should be prepared to detect and
  treat steady outflows.
\item Since most of the  angular momentum is lost with the ejected
  material, the magnitude  of the angular velocity in  the envelope is
  likely to be very low. Hence it is rather insignificant to affect either
  internal structure or provide frictional heating.
\end{itemize}

In this  study, we  only considered the 
specific case of a CEE where the donor star is a low-mass giant, and
all the conclusions are made for this case of the donor.
While we may expect that the characteristic behaviors we found will also occur in CEEs
with more massive  donors,  future studies and comparisons of
3D simulations with 1D  studies using other donors,  especially more
massive donors, are highly needed.

%% file: appendix_am.tex
\appendix 

\section{Angular momentum in 3D codes}

\label{ap:am}

The angular momentum of a particle $k$ in SPH is 
\begin{equation}
 \mathbf{L}_k=\mathbf{r}_{k}\times(m_k\mathbf{v}_{k})=L_{x,k}\hat{x}+L_{y,k}\hat{y}+L_{z,k}\hat{z},
\end{equation}
where the components of the angular momentum for $x$, $y$, and $z$ are given by
\begin{eqnarray*}
 L_{x,k}&=&m_k(y_kv_{z,k}-z_kv_{y,k}),\\
 L_{y,k}&=&m_k(z_kv_{x,k}-x_kv_{z,k}),\\
 L_{z,k}&=&m_k(x_kv_{y,k}-y_kv_{x,k}),
\end{eqnarray*}
respectively. The total angular momentum for the system is,
\begin{equation}
 \mathbf{L}=\sum_k^N \mathbf{L}_k,
\end{equation}
and its magnitude is 
\begin{equation}
 L=|\mathbf{L}|=\sqrt{L_x^2+L_y^2+L_z^2} \ , 
 \label{eq:amtotappA}
\end{equation}
\noindent where 
\begin{eqnarray*}
 L_x&=&\sum_k^N m_k(y_kv_{z,k}-z_kv_{y,k}),\\
 L_y&=&\sum_k^N m_k(z_kv_{x,k}-x_kv_{z,k}),\\
 L_z&=&\sum_k^N m_k(x_kv_{y,k}-y_kv_{x,k}).
\end{eqnarray*}

The angular momentum (by its components) around a particular point $p$ can be calculated as,
\begin{eqnarray*}
 L_{x,k}^p&=&m_k\left[(y_k-y^p)(v_{z,k}-v_{z}^p)-(z_k-z^p)(v_{y,k}-v_{y}^p)\right],\\
 L_{y,k}^p&=&m_k\left[(z_k-z^p)(v_{x,k}-v_x^p)-(x_k-x^p)(v_{z,k}-v_z^p)\right],\\
 L_{z,k}^p&=&m_k\left[(x_k-x^p)(v_{y,k}-v_y^p)-(y_k-y^p)(v_{x,k}-v_x^p)\right].
\end{eqnarray*}